\documentclass[aps,reprint,floatfix,groupedaddress,showpacs]{revtex4-2}
\usepackage{natbib}
\usepackage{amsfonts,amsmath,amssymb}
\usepackage{bm}
\usepackage{pgf}
\usepackage{dcolumn}
\graphicspath{{./mpl_plots/}}

\begin{document}
\title{Cluster Percolation in the Two-Dimensional Ising Spin Glass}
\author{L. M\"unster}
\email{lambert.muenster@physik.tu-chemnitz.de}
\author{M. Weigel}
\email{martin.weigel@physik.tu-chemnitz.de}
\affiliation{
Institut f\"ur Physik, Technische Universit\"at Chemnitz, 09107 Chemnitz, Germany
}

\date{\today}
\begin{abstract}
Suitable cluster definitions have allowed researchers to describe many ordering transitions in spin systems as geometric phenomena related to percolation.For spin glasses and some other systems with quenched
  disorder, however, such a connection has not been fully established, and the numerical evidence remains incomplete. Here we use Monte Carlo simulations to study the percolation properties of several classes of clusters occurring in the Edwards-Anderson Ising spin-glass model in two dimensions. The Fortuin-Kasteleyn--Coniglio-Klein clusters originally defined for the ferromagnetic problem do percolate at a temperature that remains non-zero in the thermodynamic limit. On the Nishimori line, this location is accurately predicted by an argument due to Yamaguchi. More relevant for the spin-glass transition are clusters defined on the basis of the overlap of several replicas. We show that various such cluster types have percolation thresholds that shift to lower temperature by increasing the system size, in agreement with the zero-temperature spin-glass transition in two dimensions. The overlap is linked to the difference in density of the two largest clusters, thus supporting a picture where the spin-glass transition corresponds to an emergent density difference of the two largest clusters inside the percolating phase.
\end{abstract}
\maketitle

\section{Introduction}\label{sec:introduction}
Cluster representations and droplet models provide a framework to study critical phenomena from the geometrical perspective of percolation \cite{StaufferAharony1994IntroductionToPercolationTheory,Grimmett2004BookTheRandomClusterModel,SahimiHunt2021BookComplexMediaAndPercolationTheory}. For the Ising ferromagnet the most prominent schemes are the Fortuin-Kasteleyn cluster representation \cite{FortuinKasteleyn1972OnTheRandomClusterModel1,Grimmett2004BookTheRandomClusterModel} as well as a microscopic definition of Fisher droplets \cite{Fisher1967TheTheoryOfCondensationAndTheCriticalPoint,Fisher1967TheTheoryOfCondensationAndTheCriticalPoint} introduced by Coniglio and Klein \cite{ConiglioKlein1980ClustersAndCriticalDroplets,ConiglioFierro2021CorrelatedPercolation}. Both approaches eventually lead to the same cluster definition. These Fortuin-Kasteleyn--Coniglio-Klein (FKCK) clusters represent thermal fluctuations, the percolation temperature is equivalent to the critical temperature of the ferromagnetic phase transition and the critical exponents of this thermal transition are linked to those of the percolation transition \cite{Grimmett2004BookTheRandomClusterModel,ConiglioFierro2021CorrelatedPercolation}. Furthermore, such clusters unveil interesting properties of the problem that are not accessible from the free energy of the Ising model \cite{ConiglioFierro2021CorrelatedPercolation}. Apart from these physical aspects, FKCK clusters are also the basis of powerful Monte Carlo cluster methods such as the Swendsen-Wang algorithm \cite{SwendsenWang1987NonuniversalCriticalDynamicsInMonteCarloSimulations,SwendsenWang1990ClusterMonteCarloAlgorithms} which dramatically reduces the critical slowing down observed in the vicinity of the transition that affects simulations with  purely local update schemes.

In contrast to the case of the Ising ferromagnet, FKCK clusters in models with frustration such as spin glasses \cite{BinderYoung1986SpinGlasses,NewmanStein2013SpinGlassesAndComplexity,MezardParisiVirasoro1987SpinGlassTheoryAndBeyond,BolthausenBovier2007SpinGlasses} do not have an obvious physical meaning \cite{DeArcangelis1991PercolationTransitionInSpinGlasses}, which is a consequence of the fact that by construction the growth of such clusters signals the increase of ferro/antiferromagnetic correlations and not ordering of the spin-glass nature. Other types of clusters may hence show more interesting behavior when studying such frustrated systems. The order parameter of the spin-glass transition, the overlap, is defined with respect to two replicas. Therefore, it seems natural to consider cluster definitions which include multiple replicas. Here we investigate in detail three different types of two-replica clusters, each of which can be linked to the overlap. The simplest of them just groups together spin sites with identical value of the overlap. These clusters are the basis of the Houdayer cluster algorithm \cite{Houdayer2001ClusterMonteCarloAlgorithmFor2DimensionalSpinGlasses}, which is why we denote them as Houdayer clusters. 
A more elaborate cluster definition can be extracted from a graphical representation initially proposed by Chayes, Machta and Redner for spin systems in external fields \cite{CMR1998GraphicalRepresentationsForIsingSystemsInExternalFields,CMR1998GraphicalRepresentationsAndClusterAlgorithmsForCriticalPointsWithFields} that also allows for a cluster representation of the Ising spin-glass model \cite{MachtaNewmanStein2008ThePercolationSignatureOfTheSpinGlassTransition}. Due to an additional connection to another cluster algorithm for (dilute) spin glasses proposed by J\"org \cite{Joerg2005ClusterMonteCarloAlgorithmsForDilutedSG}, we refer to these clusters as Chayes-Machta-Redner-J\"org (CMRJ) clusters. And thirdly, we study a cluster definition which was introduced by Newman and Stein as a generalization of the FKCK clusters to more than one replica \cite{NewmanStein2007ShortRangeSpinGlassesResultsAndSpeculations}. These structures we denote as two-replica FKCK clusters.

The CMRJ and two-replica FKCK clusters were studied numerically for the three-dimensional Ising spin glass and analytically for the Sherrington-Kirkpatrick (SK) model which corresponds to the mean-field limit of spin glasses \cite{MachtaNewmanStein2008ThePercolationSignatureOfTheSpinGlassTransition}. In three dimensions, the CMRJ and two-replica FKCK clusters are found to percolate at a temperature above the spin-glass transition, while the spin-glass transition itself can be related to an emergent density difference of the two largest clusters \cite{MachtaNewmanStein2008ThePercolationSignatureOfTheSpinGlassTransition}. In contrast, in two dimensions the spin-glass transition occurs at zero-temperature, so it will be interesting to see how the percolation behavior changes in this scenario. A better understanding of cluster structures in spin glasses might help to develop more efficient cluster Monte Carlo algorithms for these systems, hence bringing equilibrium studies of larger systems into the reach of numerical simulation methods \cite{KumarEtALMassivelyParallelSimulationsForDisorderedSystems} which are urgently needed due to the strong finite-size effects in spin-glass physics \cite{HasenbuschEtAl2008TheCriticalBehaviorOf3DIsingGlassModels}.

The rest of this paper is organized  as follows. In Sec.~\ref{sec:model_and_numerical_methods} the Ising spin-glass model is introduced and the simulation methods are described. In Sec.~\ref{sec:fkck_percolation_transition} we report on our study of the FKCK clusters in a spin glass with an interaction distribution that is symmetric around zero as well as on the Nishimori line. In particular, we compare our results to a prediction regarding the exact location of the transition point~\cite{Yamaguchi2013ConjecturedExactPercolationThresholdsOfTheFortuinKasteleynClusters}. In the following Sections we present our numerical results for the two-replica cluster definitions, namely for the CMRJ clusters in Sec.~\ref{sec:CMRJ_clusters}, the two-replica FKCK clusters in Sec.~\ref{sec:2r_fkck_clusters}, as well as the Houdayer clusters in Sec.~\ref{sec:houdayer_clusters}. Finally, Sec.~\ref{sec:discussion} contains our conclusions.

\section{Model and numerical methods}\label{sec:model_and_numerical_methods}
We consider the two-dimensional Ising spin glass with Gaussian interactions. The Hamiltonian is given by
\begin{align}
H_{\bm{J}}(\bm{S})=-\sum_{\langle \bm{x} , \bm{y} \rangle} J_{\bm{x}\bm{y}} s_{\bm{x}}s_{\bm{y}}\,.
\label{eq:sg_hamiltonian}
\end{align}
The system contains $N$ spins which sit on the sites of a square lattice of linear size $L$, such that $N=L^2$. The sum runs over all nearest-neighbor spin pairs as indicated by the notation $\langle \bm{x} , \bm{y} \rangle$ for lattice vectors $\bm{x}$ and $\bm{y}$. In this study only fully periodic boundary conditions are used. $\bm{S}$ denotes a configuration of Ising spins $s_{\bm{x}}=\pm 1$, i.e., $\bm{S}\in \{ \pm 1 \}^N$. The quenched interactions between the spins are represented by bonds $J_{\bm{x}\bm{y}}$ that are drawn from a Gaussian distribution with standard deviation $\sigma_J$ and mean $J_0$; we write $\bm{J} = \{J_{\bm{x}\bm{y}}\}$ for the coupling realization. As a consequence bonds can be ferromagnetic (positive) or antiferromagnetic (negative). A bond is said to be satisfied if $J_{\bm{x}\bm{y}} s_{\bm{x}}s_{\bm{y}}>0$ and broken if $J_{\bm{x}\bm{y}}s_{\bm{x}}s_{\bm{y}}<0\,$. If there does not exist a spin configuration such that all bonds are satisfied simultaneously, the system is referred to as frustrated. The two-dimensional model undergoes a zero-temperature spin-glass transition \cite{YoungStinchcombe1976RGRenormalizationGroupForSpinGlassesAndDilutedMagnets,BrayMoore1987ScalingTheoryOfTheOrderedPhaseOfSpinGlasses,HartmannYoung2001LowerCriticalDimensionOfIsingSpinGlasses,HoudayerHartmann2004LowTemperatureBehaviorOfTwoDimensionalGaussianIsingSpinGlasses,KatzgraberLeeYoung2004CorrelationLengthOftheTwoDimensionalIsingSpinGlassWithGaussianInteractions,FernandezEtAl2016UniversalCriticalBehaviorOfTheTwoDimensionalIsingSpinGlass}. The order parameter of the spin-glass transition, the Parisi overlap parameter~\cite{BinderYoung1986SpinGlasses}, is defined with respect to two replicas $\bm{S}^{(1)}$ and $\bm{S}^{(2)}$,
\begin{align}
	q(\bm{S}^{(1)},\bm{S}^{(2)})=\frac{1}{N} \sum_{\bm{x}}s_{\bm{x}}^{(1)} s_{\bm{x}}^{(2)}=\frac{1}{N} \sum_{\bm{x}} q_{\bm{x}} \, ,
\end{align}
where $q_{\bm{x}}=s_{\bm{x}}^{(1)} s_{\bm{x}}^{(2)}$. Replicas are spin configurations $\bm{S}^{(1)},\bm{S}^{(2)},...$ of a system at the same inverse temperature $\beta$ which evolve independently in time but share the same realization of bonds. In the high-temperature phase the absolute value of the overlap approaches zero and its distribution in the thermodynamic limit is a delta peak at the origin. At low temperature spins have the tendency to point in a direction such that the bonds are satisfied in a similar way in different replicas. As a consequence, below the spin-glass temperature the distribution of the absolute value of the overlap has a mean which is larger than zero. 

\begin{figure}
	\begin{center}
		\includegraphics{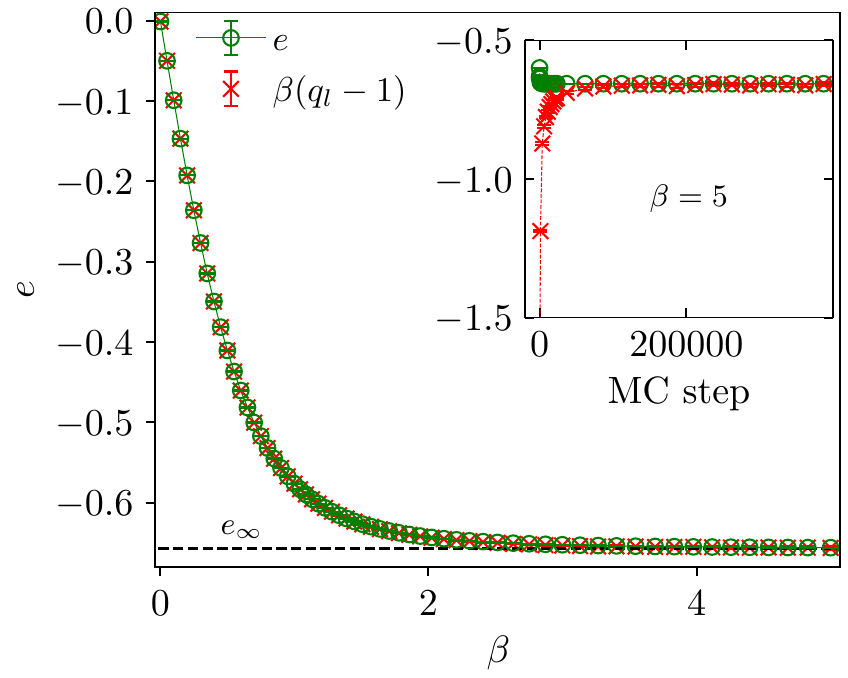}
		\caption{Internal energy per bond, $e$, of the 2D Ising spin glass of Eq.~\eqref{eq:sg_hamiltonian} as a function of inverse temperature $\beta$ at system size $L=128$, averaged over 500 realizations of the bonds. The data for the shifted and scaled link overlap $\beta(q_l-1)$ demonstrates that the equilibration condition of Eq.~\eqref{eq:energy_and_link_overlap} is fulfilled within error bars at all temperatures. The inset shows the convergence to equilibrium at the lowest temperature ($\beta = 5$) for an average over 50 disorder realizations. To ensure that the system is in equilibrium we start sampling after $4\times 10^{5}$ MC steps. The dotted line illustrates the value of the ground state energy $e_\infty=-0.657\,393\,8(4)$ according to Ref.~\cite{KhoshbakhtWeigel2018DomainWallExcitationsInTheTwoDimensionalIsingSpinGlass}. }
		\label{fig:equlibration}
	\end{center}
\end{figure}

To study the model at a range of temperatures we employ Monte Carlo simulation techniques. To obtain reliable numerical results it is important to ensure that the system is in equilibrium. As a reliable indicator to signal equilibration we use a relation that was established in Ref.~\cite{KatzgraberPalassiniYoung2001MonteCarloSimulationsOfSpinGlassesAtLowTemperatures} for short-range spin glasses. It is based on the fact that mathematically speaking Eq.\,\eqref{eq:sg_hamiltonian} defines a Gaussian 
variable with a covariance that is proportional to the link overlap, $q_l$, i.e.
\begin{equation}
[ H_{\bm{J}}(\bm{S}^{(1)})H_{\bm{J}}(\bm{S}^{(2)}) ]_J=N_b q_l(\bm{S}^{(1)},\bm{S}^{(2)})
\label{eq:covariance_and_link_overlap}
\end{equation}
for $J_0=0$ and $\sigma_J=1$ \cite{Contucci2003ReplicaEquivalenceInTheEaModel}. $N_b$ is the number of bonds and $[\ldots]_J$ denotes the disorder average with respect to the bond distribution. The link overlap is given by
\begin{equation}
	q_l(\bm{S}^{(1)},\bm{S}^{(2)})=\frac{1}{N_b}\sum_{\langle \bm{x} , \bm{y} \rangle} s_{\bm{x}}^{(1)} s_{\bm{y}}^{(1)}s_{\bm{x}}^{(2)} s_{\bm{y}}^{(2)} \,.
\end{equation}
This connection between covariance of the Hamiltonian and the link overlap has important implications. One consequence is that the energy per bond can be expressed in terms of the link overlap \cite{KatzgraberPalassiniYoung2001MonteCarloSimulationsOfSpinGlassesAtLowTemperatures,Contucci2003ReplicaEquivalenceInTheEaModel}   
\begin{align}
	e=\beta(q_l-1)\, , 
	\label{eq:energy_and_link_overlap}
\end{align}
where $e=[ \langle H_{\bm{J}}(\bm{S}) \rangle_S ]_J/N_b$~\footnote{Note that similar properties can be derived for the SK-model, but in case of the SK-model the covariance is given by the square power of the overlap \cite{BrayMoore1980SomeObservationsOnTheMeanFieldTheoryOfSG,ContucciEtAl2003ThermodynamicalLimitForCorrelatedGaussianRandomEnergyModels}.}. Keep in mind that here it is necessary to consider the disorder average with respect to the bond distribution $[\ldots]_J$ as well as the configurational average with respect to the Boltzmann distribution $\langle\ldots\rangle_S$.
When the system is initialized randomly, the energy decreases during the equilibration process and the value of the link overlap increases as is shown in the inset of Fig.~\ref{fig:equlibration}, and in equilibrium Eq.~\eqref{eq:energy_and_link_overlap} is satisfied. Note that as demonstrated by Contucci \emph{et al.} \cite{ContucciGiardina2005SpinGlassStochasticStability,Contucci2003ReplicaEquivalenceInTheEaModel,Contucci2005FactorizationPropertiesInTheThreeDimensionalEAModel,Contucci2006OverlapEquivalenceInTheEAModel}, it is possible to derive further important properties from relation \eqref{eq:covariance_and_link_overlap}, which underlines the significance of the link overlap in short-range spin glasses \cite{NewmanStein2007LocalVsGlobalVariablesForSpinGlasses}.

In order to determine the overlap, we simulate in parallel two replicas at each temperature. A Monte Carlo step of our procedure consists of four individual parts. A single spin-flip Metropolis sweep, an FKCK cluster move with a Wolff update \cite{Wolff1988CollectiveMCUpdatingForSpinSystems}, and alternatingly a Houdayer \cite{Houdayer2001ClusterMonteCarloAlgorithmFor2DimensionalSpinGlasses} or a Jörg cluster move \cite{Joerg2005ClusterMonteCarloAlgorithmsForDilutedSG} at each even or odd Monte Carlo time step, respectively. For the latter we use a Swendsen-Wang type update rule \cite{Wolff1988CollectiveMCUpdatingForSpinSystems} which ensures that also smaller clusters are flipped in the percolating phase. At the end of each Monte Carlo step we perform an exchange Monte Carlo move \cite{SwendsenWang1986replicaReplicaMonteCarloSimulationOfSpinGlasses} of replicas at neighboring temperatures. The details of the different cluster moves will be described in the Sections below. While the replica-exchange component is mandatory to achieve equilibration, the specific mix of different cluster and single-spin flip moves chosen here is empirically found to perform well, but we do not claim that this is the optimal protocol for the problem. Figure~\ref{fig:equlibration} demonstrates that the energy monotonously decreases with a declining slope when lowering the temperature. At the lowest considered temperature the energy $e(\beta=5)=-0.655\,91(13)$ at system size $L=128$ is already close to that of the ground state of the infinite system, $e_\infty=-0.657\,393\,8(4)$ \cite{KhoshbakhtWeigel2018DomainWallExcitationsInTheTwoDimensionalIsingSpinGlass}. 

To the extent that self-averaging is present in the model, it is possible to reduce the number of bond realizations to compute the disorder averages by increasing the system size. At the FKCK percolation transition the Wolff cluster updates are extremely effective and we equilibrate 2000 bond realizations
for the largest considered system size, $L=512$, and up to 12\,500 bond realizations for the smallest system size, $L=64$. In the lower temperature region down to $\beta=5$ the equilibration process needs much more time. In this case we use between 500 bond realizations for the largest system size, $L=128$, and 7000 bond realizations for the smallest size $L=16$.

\section{The FKCK percolation transition}\label{sec:fkck_percolation_transition}

In this Section we introduce essential observables which characterize percolation using the example of the FKCK percolation transition in the two-dimensional Ising spin glass with Gaussian interactions. Furthermore we numerically test a prediction of Yamaguchi \cite{Yamaguchi2010PercolationThresholdsOfTheFortuinKasteleynClusterForTheEAIsingModelOnComplexNetworks,Yamaguchi2013ConjecturedExactPercolationThresholdsOfTheFortuinKasteleynClusters} who derived a critical temperature for the percolation transition on the Nishimori line.  

\begin{figure}
	\begin{center}
		\includegraphics{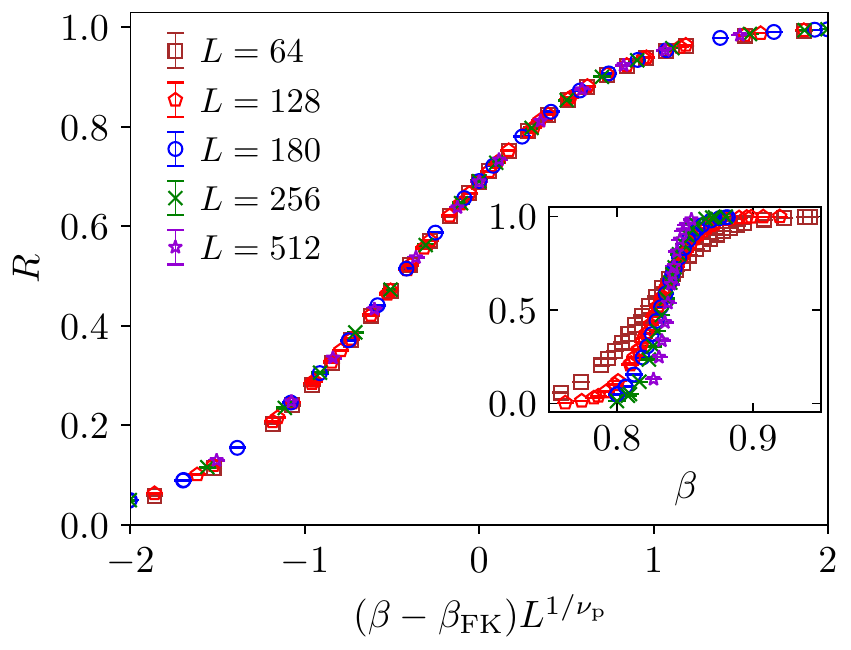}
		\caption{Scaling of the wrapping probability in case of the FKCK percolation transition of the standard 2D Ising spin glass with $J_0=0$ and $\sigma_J=1$. The data collapse is achieved according to Eq.~\eqref{eq:FSSPercolationProb} with $\beta_{\text{FK}}=0.84079(17)$ and $1/\nu_{\mathrm{p}}=0.749(4)$. The inset shows the unscaled data for the different system sizes.}
		\label{fig:FK1FssR}
	\end{center}
\end{figure}

In standard random-bond percolation all bonds are occupied independently with the same probability $p_{\bm{x}\bm{y}}=p$. In the correlated FKCK percolation problem, on the other hand, bonds are occupied with probability 
\begin{align}
	p_{\bm{x}\bm{y}}=
	\begin{cases}
	1-e^{- 2 \beta \vert J_{\bm{x}\bm{y}} \vert} &\text{if }J_{\bm{x}\bm{y}} s_{\bm{x}}  s_{\bm{y}} > 0   \\
	0&\text{else}
	\end{cases}.
	\label{eq:fkck_occ_probability}
\end{align}
Thus, the occupation probability depends on the spin configuration. 
Occupied bonds connect spin sites which group together in clusters. The smallest possible cluster contains a single spin site. These clusters are denoted in the following as FKCK clusters. Depending on the literature they are also called CK droplets \cite{ConiglioKlein1980ClustersAndCriticalDroplets} or FK clusters \cite{FortuinKasteleyn1972OnTheRandomClusterModel1}. Starting from this it is possible to define cluster updates. Flipping each cluster with probability $\tfrac{1}{2}$ corresponds to the Swendsen-Wang update rule \cite{SwendsenWang1990ClusterMonteCarloAlgorithms}. Constructing only one cluster from a randomly chosen seed site by adding bonds with the probability given in Eq.\,\eqref{eq:fkck_occ_probability} and always flipping it, gives the Wolff update rule \cite{Wolff1988CollectiveMCUpdatingForSpinSystems,Wolff1989ComparisonBetweenClusterMCAlgorithmsInTheIsingModel}. By flipping a cluster we mean that each spin inside the cluster is reversed in sign. Both, the Swendsen-Wang as well as the Wolff cluster update define ergodic Monte Carlo algorithms that satisfy the detailed balance condition with respect to the Boltzmann distribution \cite{NewmanBarkema1999MonteCarloMethodsInStatisticalPhysics}.   

\begin{figure}
	\begin{center}
		\includegraphics{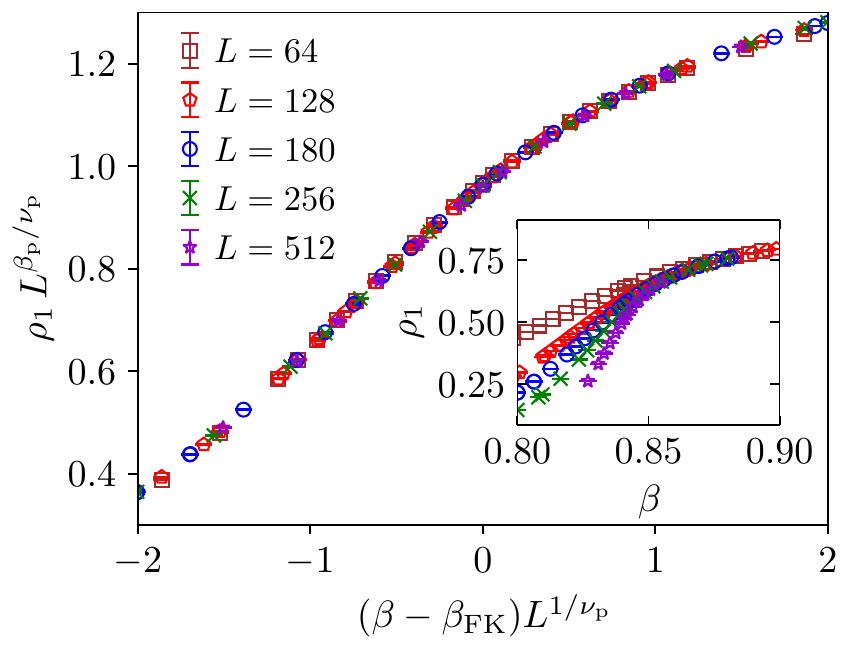}
		\caption{The density of the largest cluster as a function of inverse temperature for the FKCK percolation transition of the standard 2D Ising spin glass with $J_0=0$ and $\sigma_J=1$. The main plot shows a data collapse according to Eq.~\eqref{eq:FSSFractionInLargestCluster} with $\beta_{\text{FK}}=0.8411(5)$, $1/\nu_{\mathrm{p}}=0.754(18)$ and $\beta_\mathrm{p}/\nu_{\mathrm{p}}=0.101(6)$. The inset shows the unscaled data for the different system sizes.}
		\label{fig:Fk1FssP}
	\end{center}
\end{figure}

In case of the Ising ferromagnet the probability of two spins pointing in the same direction at two different lattice sites, $\bm{x}$, $\bm{y}$, is equal to the probability that the two lattice sites are connected by a path of occupied bonds, 
\begin{align}
    \langle s_{\bm{x}} s_{\bm{y}} \rangle_S &= \mathrm{Prob}(\text{$\bm{x}$ and $\bm{y}$ are connected} \nonumber \\ 
    &~~~~~~~~~~~~\text{by occupied bonds}) \, .
    \label{eq:correlation_connectivity}
\end{align}
In other words, the two-spin correlation function is equal to the pair-connectness function of the FKCK percolation problem \cite{Grimmett2004BookTheRandomClusterModel,ConiglioFierro2021CorrelatedPercolation}.
Thus, the  percolation transition must coincide with the ferromagnetic phase transition. Due to frustration this relation is absent in spin glasses \cite{ConiglioEtAl1991ClusterApproachToSpinGlassesAndTheFrustratedPercolationProblem,CataudellaEtAL1994CriticalClustersAndEfficientDynamicsForFrustratedSpinModels} and the percolation transition has no obvious physical interpretation. (Note that these clusters are areas of satisfied bonds and not of parallel spins, so they do not represent ferromagnetic order.) It occurs at finite temperature for dimension $d\geq2$ \cite{DeSantis1999BondPercolationInFrustratedSystems} in the vicinity of the dynamic damage spreading transition~\cite{LundowCampbell2012FortuinKasteleynAndDamageSpreadingTransitionsInRandomBondIsingLattices}. Furthermore, it takes place at a higher temperature than the spin-glass transition~\cite{ConiglioEtAl1991ClusterApproachToSpinGlassesAndTheFrustratedPercolationProblem,DeArcangelis1991PercolationTransitionInSpinGlasses} and shows the characteristics of a random (uncorrelated) percolation transition~\cite{DeArcangelis1991PercolationTransitionInSpinGlasses}. Just above the temperature $1/\beta_{\text{FK}}$, a single giant cluster begins to form, the so-called incipient infinite cluster. Here, $\beta_{\text{FK}}$ denotes the inverse temperature at which percolation emerges in an infinitely large system. In the numerically studied finite systems with fully periodic boundaries the condition of percolation is satisfied if there exits a path of connected occupied bonds that wraps around the boundary in the horizontal direction, in the vertical direction, or in both directions. The probability $R$ for such wrapping to occur thus provides information about the location of the percolation transition. Finite-size scaling (FSS) implies that it behaves as~\cite{StaufferAharony1994IntroductionToPercolationTheory}
\begin{align}
	R(\beta,L)=f_R \left( \left( \beta-\beta_{\text{FK}} \right) L^{1/\nu_\mathrm{p}} \right)
	\label{eq:FSSPercolationProb}
\end{align}
close to criticality, where $f_R$ is a scaling function. 
The order parameter of the percolation transition can be defined as the density of the largest cluster $\rho_1$ (often denoted as $\rho_{\infty}$) which is given by the fraction of sites in the cluster that contains the most spin sites. Inside the percolating phase the largest cluster corresponds to the infinite cluster for $L\to \infty$. The density of the largest cluster satisfies the scaling form~\cite{StaufferAharony1994IntroductionToPercolationTheory}
\begin{align}
	\rho_1(\beta,L)= L^{-\beta_{\mathrm{p}}/\nu_{\mathrm{p}}}f_{\rho_1} \left( \left( \beta-\beta_{\text{FK}} \right) L^{1/\nu_{\mathrm{p}}} \right).
	\label{eq:FSSFractionInLargestCluster}
\end{align}
In the Ising ferromagnet this observable corresponds to the absolute value of the magnetization per site \cite{ConiglioFierro2021CorrelatedPercolation}. Another quantity of interest is the mean cluster size \cite{ConiglioFierro2021CorrelatedPercolation}, 
\begin{align}
	\chi_{\mathrm{p}}=\sum_{s=1} s^2 n(s) \, .
\end{align}
In the paramagnetic phase of the Ising ferromagnet this quantity is equal to the magnetic susceptibility \cite{ConiglioFierro2021CorrelatedPercolation}. The sum runs over all sizes (masses) of clusters, $s$, except the infinite cluster; $n(s)$ is the cluster number per site which equals the average number of clusters of size $s$ divided by $N$. 
At the critical point the mean cluster size diverges as $\chi_{\mathrm{p}}(\beta_{\text{FK}})\sim L^{\gamma_{\mathrm{p}}/\nu_{\mathrm{p}}}$, described by the exponent $\gamma_{\mathrm{p}}/\nu_{\mathrm{p}}$ \cite{Binder2010FSSOfPercolation}.

\begin{table}[tb]
    \renewcommand{\arraystretch}{1.25}
	\begin{tabular}{|c|c|c|c|c|c|}
		\hline
		$J_0$ & $\sigma_J$ & $1/\nu_{\mathrm{p}}$ &$\beta_{\mathrm{p}}/\nu_{\mathrm{p}}$ & $\gamma_{\mathrm{p}}/\nu_{\mathrm{p}}$ & $\beta_{\text{FK}}$ \\
		\hline 
		 $0$ &$1$ &   $0.749(4)$ & $0.101(6)$ &$1.7920(8)$ & $0.84079(17)$  \\
		\hline
		$\mathrm{erf}^{-1}(1/2)\sqrt{2}$& $1$	& $0.750(7)$  & $0.102(5)$ &$1.7904(7)$ &$0.67447(8)$\\ 
		\hline
		\hline
		\multicolumn{2}{|c|}{bond percolation} &  $3/4$ & $5/48$ & $43/24$ & $p_{\text{th}}=1/2$  \\
		\multicolumn{2}{|c|}{(numerical)} & $0.75$ & $0.1041 \overline{6}$  & $1.791\overline{6}$ &$p_{\text{th}}=0.5$ \\
		\hline
	\end{tabular}
	\caption{Overview of the critical exponents and inverse percolation temperatures in case of the FKCK percolation transition. The table shows the results for the standard Ising spin glass $J_0=0$, $\sigma_J=1$ as well as for $J_0=\mathrm{erf}^{-1}(1/2)\sqrt{2}$, $\sigma_J=1$. In the bottom line we show the exact values for random bond percolation in two dimensions. $p_{\text{th}}$ denotes the percolation threshold. The data collapse of the FSS analysis is performed with the tool given in \cite{Melchert2009Autoscale}.}
 \label{tab:Fk1YamaguchiCriticalExponents}
\end{table}

In case of the standard Ising spin glass with Gaussian interactions, $J_0=0$ and $\sigma_J=1$, we perform Monte Carlo simulations and measure the observables
$R$, $\rho_1$ and $\chi_{\mathrm{p}}$ of the FKCK clusters. The data collapses of $R$ and $\rho_1$ are illustrated in Figs.~\ref{fig:FK1FssR} and \ref{fig:Fk1FssP}, respectively. The results for the critical exponents are $1/\nu_{\mathrm{p}}=0.749(4)$ and $\beta_{\mathrm{p}}/\nu_{\mathrm{p}}=0.101(6)$. The estimated critical temperature of the percolation transition, extracted from the data of $R$, is $\beta_{\text{FK}}=0.84079(17)$. To extract $\gamma_{\mathrm{p}}/\nu_{\mathrm{p}}$, we fit a power law to the data of the mean cluster size for different system sizes at inverse temperature $\beta=0.84085$ which corresponds to an inverse temperature of the Monte Carlo simulation between the estimates of $\beta_{\text{FK}}$ from $R$ and $\rho_1$ \footnote{Note that we could have used such estimates from different observables for improving the accuracy of the estimates for $\beta_\mathrm{FK}$ and the critical exponents by taking the cross correlations into account~\cite{weigel:09}, but we found that even without such extra effort the agreement with the percolation universality class is rather clear-cut.}. This leads to the critical exponent$\gamma_{\mathrm{p}}/\nu_{\mathrm{p}}=1.7920(8)$. Note that in order to compute $\chi_{\mathrm{p}}$ we do {\em not\/} exclude the percolating cluster as this was  found to reduce finite-size effects, a phenomenon which was also observed in previous work \cite{JankeSchakel2005FractalStructureOfSpinClustersAndDomainWallsInThe2DIsingModel,AkritidisFytasWeigel2022CorrectionsToScalingInGeometricalClustersOfThe2DIsingModel}. The critical parameters extracted from the data via FSS are collected in Table \ref{tab:Fk1YamaguchiCriticalExponents}. The data collapses of the FSS analysis are performed with the tool provided in Ref.~\cite{Melchert2009Autoscale}. The statistical error is computed by generating 150 bootstrap samples of the data and individually performing a data collapse for each sample. The systematic error is estimated by varying the range of the considered data which is used to perform the collapse from $\left( \beta-\beta_{\text{FK}} \right) L^{1/\nu_{\mathrm{p}}}\in [-0.35,0.35]$ to $\left( \beta-\beta_{\text{FK}} \right) L^{1/\nu_{\mathrm{p}}}\in [-1,1]$. Both, the statistical and the systematic contribution are summed to give the final estimate of the error. 

As is apparent from Table \ref{tab:Fk1YamaguchiCriticalExponents}, the critical exponents are consistent with the universality class of random percolation. The correlations which arise due to the fact that only satisfied bonds can be occupied and the influence of the varying bond strength do not affect the universality class of the percolation transition. This result is expected to be stable also when $J_0\neq 0$ \cite{ImaokaEtAl1997PercolationTransitionIn2DJIsingSG,LundowCampbell2012FortuinKasteleynAndDamageSpreadingTransitionsInRandomBondIsingLattices,FajenHartmannYoung2020percolationPercolationOfFortuinKasteleynClustersForTheRandomBondIsingModel}. 
\begin{figure}
	\begin{center}
		\includegraphics{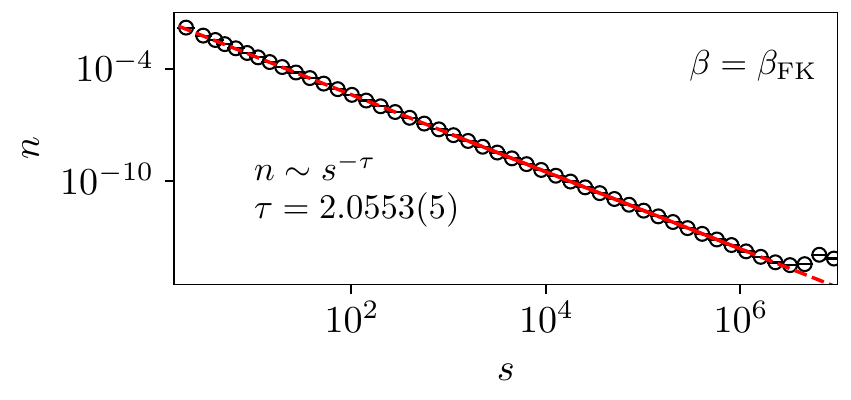}
		\caption{The cluster number $n$ as a function of cluster size $s$ at the percolation threshold on the Nishimori line, $\beta_{\text{FK}}=\mathrm{erf}^{-1}(1/2)\sqrt{2}=0.67448975\ldots$. The system size is $L=4096$. On the Nishimori line the energy is 
			$e(\beta_{\text{FK}})=-J_0$ \cite{Nishimori1981IInternalEnergySpecificHeatAndCorrelationFunctionOftheBondRandomIsingModel}. The simulation gives $e(\beta_{\text{FK}})=-0.674490(11)$, a sign that the system is in equilibrium. The disorder average is computed with respect to 125 bond samples.}
		\label{fig:YamaguchiCriticalClusterSizeDistribution}
	\end{center}
\end{figure}

For the special case of the Nishimori line, the FKCK percolation transition has been thoroughly investigated ~\cite{Yamaguchi2010PercolationThresholdsOfTheFortuinKasteleynClusterForTheEAIsingModelOnComplexNetworks,Yamaguchi2013ConjecturedExactPercolationThresholdsOfTheFortuinKasteleynClusters,Mazza1999GaugeSymmetriesAndPercolationInJIsingSpinGlasses,Gandolfi1999ARemarkOnGaugeSymmetriesInIsingSGs}. The Nishimori line is a certain set in parameter space of $\beta$ and $J_0/\sigma_J$ at which it is possible to analytically compute the internal energy and other quantities via a gauge transformation~\cite{Nishimori1980ExactResultsAndCriticalPropertiesOfTheIsingModelWithCompetingInteractions}. Yamaguchi was able to derive a condition for the critical bond occupation probability of the FKCK clusters~\cite{Yamaguchi2010PercolationThresholdsOfTheFortuinKasteleynClusterForTheEAIsingModelOnComplexNetworks,Yamaguchi2013ConjecturedExactPercolationThresholdsOfTheFortuinKasteleynClusters} which results in a prediction for the critical inverse temperature of the percolation transition,
\begin{align}
	\beta_{\text{FK}}=J_0/\sigma_J=\mathrm{erf}^{-1}(1/2)\sqrt{2}=0.67448975... \, .
	\label{eq:YamaguchiThreshhold}
\end{align}
This critical inverse temperature is exact under the condition that the FKCK percolation transition on the Nishimori line is of the pure random-bond percolation type~\footnote{For the $\pm J$ Ising spin glass it was proven by Gandolfi that FKCK percolation on the Nishimori line reduces to a random-bond percolation problem~\cite{Gandolfi1999ARemarkOnGaugeSymmetriesInIsingSGs}, such that the percolation temperatures predicted by Yamaguchi are exact in this case.}. We numerically test the prediction \eqref{eq:YamaguchiThreshhold} by performing simulations on the Nishimori line with $J_0=\mathrm{erf}^{-1}(1/2)\sqrt{2}$ and $\sigma_J=1$. We carry out the same FSS scaling analysis for the data of $R$, $\rho_1$ and $\chi_{\mathrm{p}}$ at the critical temperature, $\beta=\mathrm{erf}^{-1}(1/2)\sqrt{2}$, as previously discussed for the standard Ising spin glass.  All the results are compiled in the third line of Table \ref{tab:Fk1YamaguchiCriticalExponents}. Our estimate of the critical temperature $\beta_{\text{FK}}=0.67447(8)$ is consistent with the value in  Eq.\,\eqref{eq:YamaguchiThreshhold}.

\begin{figure}
	\begin{center}
		\includegraphics{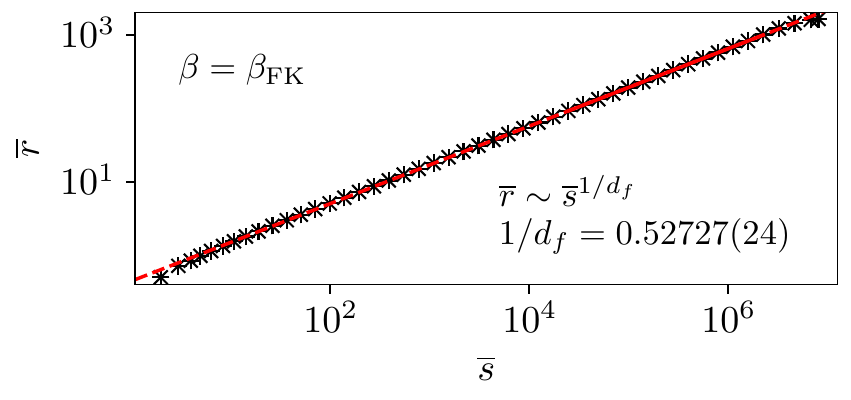}
		\caption{The cluster radius as a function of cluster size on the Nishimori line with $\beta_{\text{FK}}=\mathrm{erf}^{-1}(1/2)\sqrt{2}$.
        For the data analysis we used a logarithmic binning of the cluster sizes. The average size of the clusters within a bin is denoted as $\overline{s}$ and the corresponding average radius is given by $\overline{r}$. The system size is $L=4096$. The disorder average is computed with respect to 125 bond samples.}
		\label{fig:YamaguchiCriticalRadiusDistribution}
	\end{center}
\end{figure}

At the critical point the cluster number $n(s)$ as well as the cluster radius $r(s)$ are expected to follow power laws $n \sim s^{-\tau}$ and $r\sim s^{1/d_f}$, respectively. The radius of a cluster is defined as \cite{StaufferAharony1994IntroductionToPercolationTheory}
\begin{align}
	r(s)=\sqrt{\sum_{k=1}^s \frac{\vert \bm{r}_k-\bm{r}_c  \vert^2}{s} }\, .
\end{align}
The sum corresponds to the average Euclidean distance of all sites $\bm{r}_k$ of the cluster from its center of mass $\bm{r}_c$. For non-percolating clusters the center of mass can simply be derived from the average distance to the origin of the coordinate system. As the origin we chose the root of the cluster. For percolating clusters we used the algorithm proposed in Ref.~\cite{BaiBreen2008CalculatingCenterOfMassInAnUnbounded2DEnvironment}. Figures~\ref{fig:YamaguchiCriticalClusterSizeDistribution} and \ref{fig:YamaguchiCriticalRadiusDistribution} show the cluster number and the cluster radius at $\beta_{\text{FK}}=\mathrm{erf}^{-1}(1/2)\sqrt{2}$, respectively. The values for $\tau=2.0552(4)$ and for the fractal surface dimension $d_f=1.8966(9)$ are consistent with random percolation for which the exact values are $\tau=187/91=2.\overline{054945}$ and $d_f=91/48=1.8958\overline{3}$, respectively. The data numerically verifies the prediction from Yamaguchi of a random-bond percolation transition at $\beta_{\text{FK}}=\mathrm{erf}^{-1}(1/2)\sqrt{2}$. Altogether the results are in agreement with the idea that the FKCK transition in Ising spin glasses belongs to the random-percolation universality class. 

\section{CMRJ Clusters}\label{sec:CMRJ_clusters}
	\begin{figure}
		\begin{center}
			\includegraphics[width=0.99\linewidth]{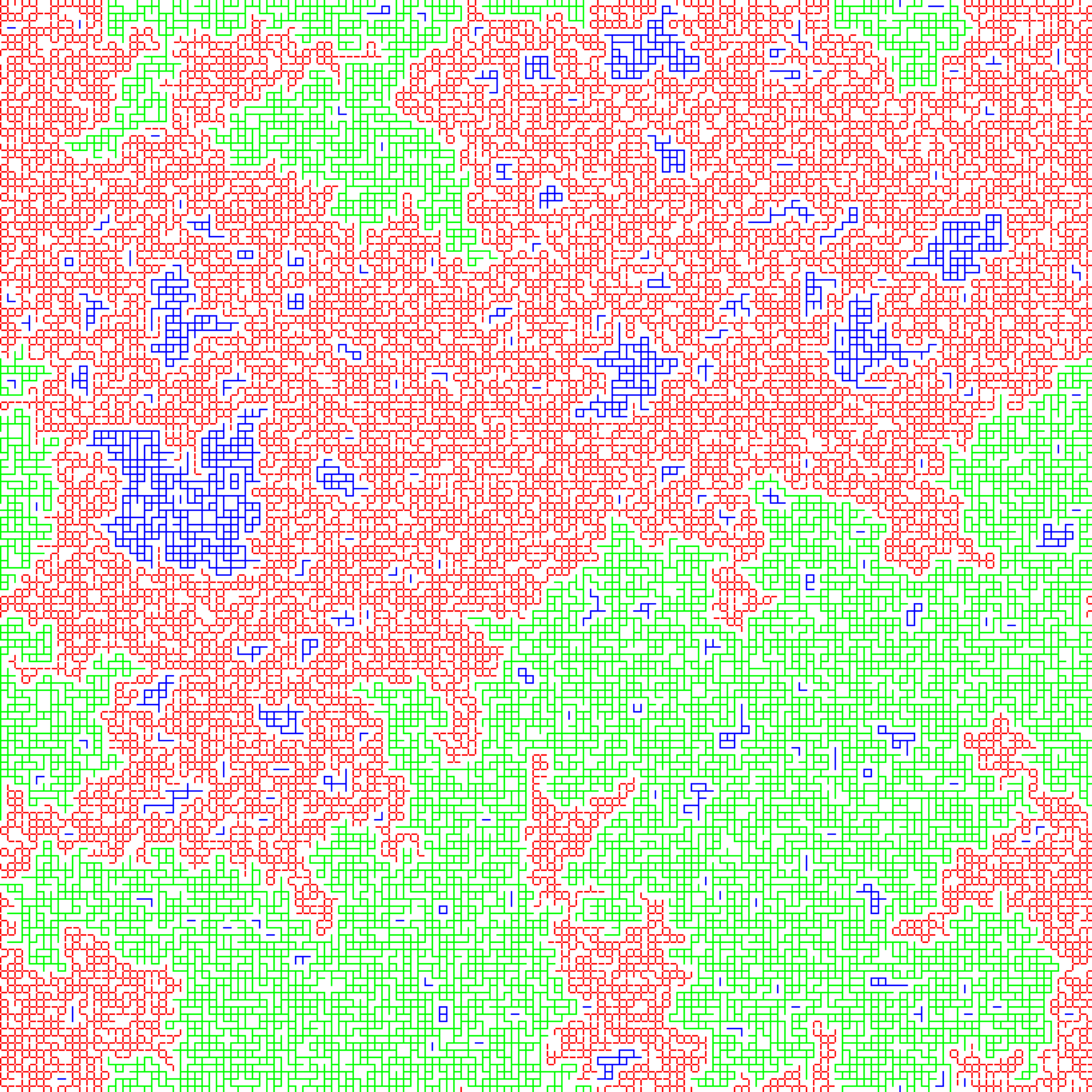}
			\caption{Typical example of CMRJ clusters at $\beta=3\,$ in thermal equilibrium, extracted from a simulation of a $152\times 152$ sample of the Gaussian 2D Ising spin glass using two replicas ($I=2$). The red bonds belong to the largest cluster and and the green bonds to the second largest cluster. Both clusters have an opposite sign of the overlap. The blue bonds are part of smaller clusters. Unoccupied bonds are white.}
			\label{fig:BlueOccupationMap}
		\end{center}
	\end{figure}
	\begin{figure}
		\includegraphics{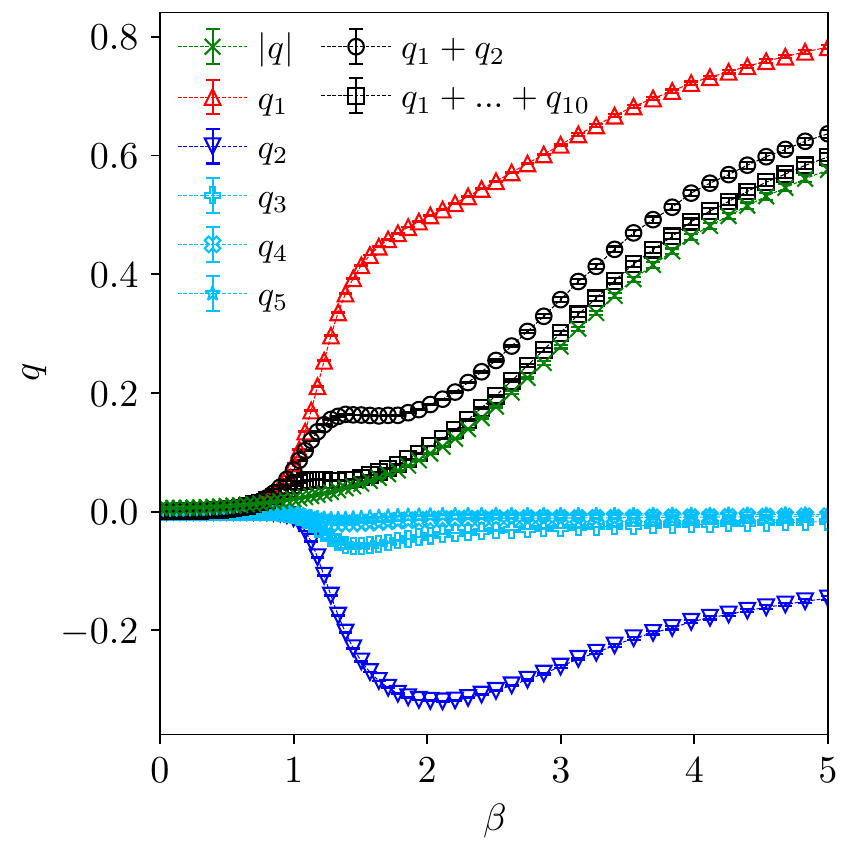}
		\caption{Overlap of the CMRJ clusters at system size $L=128$ averaged over 500 bond realizations. $\vert q \vert$ denotes the average value of the overlap per site which is the order parameter of the spin-glass transition. $q_k$ is the overlap density of the $k$-th largest cluster, corresponding to  the density of the cluster multiplied by the sign of its overlap. The relative spin orientation of the replicas is chosen such that the sign of the overlap of the largest cluster is always positive. We see that the largest and the second largest cluster are anticorrelated with respect to the sign of the overlap. The sum of the overlap densities of the largest clusters approaches $\vert q\vert\,$. Already $q_1+q_2$ is almost parallel to $\vert q \vert$ for $\beta>2\,$.
		}
		\label{fig:BlueClusterOverlap}
	\end{figure} 
	\begin{figure}
		\includegraphics{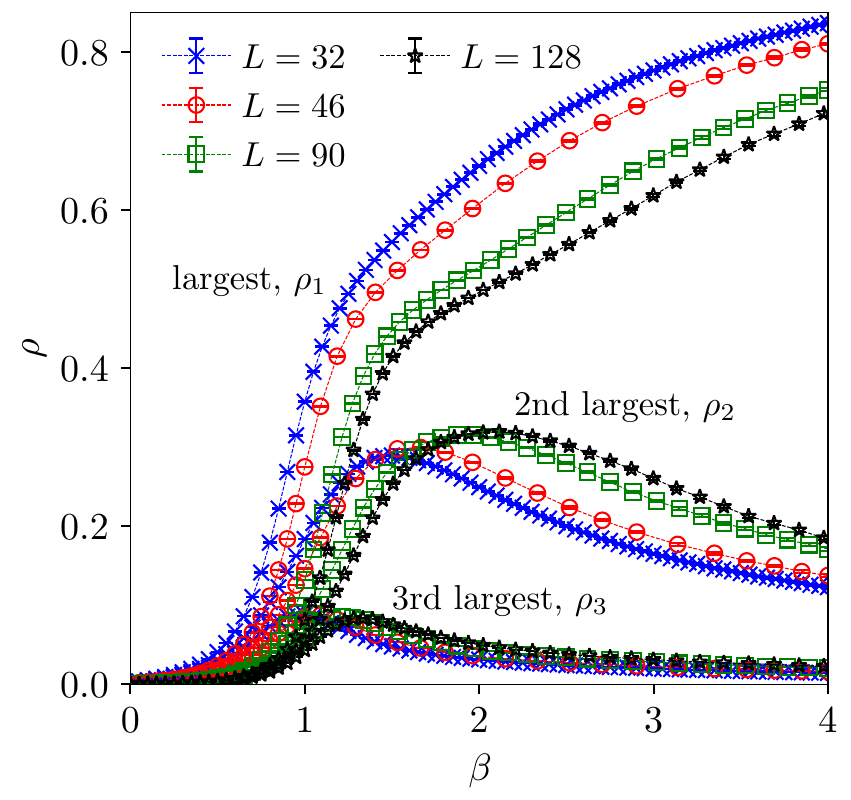}
		\caption{Density $\rho=s/N$ of the three largest CMRJ clusters for different system sizes $L$. The curves shift to lower temperatures when the system size is increased. The height of the peak of the second largest cluster increases whereas the peak of the third largest cluster decreases.}
		\label{fig:BlueClusterPercolation}
	\end{figure} 
	\begin{figure}
		\includegraphics{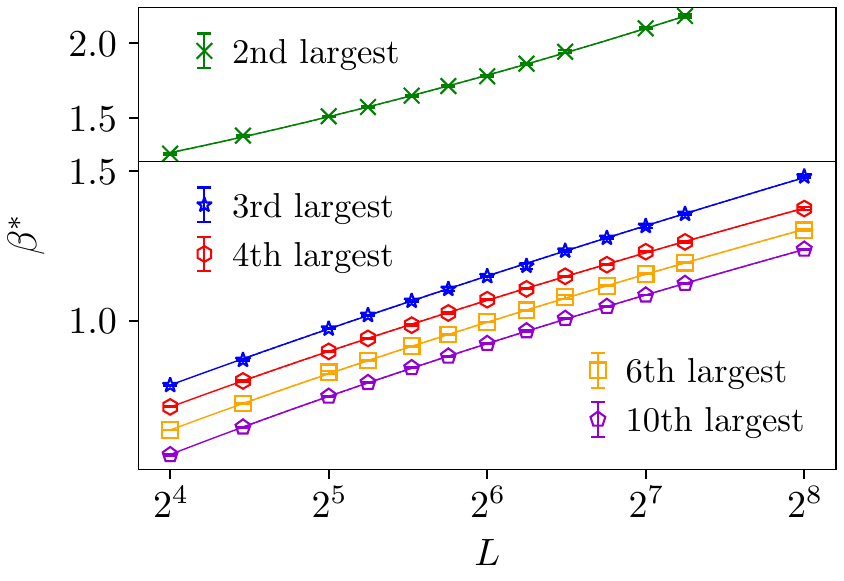}
		\includegraphics{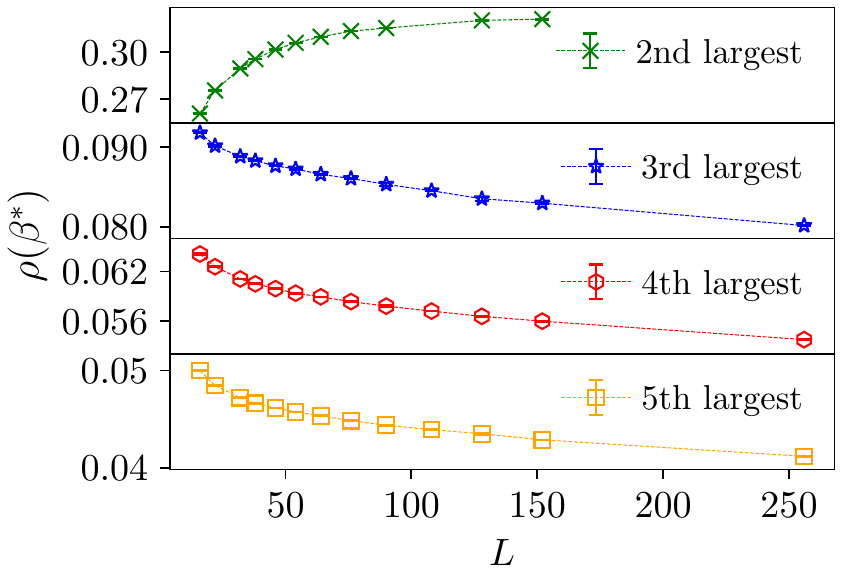}
		\caption{Scaling of the peaks of the second largest and smaller CMRJ clusters. The upper panel shows the location of the peak, $\beta^*$, of the $k$-th largest clusters as a function of system size. The lines are fits of the functional form $\beta^*(L)=c_1\ln(L)^{c_2}+c_3$, where $c_1$,$c_2$, and $c_3$ are fit parameters. The lower panel shows the density of the corresponding clusters at their peak location. The density of the second largest cluster increases in contrast to that of the smaller clusters. The dotted lines are guides to the eye.}
		\label{fig:BlueClusterPeakProperties}
	\end{figure} 
	\begin{figure}
		\includegraphics{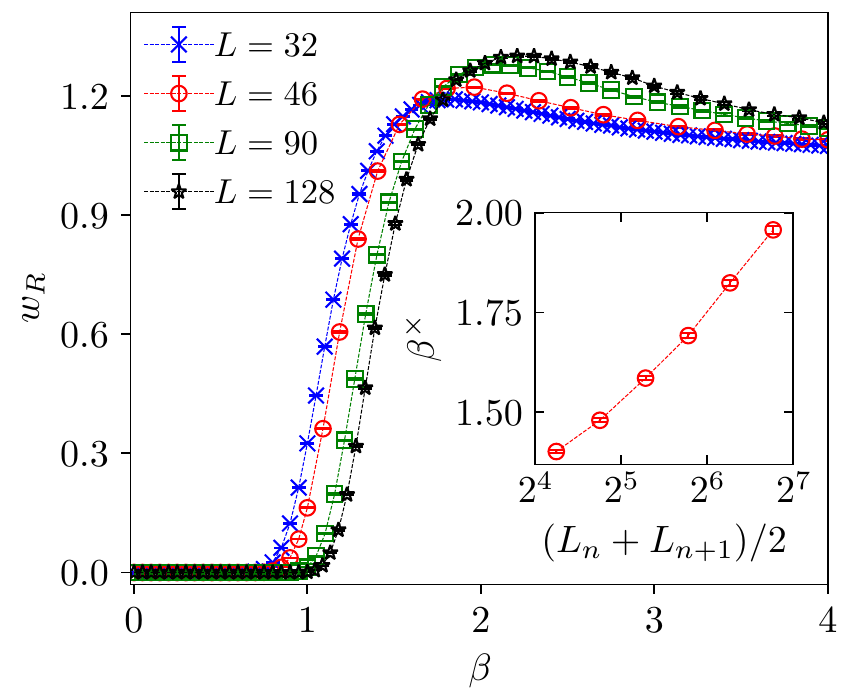}
		\caption{Number of wrapping CMRJ clusters $w_R$ as a function of inverse temperature for different system sizes. The inset shows the crossing points $\beta^\times$ of two neighboring system sizes with $L\in\{16,22,32,46,64,90,128\}$. The lines are guides to the eyes only.}
		\label{fig:BlueNrWrappingClusters}
	\end{figure} 
	\begin{figure}
		\includegraphics{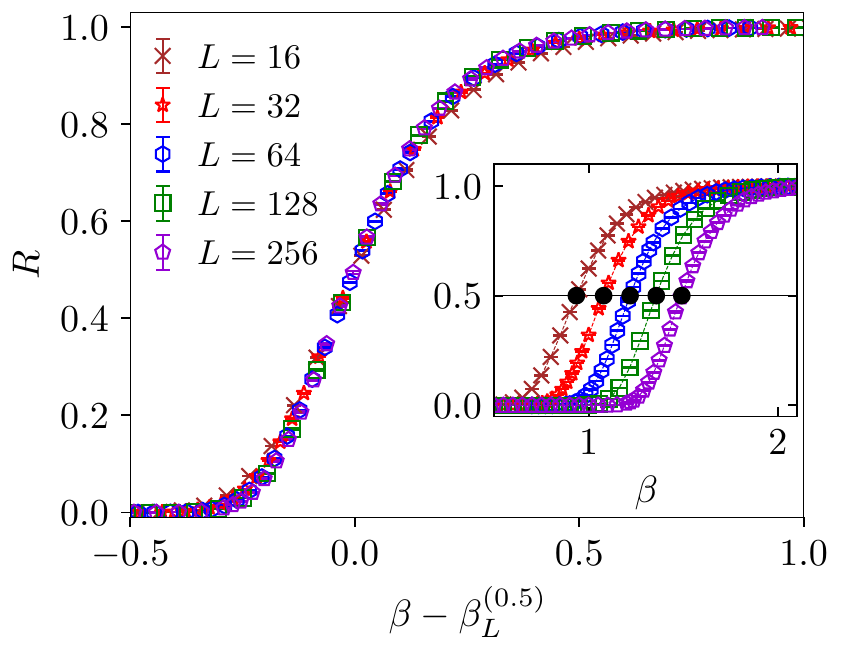}
		\caption{Wrapping probability of the CMRJ clusters for different system sizes. The inset shows the unscaled data. The solid black points correspond to the inverse temperatures at which the wrapping probability is one half, $R(\beta_L^{(0.5)})=0.5$. The main plot shows a collapse of the data onto a single curve obtained by a shift of the data along the $x$-axis by the corresponding value of $\beta_L^{(0.5)}$.}
		\label{fig:BlueClusterWrappingProb}
\end{figure}
\begin{figure}
		\includegraphics{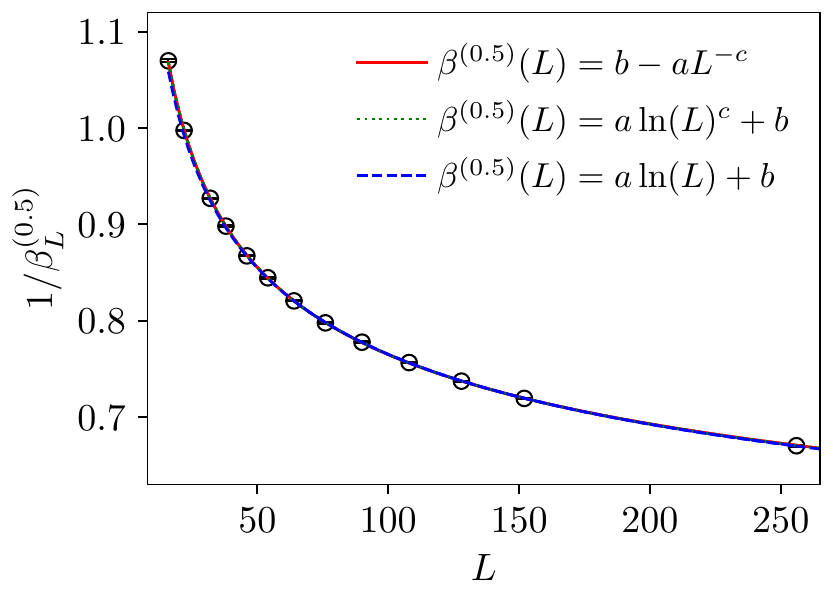}
		\caption{The plot demonstrates how the percolation transition shifts to lower temperature $T=1/\beta$ as the system size is increased.
		The temperature at which the wrapping probability is one half, $1/\beta_L^{(0.5)}$, is plotted as a function of $L$. The lines 
		are fits of different type. The simple logarithmic law $\beta^{(0.5)}(L)=a \ln(L)+b$ is only consistent with the data for large system sizes (see main text).}
		\label{fig:BlueClusterWrappingProbShift}
\end{figure} 

The order parameter of the spin-glass transition, the overlap, is defined with respect to two replicas. It is hence natural to construct clusters related to the spin-glass transition by considering the overlap of several replicas. Such clusters might additionally turn out to be useful for designing effective Monte Carlo cluster algorithms. Proposals along these lines were already put forward rather early on by Swendsen and Wang in Ref.~\cite{SwendsenWang1986replicaReplicaMonteCarloSimulationOfSpinGlasses}. In the following, we investigate the properties of a particular type of multiple-replica clusters. The occupation probability of the clusters for $I$ replicas is given by
\begin{align}
		p_{\bm{x}\bm{y}}&=
		\begin{cases}
			1-\text{exp}{ \left(- 2 \beta J_{\bm{x}\bm{y}} \tilde{s}_{\bm{x}} \tilde{s}_{\bm{y}} \right) } &\text{if}~~J_{\bm{x}\bm{y}} \tilde{s}_{\bm{x}}  \tilde{s}_{\bm{y}}>0  \\
			0~~&\text{else} 
		\end{cases}  \nonumber \\ 
		&~~~\text{ with } \tilde{s}_{\bm{x}} \tilde{s}_{\bm{y}}=\sum_{i=1}^I s_{\bm{x}}^{(i)} s_{\bm{y}}^{(i)}
\end{align}   
where $i$ is the replica index and  $\tilde{s}_{\bm{x}}=(s_{\bm{x}}^{(1)},...,s_{\bm{x}}^{(I)} )$ denotes an $I$-component spin. This is a generalization of the cluster definition proposed by Chayes, Machta and Redner~\cite{CMR1998GraphicalRepresentationsForIsingSystemsInExternalFields,CMR1998GraphicalRepresentationsAndClusterAlgorithmsForCriticalPointsWithFields,MachtaNewmanStein2008ThePercolationSignatureOfTheSpinGlassTransition} as well as by Jörg~\cite{Joerg2005ClusterMonteCarloAlgorithmsForDilutedSG} to more than two replicas.
The condition $J_{\bm{x}\bm{y}} \tilde{s}_{\bm{x}} \tilde{s}_{\bm{y}}>0$ enforces the constraint that only those bonds are occupied which are satisfied simultaneously in the majority of the replicas~\footnote{Note that this implies that they are satisfied in {\em all\/} (i.e., both) replicas for $I=2$.}. Clusters are defined as spin sites which are connected by a path of occupied bonds. A cluster move can then be realized by flipping all $I$-component spins $\tilde{s}_{\bm{x}}$ within the same cluster, i.e., $\tilde{s}_{\bm{x}}\to -\tilde{s}_{\bm{x}}$ for all $\bm{x}$ inside the cluster. Like for the FKCK clusters, this can be done in a single-cluster or multi-cluster fashion as in the Wolff \cite{Wolff1988CollectiveMCUpdatingForSpinSystems} and Swendsen-Wang \cite{SwendsenWang1987NonuniversalCriticalDynamicsInMonteCarloSimulations} algorithms, respectively. Such a procedure aims to generate equilibrium states according to the $I$-replica Boltzmann distribution
	\begin{align}
		\mathcal{P}(\bm{S}^{(1)},...,\bm{S}^{(I)}|\bm{J} )&=Z_{\bm{J}}^{-I}\exp \left( -\beta H_{\bm{J}}^{(I)}\right)  \nonumber \\
\text{ with } H_{\bm{J}}^{(I)}( \bm{S}^{(1)},...,\bm{S}^{(I)})&=-\sum_{\langle \bm{x}, \bm{y} \rangle} J_{\bm{x}\bm{y}} \tilde{s}_{\bm{x}} \tilde{s}_{\bm{y}}   
		\label{eq:i_replica_boltzmann}
	\end{align}
where $Z_{\bm{J}}$ is the partition function of a single replica for a given realization of the bonds $\bm{J}$.

To check for the convergence of the corresponding Markov chain, we need to investigate (detailed) balance and ergodicity \cite{NewmanBarkema1999MonteCarloMethodsInStatisticalPhysics}. That detailed balance holds with respect to the distribution \eqref{eq:i_replica_boltzmann} can be derived from the cluster surface energy with respect to the $I$-replica Hamiltonian, $H_{\bm{J}}^{(I)}$, that is the sum of the energies of the individual replicas. We provide this derivation in Appendix \ref{app:i_replica_mc}. 
Since during cluster moves the overlap of any two replicas, $q_{\bm{x}}=s_{\bm{x}}^{(i)} s_{\bm{x}}^{(j)}$, $i,j=1,2,...,I$, $i\neq j$, is conserved, such cluster flips are not ergodic, however. In case of $I=2$ the described cluster algorithm corresponds to the method proposed by Jörg \cite{Joerg2005ClusterMonteCarloAlgorithmsForDilutedSG} which has been successfully applied to simulate diluted spin glasses \cite{Joerg2006CriticalBehaviorOfThe3DBondDilutedIsingSG}. Since an equivalent cluster definition does also emerge from the graphical representation of Chayes, Machta and Redner \cite{CMR1998GraphicalRepresentationsForIsingSystemsInExternalFields,CMR1998GraphicalRepresentationsAndClusterAlgorithmsForCriticalPointsWithFields,MachtaNewmanStein2008ThePercolationSignatureOfTheSpinGlassTransition} connected occupied components of this kind are denoted here as CMRJ clusters. In the CMRJ framework the above clusters constructed from ``blue'' bonds that are satisfied in both replicas that are augmented by clusters constructed of singly-satisfied ``red'' bonds to form an overall ergodic cluster update, cf. Ref.~\cite{MachtaNewmanStein2008ThePercolationSignatureOfTheSpinGlassTransition}. In the following we focus on the case of $I=2$ which is, of course, of special significance for spin glasses. This is due to the fact that in this case all spins within the same CMRJ cluster have identical sign of the overlap. It is hence plausible to ask how these clusters are linked to the overlap, especially in connection with the spin-glass transition.
	
Figure~\ref{fig:BlueOccupationMap} illustrates an instance of CMRJ clusters of a sample of the 2D Gaussian spin glass at $\beta=3$ in thermal equilibrium. There are mainly two large clusters. As becomes clear in Fig.~\ref{fig:BlueClusterOverlap}, this is a typical constellation at low temperatures. This plot shows the overlap density of the five largest clusters as well as the mean of the absolute value of the overlap, $\vert q\vert= [ \langle \vert q(\bm{S}^{(1)},\bm{S}^{(2)}) \vert \rangle_S ]_J$. The overlap density of a cluster is defined as the density of a cluster multiplied by the sign of its overlap. The relative orientation of the spins of the replicas is chosen such that the largest cluster has positive overlap. The overlap density of the second largest cluster, $q_2\,$, is anticorrelated in sign with respect to the largest cluster. The sum of the overlap densities $q_1+q_2$ is almost parallel to $\vert q \vert$ for $\beta>2$. This observation supports the idea that the difference in density of the two largest clusters is directly linked to $\vert q\vert$ at low temperatures \cite{MachtaNewmanStein2008ThePercolationSignatureOfTheSpinGlassTransition}. 
In finite systems and for continuous distributions of the interaction, such as the Gaussian distribution considered in this work, there exists a single pair of ground states which are connected by a global spin flip \cite{vaezi:17}. As a consequence, $\vert q \vert$ would further increase by lowering the temperature \cite{HoudayerHartmann2004LowTemperatureBehaviorOfTwoDimensionalGaussianIsingSpinGlasses} and for $\beta\to\infty$ it would approach one as then the system reaches the ground state and all spins are contained in a single CMRJ cluster. In an infinitely large system this does not have to be the case since it cannot be ruled out that there exist infinitely many ground states \cite{ArguinDamron2014OnTheNumberOfGroundStatesOfTheEAModel}. For discrete interaction distributions the situation for $\beta\to\infty$ is more involved since the ground state is degenerate even in finite systems, and a temperature-dependent crossover length between a discrete and an effectively continuous behavior emerges when a careful analysis of the problem is performed \cite{CreightonHuseMiddleton2011ZeroAndLowTemperatureBehaviorOfThe2DPmJIsingSG,ParisenEtAl2011FSSIn2DIsingSG,ParisenEtAl2011FSSIn2DIsingSG,JoergEtAl2006StrongUniversalityAndAlgebraicScalingIn2DIsingSG}.

In Fig.~\ref{fig:BlueClusterPercolation} we show the density of the three largest CMRJ clusters at different system sizes. Again it is observed that the two largest clusters contain most of the spin sites at low temperature. On increasing the system size, the curves shift to lower temperature. The height of the peak of the second largest cluster increases whereas that of the of smaller clusters decrease. This is of some significance since if the contributions of the third largest and smaller clusters become negligible for $L\to \infty$, the overlap approaches the difference in the density of the two largest clusters. A scenario of this type was proposed in Ref.~\cite{MachtaNewmanStein2008ThePercolationSignatureOfTheSpinGlassTransition} which also provides a rigorous proof for the SK model.    

To better understand this aspect, the behavior of the peaks of the densities of the largest clusters are investigated in more detail. The location and height of the peak are extracted from the data by fitting parabolas in the vicinity of the peaks. Error bars are obtained via parabolic fits of 250 bootstrap samples as described in Ref.~\cite{Young2015EverythingYouWantedToKnowAboutDataAnalysis}. The results are depicted in Fig.~\ref{fig:BlueClusterPeakProperties}. The locations $\beta^*$ of the peaks shift to smaller temperatures as the system size is increased. The density of the second largest cluster increases with system size, whereas that of smaller clusters decreases. Therefore, the influence of the smaller clusters diminishes on increasing $L$, which is consistent with the idea that the overlap becomes equal to the difference in density of the two largest clusters in the thermodynamic limit~\cite{MachtaNewmanStein2008ThePercolationSignatureOfTheSpinGlassTransition}.
	
The percolation transition of the CMRJ clusters does not fit the standard template of a random percolation transition. The latter features a single incipient infinite cluster which forms at the transition point, while the former shows two giant clusters that develop a density difference at the spin-glass transition. Therefore, it is interesting to investigate whether both clusters can wrap simultaneously around the boundaries, similar to what is observed in three dimensions \cite{MachtaNewmanStein2008ThePercolationSignatureOfTheSpinGlassTransition}. The data in Fig.~\ref{fig:BlueNrWrappingClusters} demonstrate that this is the case for finite systems, since on average there is more than one wrapping cluster in the temperature region where the system starts to percolate. For $\beta \to \infty$, $w_R$ is expected to approach one since in the unique ground state of a finite system there is only one single giant CMRJ cluster left which contains all spin sites. The inset of Fig.~\ref{fig:BlueNrWrappingClusters} shows the crossing points $\beta^\times$ of $w_R$ of two adjacent system sizes from the list $L \in \{16, 22, 32, 46, 64, 90, 128\}$ which are expected to behave like pseudo-critical temperatures of the problem. The crossing points shift to larger $\beta$ in agreement with the behavior of the peak locations shown in Fig.~\ref{fig:BlueClusterPeakProperties}.
	
An alternative definition of pseudo-critical temperatures results from a consideration of the wrapping probability $R$. The inset of Fig.~\ref{fig:BlueClusterWrappingProb} depicts $R$ for different system sizes. It is observed that for increasing $L$ the curves shift to larger $\beta$ along the $x$-axis. We define pseudo-critical points as the inverse temperatures at which the wrapping probability is one half, i.e., $R(\beta_L^{(0.5)})=0.5$. These values are determined by spline interpolation of the data and their error bars are generated with bootstrapping. The main plot of Fig.~\ref{fig:BlueClusterWrappingProb} shows a data collapse of $R$ which is obtained by shifting the data along the $x$-axis by the values of $\beta_L^{(0.5)}$. 
The estimated values of $\beta_L^{(0.5)}$ are shown individually in Fig.~\ref{fig:BlueClusterWrappingProbShift}. They can be well described by the functional form $\beta^{(0.5)}(L)=a\ln(L)^c+b$ with $a=0.260(13)$ 
$b=0.289(20)$ and $c=0.894(20)$ where the minimal considered system size is $L_{\text{min}}=16$ and the quality of the fit is $Q=0.87$.
A simpler logarithmic law with $c=1$, namely $\beta^{(0.5)}=a\ln(L)+b$, only yields good results for larger system sizes $L_{\text{min}}=64$ with $a=0.1978(10)$, $b=0.396(5)$ and $Q=0.60$. These fits do suggest a zero temperature percolation transition in the thermodynamic limit, since $1/\beta_L^{(0.5)}\to 0$ when $L\to \infty$. Note that a power law of the form $\beta^{(0.5)}(L)=b-aL^{-c}$ can also be fitted to the the data, resulting in $a=8.5(1.5)$, $b=8.8(1.5)$, $c=0.026(6)$ with $L_{\text{min}}=16$ and $Q=0.81$. This would imply a finite-temperature transition \cite{PeiDiVentra2021AFiniteTemperaturePhaseTransitionForTheSpinGlassIn2D} with a critical inverse temperature of $8.8(1.5)$. However, in view of the small value $c=0.026(6)$ of the exponent, a fit of the form $a\ln(L)^c+b$ appears to be much more natural. Additionally, the pure power-law fit strongly depends on $L_{\text{min}}$, such that already for $L_{\text{min}}=46$ the error bars exceed the values of the fit parameters. Finally, the positive curvature in the peak locations of the second largest cluster and in the crossing points of the number of wrapping clusters, see Figs.~\ref{fig:BlueClusterPeakProperties} and \ref{fig:BlueNrWrappingClusters}, rather support the idea of a zero-temperature transition. To sum up, our results are in good agreement with the scenario of a zero-temperature percolation transition.

The spin-glass transition occurs at a temperature below the percolation transition, and it is connected to  
the difference in density of the two largest clusters~\cite{MachtaNewmanStein2008ThePercolationSignatureOfTheSpinGlassTransition}.
Therefore, both above discussed scenarios, a finite-temperature or a zero-temperature percolation-transition, are consistent with the zero-temperature spin-glass transition in two dimensions.
The pseudo-critical temperatures that describe the spin-glass transition shift towards zero temperature with a power-law behavior according to $T_{\mathrm{SG}}(L)\sim L^{-1/\nu}$ \cite{HoudayerHartmann2004LowTemperatureBehaviorOfTwoDimensionalGaussianIsingSpinGlasses,KatzgraberLeeYoung2004CorrelationLengthOftheTwoDimensionalIsingSpinGlassWithGaussianInteractions}, where $1/\nu=0.2793(3)$ \cite{KhoshbakhtWeigel2018DomainWallExcitationsInTheTwoDimensionalIsingSpinGlass}. This is much faster than the asymptotically logarithmic scaling of the  pseudo-critical temperatures of the percolation transition as shown in Fig.~\ref{fig:BlueClusterWrappingProbShift}.     

%
%
%
\section{Two-Replica FKCK clusters}\label{sec:2r_fkck_clusters}
\begin{figure}
		\includegraphics{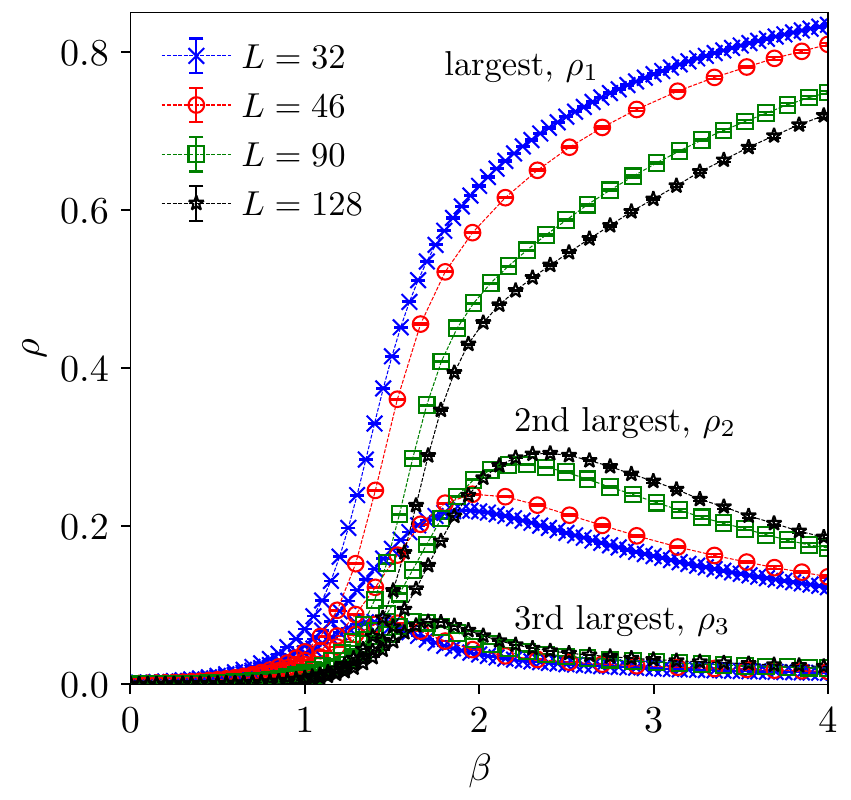}
		\caption{Density of the three largest two-replica FKCK clusters at different system sizes. The curves shift to lower temperature when the system size is increased.}
		\label{fig:2FKClusterPercolation}
	\end{figure} 
	\begin{figure}
		\includegraphics{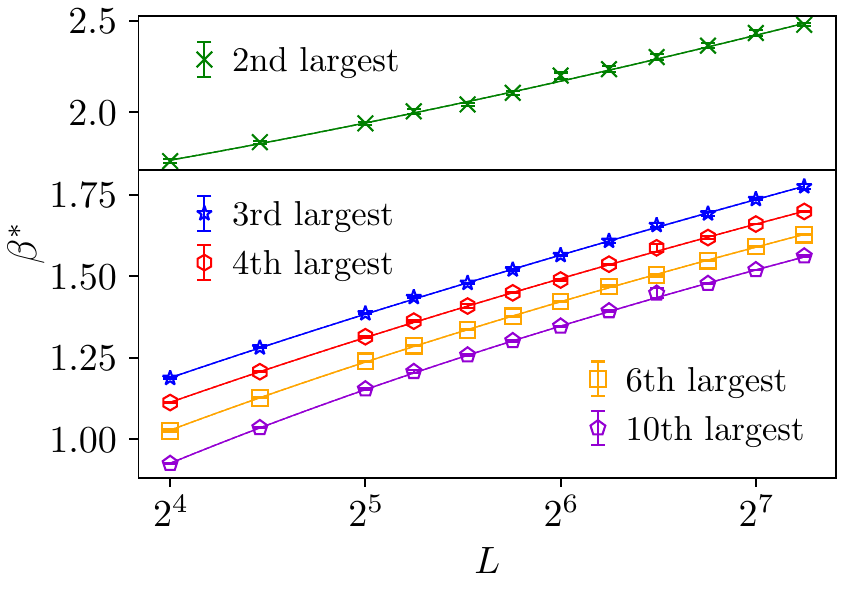}
		\includegraphics{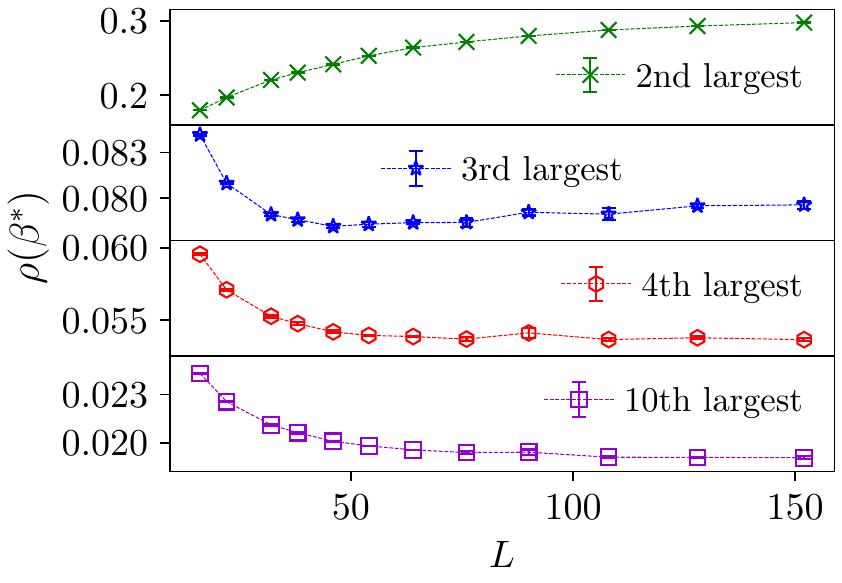}
		\caption{Properties of the peaks of the second largest and smaller two-replica FKCK clusters. The upper panel shows the location of the peak $\beta^*$ of the largest clusters as a function of system size. The lines are fits of the form $\beta^*(L)=c_1 \ln(L)^{c_2}+c_3$ to the data. The lower panel shows the density of the clusters at the peak locations. The density of the second largest cluster increases whereas the density of the smaller clusters show no clear trend for large system sizes.
		}
		\label{fig:2FKClusterPeakProperties}
	\end{figure}
	\begin{figure}
		\includegraphics{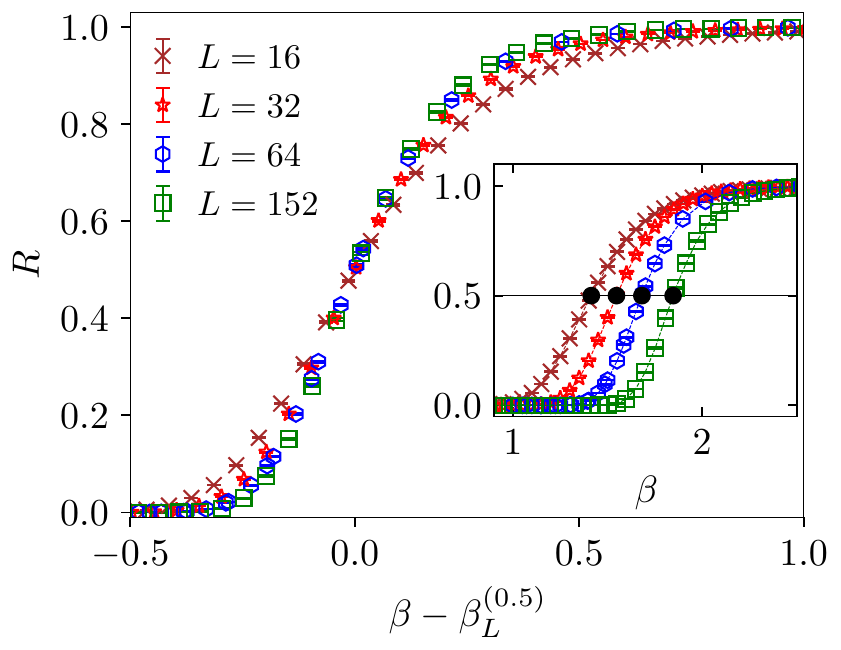}
		\includegraphics{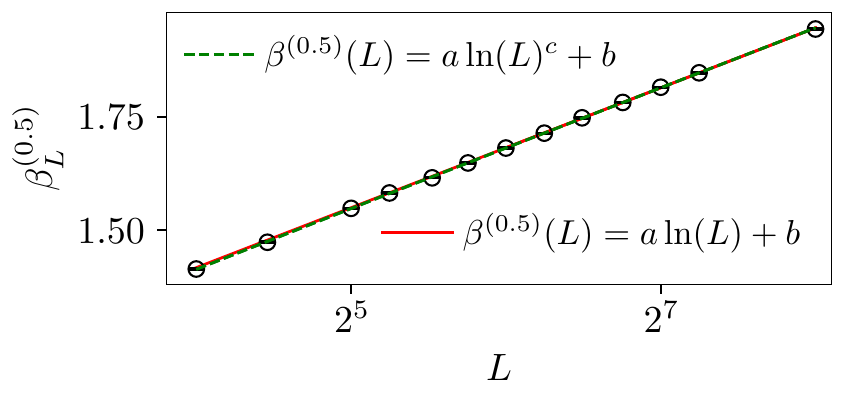}
		\caption{Wrapping probability of the two-replica FKCK clusters for different system size. The upper panel shows the wrapping probability as a function of inverse temperature $\beta$ for a number of the considered system sizes $L$. The inset contains the curves of the unscaled data. In the main plot there is a collapse of the data onto a single curve, which was obtained by a shift of the curves along the $x$-axis by the values of $\beta_L^{(0.5)}$. In the lower panel, these shifts are shown as a function of $L$; they can be described by a logarithmic law of the type $a\ln(L)^c+b$ where $c$ is close to one.}
		\label{fig:2FKClusterWrappingProbShift}
	\end{figure} 
	
	Another approach towards a multiple-replica cluster definition is to connect FKCK clusters of different replicas. A straightforward definition of such clusters is given by the occupation probability       
	\begin{align}
		p_{\bm{x}\bm{y}}&=
		\begin{cases}
			\left(1-e^{-2\beta \vert J_{\bm{x}\bm{y}}\vert }\right)^I~\text{if}~J_{\bm{x}\bm{y}} \tilde{s}_{\bm{x}}  \tilde{s}_{\bm{y}}>0~\text{and}~\vert \tilde{s}_{\bm{x}}  \tilde{s}_{\bm{y}} \vert =I \\
			0~~~\text{else}
		\end{cases}
		\nonumber \\ 
		&~~~\text{ where } \tilde{s}_{\bm{x}} \tilde{s}_{\bm{y}}=\sum_{i=1}^I s_{\bm{x}}^{(i)} s_{\bm{y}}^{(i)} \,.
		\label{eq:i_replica_fkck}
	\end{align}
	This means only those bonds can be occupied which are satisfied in all $I$ replicas simultaneously. The occupation probability is equal to the event 
	that in all $I$ replicas the bond is occupied individually with the probability of the FKCK clusters, see Eq.~\eqref{eq:fkck_occ_probability}. In the present paper, we focus on the scenario of $I=2$ and connected components are denoted as two-replica FKCK clusters~\footnote{Note that using more replicas in Eq.~\eqref{eq:i_replica_fkck} notably reduces the occupation probability and thus lowers the onset temperature for percolation. As a consequence, it might be interesting to see if for a certain $I$ the percolation transition is directly linked to the finite temperature spin-glass transition for $d>2$.}. Spins within the same two-replica FKCK cluster have the same overlap. 
	These two-replica FKCK clusters were initially proposed by Newman and Stein with the aim of providing tools to mathematically show broken spin-flip symmetry in short-range spin glasses at non-zero temperature \cite{NewmanStein2007ShortRangeSpinGlassesResultsAndSpeculations}.
	
	Although there is rather strong numerical evidence for a finite-temperature spin glass phase for $d\geq3$ \cite{BrayMoore1987ScalingTheoryOfTheOrderedPhaseOfSpinGlasses,HartmannYoung2001LowerCriticalDimensionOfIsingSpinGlasses,FernandezEtAl2016UniversalCriticalBehaviorOfTheTwoDimensionalIsingSpinGlass,KatzgraberEtAl2006Universality3DIsingSpinGlasses,BaityJesiEtAl2013CriticalParametersOfThe3DIsingSG} we do not have a rigorous proof. In the case of a ferromagnet it can be shown that the appearance of a unique largest percolating FKCK cluster corresponds to the onset of long-range order and broken symmetry~\cite{ConiglioFierro2021CorrelatedPercolation}. In spin glasses, FKCK percolation is a necessary condition for broken symmetry and the occurrence of a unique largest two-replica FKCK cluster is a sufficient condition \cite{MachtaNewmanStein2009APercolationTheoreticApproachToSGPhaseTransitions}. Furthermore, in the SK model the difference in density of the two largest two-replica FKCK clusters is equal to the overlap and thus directly connected to the spin-glass transition. Note that in case of the SK model this is also true for the previously discussed CMRJ clusters. In the three-dimensional Ising spin glass a similar behavior is expected. Although there is some numerical evidence in favor of this scenario~\cite{MachtaNewmanStein2009APercolationTheoreticApproachToSGPhaseTransitions}, it is not entirely clear if the difference in density of the two largest clusters is precisely equal to the overlap or the contributions of smaller clusters are also relevant. 
    In two dimensions there is no finite-temperature spin glass transition, and it is hence of particular interest to see what happens in this case. 
	
	Figure~\ref{fig:2FKClusterPercolation} shows the density of the three largest two-replica FKCK clusters. In general, the percolation transition has rather similar properties to the CMRJ one. In the vicinity of the transition there are two dominating clusters which can wrap simultaneously around the boundaries (not shown). These two largest clusters are anticorrelated with respect to the sign of the overlap and their difference in density approaches the average overlap at low temperatures. 	
	In Fig.~\ref{fig:2FKClusterPeakProperties} we present the scaling of the peaks in the density of the second largest and smaller clusters. As is shown in the upper panel, the peak positions shift to lower temperatures as the system size is increased. The lower panel visualizes the peak densities of the largest clusters. It is visible that the height of the peak for the second largest cluster increases. The heights of the third largest and smaller clusters do not show a clear trend. Therefore, the evidence for an equality between the difference in density of the two largest clusters and the overlap is less convincing for the two-replica FKCK clusters than in case of the CMRJ clusters, cf.~Fig.~\ref{fig:BlueClusterPeakProperties}.  
	
	To further investigate the properties of two-replica FKCK clusters, we also considered the wrapping probabilities which are shown in Fig.~\ref{fig:2FKClusterWrappingProbShift}. The data show a clear shift along the $x$-axis to lower temperatures as the system size is increased. The shift is well described by a logarithmic law, i.e., $\beta^{(\text{0.5})}(L)=a \ln(L)^c+b$ with $a=0.206(14)$, $b=0.859(22)$, $c=0.972(28)$ and $L_{\text{min}}=16$. If the exponent is fixed to $c=1$ and $L_{\text{min}}=64$ chosen, the result is $a=0.1908(15)$ and $b=0.888(7)$. This is similar to the situation of the CMRJ clusters where the slope is $a=0.1978(10)$. A power-law fit of the form $\beta^{(\text{0.5})}(L)=b-aL^{-c}$ does not yield a satisfactory result.
	
	The shift of the percolation transition towards zero temperature demonstrates that there is no sufficient condition for a broken spin-flip symmetry in two dimensions at finite temperatures. This observation is in agreement with previous numerical studies which show that there is no finite-temperature spin-glass phase in this case \cite{YoungStinchcombe1976RGRenormalizationGroupForSpinGlassesAndDilutedMagnets,BrayMoore1987ScalingTheoryOfTheOrderedPhaseOfSpinGlasses,HartmannYoung2001LowerCriticalDimensionOfIsingSpinGlasses,FernandezEtAl2016UniversalCriticalBehaviorOfTheTwoDimensionalIsingSpinGlass,
		KatzgraberEtAl2006Universality3DIsingSpinGlasses,BaityJesiEtAl2013CriticalParametersOfThe3DIsingSG}. Furthermore, it is consistent with the argument that is not possible to simultaneously have two infinitely large percolating clusters with opposing order parameter in two dimensions in the thermodynamic limit, simply because there is not enough space \cite{MachtaNewmanStein2009APercolationTheoreticApproachToSGPhaseTransitions,ConiglioEtAl1976PercolationAndPhaseTransitionsInTheIsingModel}. Note that the appearance of a single largest percolating CMRJ cluster would also imply broken symmetry \cite{MachtaNewmanStein2009APercolationTheoreticApproachToSGPhaseTransitions}. 
		
\section{Houdayer Clusters}\label{sec:houdayer_clusters}
	\begin{figure}
		\includegraphics{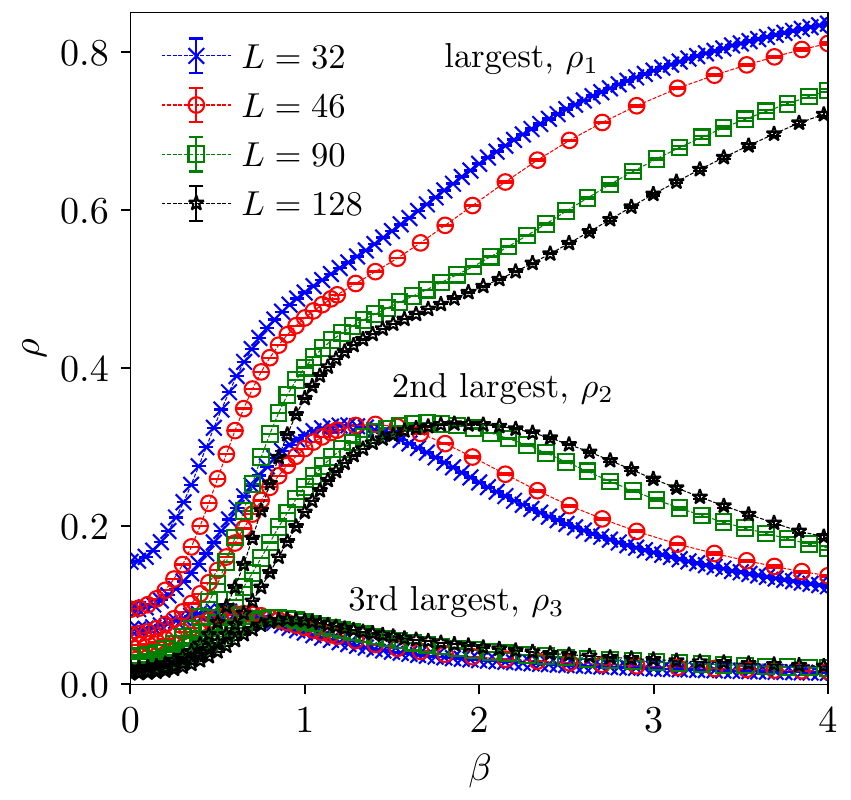}
		\caption{Density of the three largest Houdayer clusters for different system sizes.}
		\label{fig:HoudayerClusterPercolation}
	\end{figure} 
	
	\begin{figure}
		\includegraphics{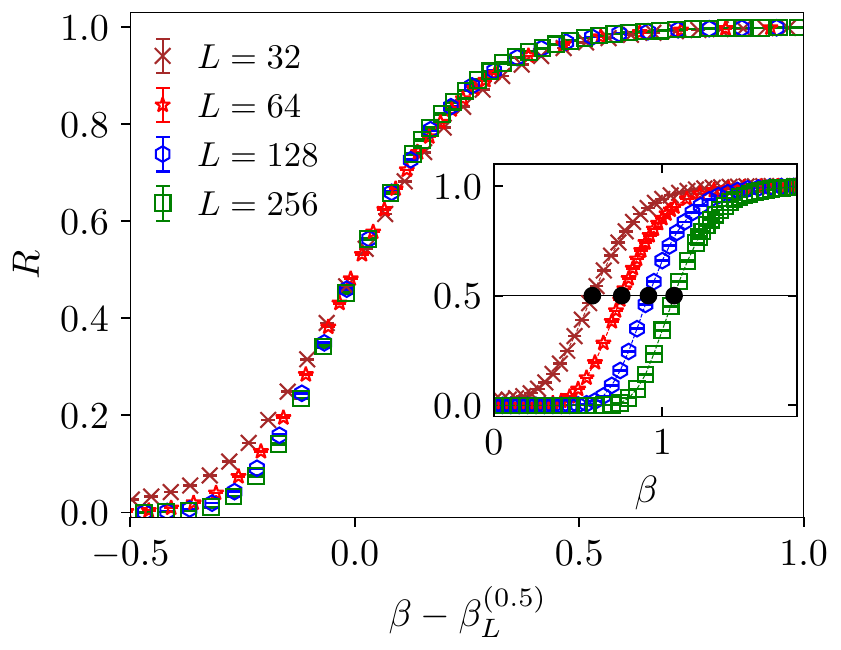}
		\includegraphics{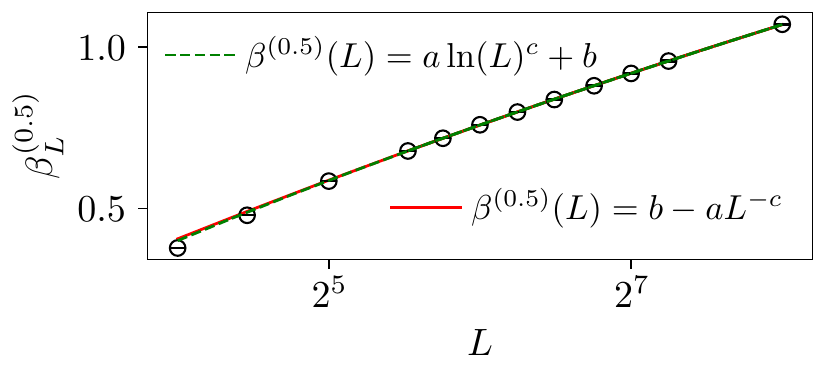}
		\caption{Wrapping probability of Houdayer clusters as a function of system size. The upper panel shows the wrapping probability for various system sizes. The inset contains the original data. In the main plot there is a collapse of the data onto a single curve, which was obtained by a shift of the curves along the $x$-axis by the values of $\beta_L^{(\text{0.5})}$. The lower panel contains the data for the shifts $\beta_L^{(\text{0.5})}$, together with a fit of the functional form $a \ln(L)^c+b$ with $L_{\text{min}}=46$ to the data. 
		}
		\label{fig:HoudayerClusterWrappingProbShift}
	\end{figure} 
	
	Finally, we also considered clusters of spin sites with the same overlap. These geometrical overlap clusters are used in Houdayer's cluster algorithm which allows one to speed up the Monte Carlo simulation of spin glasses, at least in two dimensions~\cite{Houdayer2001ClusterMonteCarloAlgorithmFor2DimensionalSpinGlasses,ZhuOchoaKatzgraber2015EfficientClusterAlgorithmForSpinGlassesInAnySpaceDimension,VandenbroucqueChiacchioMunro2022TheHoudayerAlgorithm}. These clusters are closely analogous to geometrical clusters in case of the Ising ferrogment \cite{JankeSchakel2005FractalStructureOfSpinClustersAndDomainWallsInThe2DIsingModel,akritidis:prep}. The occupation probability is given by 
	\begin{align}
		p_{\bm{x}\bm{y}}&=
		\begin{cases}
			1~~\text{if}~~\vert \tilde{s}_{\bm{x}} \tilde{s}_{\bm{y}} \vert=2 \\
			0~~~\text{else}
		\end{cases} 
	\end{align}
    and the number of replicas is two. Thus, clusters are simply defined as connected components of spin sites with identical overlap. These clusters have a vanishing surface energy within the energy of the $I=2$ Boltzmann distribution, see Eq.~\eqref{eq:i_replica_boltzmann}, since $\tilde{s}_{\bm{x}}\tilde{s}_{\bm{y}}=0$ at the cluster surface. Therefore, flipping the spins in these clusters in both replicas simultaneously, $\tilde{s}_{\bm{x}}\to -\tilde{s}_{\bm{x}}$, does not change the energy and is hence in agreement with the detailed balance condition. Because the two-replica energy and the overlap are conserved quantities, such cluster moves are not ergodic. Note that any two adjacent clusters necessarily have an opposite sign of the overlap, and the sum over all bonds along the surface of the clusters is proportional to the link overlap (up to an additive constant) \cite{NewmanStein2007LocalVsGlobalVariablesForSpinGlasses}.   
	The previously discussed CMRJ and two-replica FKCK clusters are geometric subregions of the Houdayer clusters because each Houdayer cluster includes all connected spin sites with identical overlap, irrespective of whether the underlying bonds are satisfied or not. Therefore, it might be expected that Houdayer clusters show properties similar to those of CMRJ and two-replica FKCK clusters. In fact, Fig.~\ref{fig:HoudayerClusterPercolation} demonstrates that, again, there are mainly two large clusters.  In the vicinity of the percolation transition we again find that there is more than one wrapping cluster (not shown). Since the occupation probability does not depend on temperature, the wrapping probability can already be non-zero in the high-temperature regime. The percolation transition shifts to lower temperature as is shown in Fig.~\ref{fig:HoudayerClusterWrappingProbShift}. The shift is consistent with a fit of the form $\beta^{(\text{0.5})}(L)=a \ln(L)^c+b$ with $a=0.73(12)$, $b=-0.93(15)$ and $c=0.59(6)$ with $Q=0.68$ and $L_{\text{min}}=46$. The data can also be fitted to a power law, $\beta^{(0.5)}(L)=b-aL^{-c}$ with $a=3.85(29)$, $b=3.4(4)$, $c=0.090(12)$, $L_{\text{min}}=46$ and $Q=0.56$. Again the exponent $c=0.090(12)$ is so small that the fit does not contradict a logarithmic law.
	
	\section{Discussion}\label{sec:discussion}
		\begin{figure}
		\includegraphics{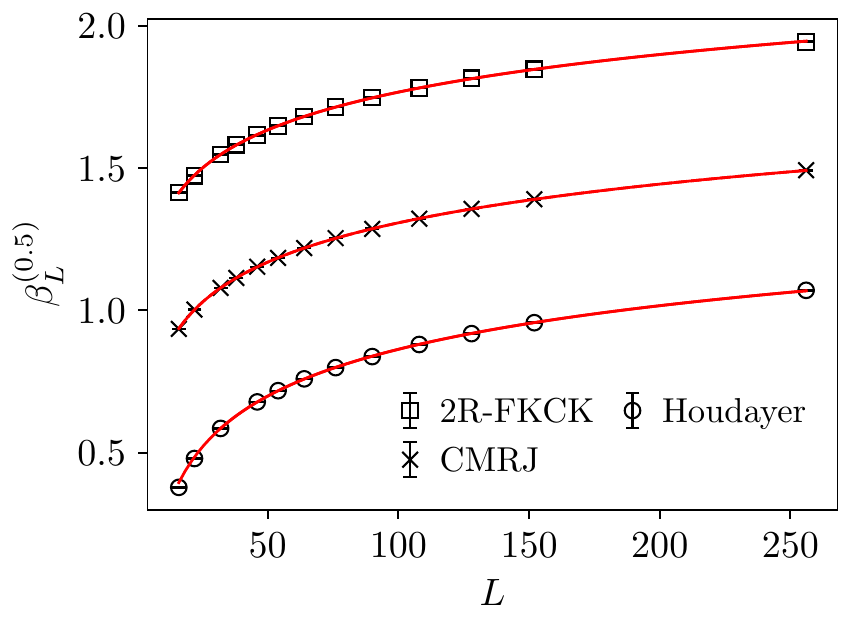}
		\caption{Overview of the shift in the wrapping probabilities for the three cluster types, namely two-replica FKCK (2R-FKCK), CMRJ and Houdayer. The plot shows the value of the inverse temperature at which the wrapping probability is one half $R(\beta_L^{(0.5)})=0.5$. The lines are fits of type $\beta^{(\text{0.5})}(L)=a \ln(L)^c+b$. }
		\label{fig:ClusterWrappingProbShiftTogether}
	\end{figure} 
	
	We have studied percolation properties of the two-dimensional Ising spin glass with Gaussian interactions by performing Monte Carlo simulations. In the Ising ferromagnet Fortuin-Kasteleyn--Coniglio-Klein (FKCK) percolation is directly linked to the thermal phase transition. In spin glasses the FKCK percolation transition occurs at a higher temperature than the spin glass transition, close to the heat-bath damage-spreading transition \cite{LundowCampbell2012FortuinKasteleynAndDamageSpreadingTransitionsInRandomBondIsingLattices}. We find that the critical inverse temerature of the percolation transition is $\beta_{\text{FK}}=0.84085(8)$ in case of the standard spin glass where the mean of the Gaussian interaction distribution is zero, $J_0=0$, and the standard deviation is one, $\sigma_J=1$. The transition belongs to the universality class of random percolation. On the Nishimori line \cite{Nishimori1980ExactResultsAndCriticalPropertiesOfTheIsingModelWithCompetingInteractions}, which is a certain line in the parameter space of $J_0$ and $\sigma_J$, Yamaguchi derived a critical inverse temperature for the FKCK percolation transition \cite{Yamaguchi2010PercolationThresholdsOfTheFortuinKasteleynClusterForTheEAIsingModelOnComplexNetworks,Yamaguchi2013ConjecturedExactPercolationThresholdsOfTheFortuinKasteleynClusters}. His prediction agrees within error bars with the numerical data of our analysis. See also \cite{LundowCampbell2012FortuinKasteleynAndDamageSpreadingTransitionsInRandomBondIsingLattices,Gandolfi1999ARemarkOnGaugeSymmetriesInIsingSGs,Mazza1999GaugeSymmetriesAndPercolationInJIsingSpinGlasses} for the case of a bimodal $\pm J$ interaction distribution.  
	
	The order parameter of the spin-glass transition, the overlap, is defined with respect to two replicas which suggests to study clusters which are defined with respect to two replicas. Thus, three different cluster types are considered, namely Chayes-Machta-Redner-Jörg (CMRJ), 2-replica-FKCK and Houdayer clusters where spins that belong to the same cluster have identical overlap. 
    In all considered cases the pseudo-critical temperature of the percolation transition shifts to lower temperatures as the system size is increased. This is demonstrated in Fig.~\ref{fig:ClusterWrappingProbShiftTogether}. The shift is well described by a functional form of $\beta^{(\text{0.5})}(L)=a \ln(L)^c+b$ as the fits demonstrate. This behavior implies a zero temperature percolation transition in the thermodynamic limit. Furthermore, it is visible that the Houdayer clusters percolate at the highest temperature followed by the CMRJ clusters and the two-replica FKCK clusters at a given system size $L$. This relative order of clusters is a consequence of the fact that the CMRJ clusters as well as the two-replica FKCK clusters are geometric subregions of the Houdayer clusters since there only bonds can be occupied which are satisfied in both replicas. The occupation probability of the CMRJ clusters is higher than that of two-replica FKCK which explains why the CMRJ clusters percolate at a higher temperature than the two-replica FKCK clusters.  

    In all considered cluster definitions percolation is characterized by two large clusters which can simultaneously wrap around the boundaries close to the transition. These two largest clusters are anticorrelated with respect to the sign of the overlap. Furthermore, at low temperatures the mean of the absolute value of the overlap is directly related to the difference in density of the two largest clusters. In case of the CMRJ clusters it is visible that the density of the second largest cluster increases with system size whereas that of the smaller clusters decreases. In this scenario, the mean of the absolute value of the overlap equals the difference in density of the two largest clusters in the thermodynamic limit as proposed and shown for the SK model in Ref.~\cite{MachtaNewmanStein2009APercolationTheoreticApproachToSGPhaseTransitions}. Thus, if smaller clusters are irrelevant, the dominance of the largest cluster over the second largest cluster is directly linked to the spin-glass transition. It is well known that  pseudo-critical temperatures of the spin-glass transition shift to zero temperature in a power-law fashion with $\beta_{\mathrm{SG}}(L)\sim L^{1/\nu}$ \cite{KatzgraberLeeYoung2004CorrelationLengthOftheTwoDimensionalIsingSpinGlassWithGaussianInteractions,HoudayerHartmann2004LowTemperatureBehaviorOfTwoDimensionalGaussianIsingSpinGlasses}, where $1/\nu=0.2793(3)$ \cite{KhoshbakhtWeigel2018DomainWallExcitationsInTheTwoDimensionalIsingSpinGlass}. Since this power law decays much faster than the above discussed logarithmic law, the percolation transitions will always appear at a higher temperature as compared to the effective spin-glass transition at a given system size.      
	
	In case of the Ising ferromagnet FKCK clusters correspond to Fisher droplets and the pair-connectness function of the percolation problem is equal to the two-spin correlation function \cite{ConiglioFierro2021CorrelatedPercolation}. Thus, the percolation transition 
	coincides with the thermal phase transition, which makes Monte Carlo cluster dynamics such as the Swendsen-Wang or Wolff algorithm very efficient close to criticality. This relation is absent in case of the considered two-replica cluster definitions and percolation takes place at a higher temperature than the spin-glass transition. At low temperatures there are mainly two large clusters which contain almost all spins. To identify clusters where the pair-connectedness function is directly related to the overlap-correlation function seems to be an important goal for further studies. In other words, $\langle q_{\bm{x}} q_{\bm{y}} \rangle_S$ has to be equal to the probability that the lattice sites $\bm{x}$ and $\bm{y}$ are connected by a path of occupied bonds for a given realization of interactions $\bm{J}$. If such clusters could be constructed in a computationally efficient way this would probably lead to an effective cluster algorithm to study the spin-glass transition. Note that it is possible to set up a connection between the overlap-correlation function and the pair-connection function within the scope of the CMR representation, but unfortunately this connection does not have the form of an equality \cite{MachtaNewmanStein2008ThePercolationSignatureOfTheSpinGlassTransition}. Interesting candidates of this type may include clusters constructed from more than two replicas. Above we have generalized the CMRJ and two-replica FKCK clusters to the case of $I$ independent replicas, and a numerical investigation of the properties of such clusters is a promising future enterprise. Furthermore, it might be useful to develop more elaborated multi-replica cluster updates which do not only satisfy detailed balance but are also ergodic. An example for such an ergodic two-replica cluster update is given in \cite{MachtaNewmanStein2008ThePercolationSignatureOfTheSpinGlassTransition}. In terms of cluster updates which reduce equilibration time and autocorrelation time it might also be useful to consider more sophisticated computational methods. An interesting ansatz would be to define multiple-replica clusters within the framework of the generalized cluster algorithm  \cite{KandelDomany1991GeneralClusterMonteCarloDynamics,Cataudella1996PercolationAndClusterMonteCarloDynamicsForSpinModels}. Another direction might be the application of machine learning techniques to identify clusters or to look at other non-local updating schemes \cite{LiuEtAl2017SelfLearningMonteCarloMethod,McNaughtonMilosevicPeraliPilati2020BoostingMCSimulationsOfSGUsingAutoregressiveNeuralNetworks,Wang2017ExploringClusterMCUpdatesWithBoltzmannMachines,PeiDiVentra2022NonEquilibriumCriticalityAndEfficientExplorationOfGlassyLandscapesWithMemoryDynamics}. Overall, it remains an attractive but so far elusive goal to arrive at a full description of the spin-glass transition in terms of the percolation of clusters that directly propagate the correlations. 
	
\acknowledgements
The authors thank M. Akritidis for useful discussions. The simulations were performed at the Coventry university on the Zeus cluster with excellent technical support of A. Pedcenko.

\appendix
\section{$I$-Replica Cluster Algorithm}\label{app:i_replica_mc}	
In this section we show that the $I$-replica cluster moves satisfy the detailed balance condition with respect to the $I$-replica Boltzmann distribution. Note that a similar derivation in case of the Ising ferromagnet can be found in \cite{Krauth2004ClusterMonteCarloAlgorithms}. The $I$-replica Hamiltonian for given a realization of bonds $\bm{J}$ is given by
\begin{align*}
	\mathcal{H}(\mathcal{S})&=H_{\bm{J}}^{(I)}(\bm{S}^{(1)},\dots,\bm{S}^{(I)})=-\sum_{\langle \bm{x}, \bm{y} \rangle} J_{\bm{x}\bm{y}} \tilde{s}_{\bm{x}} \tilde{s}_{\bm{y}} \\
			&~~~\text{ where } \tilde{s}_{\bm{x}} \tilde{s}_{\bm{y}}=\sum_{i=1}^I s_{\bm{x}}^{(i)} s_{\bm{y}}^{(i)} 
\end{align*}
and $\tilde{s}_{\bm{x}}=(s_{\bm{x}}^{(1)},...,s_{\bm{x}}^{(I)})$. We denote a certain spin state $\mathcal{S}_{\mu}=(\bm{S}_{\mu}^{(1)},\dots,\bm{S}_{\mu}^{(I)})$ with the index $\mu$ and its energy by $\mathcal{H}_\mu=\mathcal{H}(\mathcal{S}_\mu)$. The cluster algorithm works as follows. First, each bond is occupied with probability $p_{\bm{x}\bm{y}}^{(\mu)}$, where $p_{\bm{x}\bm{y}}^{(\mu)}=0$ if $J_{\bm{x}\bm{y}}\tilde{s}_{\bm{x}}^{(\mu)}\tilde{s}_{\bm{x}}^{(\mu)}\leq0$. Occupied bonds connect spin sites that group together in clusters. The smallest clusters are isolated spins and each site belongs to a exactly one cluster. Of all clusters we randomly choose a fraction $f\in[0,1]$ and we put all spin sites within these clusters into a single set called $\mathcal{A}$. Often the fraction is set to $f=0.5$. The rest of the spins belong to the complement of $\mathcal{A}$, $\mathcal{A}^c=\mathcal{A}\setminus\mathcal{N}$, where $\mathcal{N}$ is the set that contains all sites $\bm{x}$, such that $\vert \mathcal{N} \vert=N$. By this procedure we have created a partition which separates all sites into two sets. The cluster move is performed by flipping all $I$-component spins inside $\mathcal{A}$, i.e. $\tilde{s}_{\bm{x}}^{(\mu)}\to -\tilde{s}_{\bm{x}}^{(\mu)}$, $\forall \, \bm{x} \in \mathcal{A}$. Afterwards, the system is in spin state $\mathcal{S}_\nu$.

We now we derive the occupation probabilities $p_{\bm{x}\bm{y}}$ that lead to transition probabilities between the states that satisfy the property of detailed balance. The condition of detailed balance is given by \cite{NewmanBarkema1999MonteCarloMethodsInStatisticalPhysics}    
\begin{align} 
	\frac{T(\mu\to \nu)}{T(\nu \to \mu)}&=\frac{\mathcal{P}(\mathcal{S}_{\nu}\vert \bm{J})}{\mathcal{P}(\mathcal{S}_{\mu}\vert \bm{J})}=
	\exp\left\{-\beta \left( \mathcal{H}_\nu-\mathcal{H}_\mu \right) \right\} \, .
	\label{eq:detailed_balance} 
\end{align}
Here, $T(\mu\to \nu)$ denotes the transition probability from spin state $\mathcal{S}_\mu$ to spin state $\mathcal{S}_\nu$ and $T(\nu \to \mu)$ 
is the probability of the inverted transition. The weight of the states within the $I$-replica Boltzmann distribution is given by $\mathcal{P}(\mathcal{S}_{\nu}\vert \bm{J})$ and $\mathcal{P}(\mathcal{S}_{\mu} \vert\bm{J})$, respectively. The right hand side of Eq.~\eqref{eq:detailed_balance} depends on the energies $\mathcal{H}_\mu$ and $\mathcal{H}_\nu$. States $\mu$ and $\nu$ only differ from one another by the flipped spins inside of $\mathcal{A}$, i.e., $\tilde{s}_{\bm{x}}^{(\nu)}=\tilde{s}_{\bm{x}}^{(\mu)}~\forall\,\bm{x}\in \mathcal{A}^c$ and $\tilde{s}_{\bm{x}}^{(\nu)}=-\tilde{s}_{\bm{x}}^{(\mu)}~\forall\,\bm{x}\in \mathcal{A}$. Due to the global spin-flip symmetry of the Hamiltonian, $\mathcal{H}(\mathcal{S})=\mathcal{H}(-\mathcal{S})$, neighboring spins within the identical set $\mathcal{A}$ or $\mathcal{A}^c$ contribute the same energy in both states. Thus, the energies of the states are
\begin{align*}
	\mathcal{H}_\mu &= \mathcal{H}^{(\mathcal{A})}_{\mu}+\partial \mathcal{H}_{\mu}+\mathcal{H}_{\mu}^{(\mathcal{A}^c)} \, ,\\
	\mathcal{H}_\nu&=\mathcal{H}^{(\mathcal{A})}_{\nu}+\partial \mathcal{H}_{\nu}+\mathcal{H}_{\nu}^{(\mathcal{A}^c)} \\
	&=\mathcal{H}^{(\mathcal{A})}_{\mu}+\partial \mathcal{H}_{\nu}+\mathcal{H}_{\mu}^{(\mathcal{A}^c)} \,.
\end{align*}
Here, $\mathcal{H}^{(\mathcal{A})}$ is the contribution to the energy of spins inside $\mathcal{A}$, $\mathcal{H}^{(\mathcal{A}^c)}$
is the contribution from spins which belong to $\mathcal{A}^c$ and $\partial \mathcal{H}$ is the energy contribution of the surface of the partition.
This surface is defined in terms of bonds which connect spins from $\mathcal{A}$ with those of $\mathcal{A}^c$, i.e. $\partial A =\{  (\bm{x},\bm{y}) \in \langle\bm{x},\bm{y}\rangle : 
\bm{x} \in \mathcal{A} \land \bm{y}\in \mathcal{A}^c \}$. The energy difference of both states is proportional to the surface energy, 
\begin{align*}
	\left( \mathcal{H}_\nu-\mathcal{H}_\mu \right) = -2 \partial \mathcal{H}_{\mu},
\end{align*}
where we made use of the relation $\partial \mathcal{H}_{\nu}=-\partial \mathcal{H}_{\mu}$, which results from the fact that  bonds in the surface which are broken in state $\mu$ are satisfied 
in state $\nu$ and vice versa. By broken and satisfied bonds we mean that $J_{\bm{x}\bm{y}}\tilde{s}_{\bm{x}}\tilde{s}_{\bm{y}}\leq0$ and $J_{\bm{x}\bm{y}}\tilde{s}_{\bm{x}}\tilde{s}_{\bm{y}}>0$, respectively. The energy of the surface can be written in terms of broken and satisfied bonds
\begin{align*}
	\partial \mathcal{H}_{\mu}&=-\sum_{ \partial \mathcal{A}_{\mu}^{+}} \vert J_{\bm{x} \bm{y}}  \tilde{s}_{\bm{x}}^{(\mu)} \tilde{s}_{\bm{y}}^{(\mu)} \vert+\sum_{ \partial \mathcal{A}_{\mu}^{-}} \vert J_{\bm{x} \bm{y}}   \tilde{s}_{\bm{x}}^{(\mu)} \tilde{s}_{\bm{y}}^{(\mu)} \vert\\
    &=-\sum_{ \partial \mathcal{A}_{\mu}^{+}}  J_{\bm{x} \bm{y}}  \tilde{s}_{\bm{x}}^{(\mu)} \tilde{s}_{\bm{y}}^{(\mu)} + \sum_{ \partial \mathcal{A}_{\nu}^{+}}  J_{\bm{x} \bm{y}}   \tilde{s}_{\bm{x}}^{(\nu)} \tilde{s}_{\bm{y}}^{(\nu)}  \, .
\end{align*}
Here $\partial\mathcal{A}_{\mu}^{+}$ denotes the section in the surface where bonds are satisfied in state $\mu$ and $\partial \mathcal{A}_{\mu}^{-}$ the section with broken bonds. As already mentioned, $\partial\mathcal{A}_{\mu}^{+}=\partial\mathcal{A}_{\nu}^{-}$ and $\partial\mathcal{A}_{\mu}^{-}=\partial\mathcal{A}_{\nu}^{+}$.

Now the left hand side of equation \eqref{eq:detailed_balance} is considered. The transition probability from state $\mu$ to $\nu$ can be written as  
\begin{align*}
	T(\mu \to \nu)=K(\mathcal{A}\land \mathcal{A}^c \vert \mu)\prod_{ \partial\mathcal{A}_{\mu}^{+}}\left(1-p_{\bm{x}\bm{y}}^{(\mu)} \right) \,.
\end{align*}    
The first factor $K(\mathcal{A} \land \mathcal{A}^c \vert \mu)$ denotes the probability that the sets $\mathcal{A}$ and $\mathcal{A}^c$ are constructed given the state $\mu$. The second factor is the probability that satisfied bonds in the surface between $\mathcal{A}$ and $\mathcal{A}^c$ are not occupied. Bonds in the surface which are broken in state $\mu$ are not included because they are not occupied with probability one. The probability of the inverted transition from $\nu$ to $\mu$ is given by
\begin{align*}
	T(\nu \to \mu)=K(\mathcal{A}\land \mathcal{A}^c \vert \nu )\prod_{\partial \mathcal{A}_{\nu}^{+}}\left(1-p_{\bm{x}\bm{y}}^{(\nu)} \right)\,.
\end{align*} 
Except for the bonds in the surface the identical bonds are broken or satisfied in both states $\mu$ and $\nu$. As a consequence, the probability to construct the sets $\mathcal{A}$ and $\mathcal{A}^c$ starting from the state $\nu$ is equal to the probability to construct the same sets starting with $\mu$, i.e  $K(\mathcal{A}\land\mathcal{A}^c\vert \nu )=K(\mathcal{A}\land\mathcal{A}^c\vert \mu )$.    
By inserting our results for the transition probabilities as well as the surface energy into Eq.\,\eqref{eq:detailed_balance} one obtains, 
\begin{align*}
\displaystyle
	&\frac{K(\mathcal{A}\land \mathcal{A}^c\vert \mu )\prod_{ \partial\mathcal{A}_{\mu}^{+}}
 \left(1-p_{\bm{x}\bm{y}}^{(\mu)} \right)}
 {K(\mathcal{A}\land \mathcal{A}^c\vert \mu )\prod_{\partial \mathcal{A}_{\nu}^{+}}\left(1-p_{\bm{x}\bm{y}}^{(\nu)} \right)}= \\
	&\exp\left\{-2\beta \left( 
 \sum_{ \partial \mathcal{A}_{\mu}^{+}}  J_{\bm{x} \bm{y}}  \tilde{s}_{\bm{x}}^{(\mu)} \tilde{s}_{\bm{y}}^{(\mu)} - \sum_{ \partial \mathcal{A}_{\nu}^{+}}  J_{\bm{x} \bm{y}}   \tilde{s}_{\bm{x}}^{(\nu)} \tilde{s}_{\bm{y}}^{(\nu)}
 \right) \right\} \, ,
\end{align*}
which can be rewritten as
\begin{align*}
 \displaystyle
	\frac
	{\prod_{\partial\mathcal{A}_{\mu}^{+}} \left(1-p_{\bm{x}\bm{y}}^{(\mu)} \right)}
	{\exp\left\{-2\beta \sum_{\partial \mathcal{A}_{\mu}^{+}}  J_{\bm{x} \bm{y}}\tilde{s}_{\bm{x}}^{(\mu)} \tilde{s}_{\bm{y}}^{(\mu)}  \right\} }
	 =  \\
	\frac
	{\prod_{\partial\mathcal{A}_{\nu}^{+}} \left(1-p_{\bm{x}\bm{y}}^{(\nu)} \right)}
	{\exp\left\{-2\beta \sum_{ \partial \mathcal{A}_{\nu}^{+}}  J_{\bm{x} \bm{y}}\tilde{s}_{\bm{x}}^{(\nu)} \tilde{s}_{\bm{y}}^{(\nu)} \right\}} \, .
\end{align*}
This equation is satisfied if 
\begin{align*}
p_{\bm{x}\bm{y}}=1-\exp( -2\beta J_{\bm{x}\bm{y}}\tilde{s}_{\bm{x}}\tilde{s}_{\bm{y}}) 
\end{align*}
and the condition of detailed balance is fulfilled. For $I=2$ the algorithm is equivalent to the algorithm of Jörg \cite{Joerg2005ClusterMonteCarloAlgorithmsForDilutedSG} and the occupied bonds are equal to the blue bonds in the CMR representation \cite{MachtaNewmanStein2008ThePercolationSignatureOfTheSpinGlassTransition}. 

\section*{ }

\bibliography{perc_sg}

\begin{thebibliography}{86}%
\makeatletter
\providecommand \@ifxundefined [1]{%
 \@ifx{#1\undefined}
}%
\providecommand \@ifnum [1]{%
 \ifnum #1\expandafter \@firstoftwo
 \else \expandafter \@secondoftwo
 \fi
}%
\providecommand \@ifx [1]{%
 \ifx #1\expandafter \@firstoftwo
 \else \expandafter \@secondoftwo
 \fi
}%
\providecommand \natexlab [1]{#1}%
\providecommand \enquote  [1]{``#1''}%
\providecommand \bibnamefont  [1]{#1}%
\providecommand \bibfnamefont [1]{#1}%
\providecommand \citenamefont [1]{#1}%
\providecommand \href@noop [0]{\@secondoftwo}%
\providecommand \href [0]{\begingroup \@sanitize@url \@href}%
\providecommand \@href[1]{\@@startlink{#1}\@@href}%
\providecommand \@@href[1]{\endgroup#1\@@endlink}%
\providecommand \@sanitize@url [0]{\catcode `\\12\catcode `\$12\catcode
  `\&12\catcode `\#12\catcode `\^12\catcode `\_12\catcode `\%12\relax}%
\providecommand \@@startlink[1]{}%
\providecommand \@@endlink[0]{}%
\providecommand \url  [0]{\begingroup\@sanitize@url \@url }%
\providecommand \@url [1]{\endgroup\@href {#1}{\urlprefix }}%
\providecommand \urlprefix  [0]{URL }%
\providecommand \Eprint [0]{\href }%
\providecommand \doibase [0]{https://doi.org/}%
\providecommand \selectlanguage [0]{\@gobble}%
\providecommand \bibinfo  [0]{\@secondoftwo}%
\providecommand \bibfield  [0]{\@secondoftwo}%
\providecommand \translation [1]{[#1]}%
\providecommand \BibitemOpen [0]{}%
\providecommand \bibitemStop [0]{}%
\providecommand \bibitemNoStop [0]{.\EOS\space}%
\providecommand \EOS [0]{\spacefactor3000\relax}%
\providecommand \BibitemShut  [1]{\csname bibitem#1\endcsname}%
\let\auto@bib@innerbib\@empty
\bibitem [{\citenamefont {Stauffer}\ and\ \citenamefont
  {Aharony}(1994)}]{StaufferAharony1994IntroductionToPercolationTheory}%
  \BibitemOpen
  \bibfield  {author} {\bibinfo {author} {\bibfnamefont {D.}~\bibnamefont
  {Stauffer}}\ and\ \bibinfo {author} {\bibfnamefont {A.}~\bibnamefont
  {Aharony}},\ }\href@noop {} {\emph {\bibinfo {title} {{Introduction to
  Percolation Theory}}}},\ \bibinfo {edition} {2nd}\ ed.\ (\bibinfo
  {publisher} {Taylor \& Francis},\ \bibinfo {address} {London},\ \bibinfo
  {year} {1994})\BibitemShut {NoStop}%
\bibitem [{\citenamefont
  {Grimmett}(2006)}]{Grimmett2004BookTheRandomClusterModel}%
  \BibitemOpen
  \bibfield  {author} {\bibinfo {author} {\bibfnamefont {G.}~\bibnamefont
  {Grimmett}},\ }\href@noop {} {\emph {\bibinfo {title} {{The Random-Cluster
  Model}}}}\ (\bibinfo  {publisher} {Springer},\ \bibinfo {address}
  {Berlin,Heidelberg},\ \bibinfo {year} {2006})\BibitemShut {NoStop}%
\bibitem [{\citenamefont {Sahimi}\ and\ \citenamefont
  {Hunt}(2021)}]{SahimiHunt2021BookComplexMediaAndPercolationTheory}%
  \BibitemOpen
  \bibinfo {editor} {\bibfnamefont {M.}~\bibnamefont {Sahimi}}\ and\ \bibinfo
  {editor} {\bibfnamefont {A.~G.}\ \bibnamefont {Hunt}},\ eds.,\ \href@noop {}
  {\emph {\bibinfo {title} {{Complex Media and Percolation Theory}}}}\
  (\bibinfo  {publisher} {Springer},\ \bibinfo {address} {New York},\ \bibinfo
  {year} {2021})\BibitemShut {NoStop}%
\bibitem [{\citenamefont {Fortuin}\ and\ \citenamefont
  {Kasteleyn}(1972)}]{FortuinKasteleyn1972OnTheRandomClusterModel1}%
  \BibitemOpen
  \bibfield  {author} {\bibinfo {author} {\bibfnamefont {C.~M.}\ \bibnamefont
  {Fortuin}}\ and\ \bibinfo {author} {\bibfnamefont {P.~W.}\ \bibnamefont
  {Kasteleyn}},\ }\bibfield  {title} {\bibinfo {title} {On the random-cluster
  model: I. introduction and relation to other models},\ }\href@noop {}
  {\bibfield  {journal} {\bibinfo  {journal} {Physica}\ }\textbf {\bibinfo
  {volume} {{57}}},\ \bibinfo {pages} {536} (\bibinfo {year}
  {1972})}\BibitemShut {NoStop}%
\bibitem [{\citenamefont
  {Fisher}(1967)}]{Fisher1967TheTheoryOfCondensationAndTheCriticalPoint}%
  \BibitemOpen
  \bibfield  {author} {\bibinfo {author} {\bibfnamefont {M.~E.}\ \bibnamefont
  {Fisher}},\ }\bibfield  {title} {\bibinfo {title} {{The theory of
  condensation and the critical point}},\ }\href@noop {} {\bibfield  {journal}
  {\bibinfo  {journal} {Physics}\ }\textbf {\bibinfo {volume} {{3}}},\ \bibinfo
  {pages} {255} (\bibinfo {year} {1967})}\BibitemShut {NoStop}%
\bibitem [{\citenamefont {Coniglio}\ and\ \citenamefont
  {Klein}(1980)}]{ConiglioKlein1980ClustersAndCriticalDroplets}%
  \BibitemOpen
  \bibfield  {author} {\bibinfo {author} {\bibfnamefont {A.}~\bibnamefont
  {Coniglio}}\ and\ \bibinfo {author} {\bibfnamefont {W.}~\bibnamefont
  {Klein}},\ }\bibfield  {title} {\bibinfo {title} {Clusters and {I}sing
  critical droplets: a renormalisation group approach},\ }\href@noop {}
  {\bibfield  {journal} {\bibinfo  {journal} {J. Phys. A}\ }\textbf {\bibinfo
  {volume} {{13}}},\ \bibinfo {pages} {2775} (\bibinfo {year}
  {1980})}\BibitemShut {NoStop}%
\bibitem [{\citenamefont {Coniglio}\ and\ \citenamefont
  {Fierro}(2021)}]{ConiglioFierro2021CorrelatedPercolation}%
  \BibitemOpen
  \bibfield  {author} {\bibinfo {author} {\bibfnamefont {A.}~\bibnamefont
  {Coniglio}}\ and\ \bibinfo {author} {\bibfnamefont {A.}~\bibnamefont
  {Fierro}},\ }\bibinfo {title} {Correlated percolation},\ in\ \href@noop {}
  {\emph {\bibinfo {booktitle} {{Complex Media and Percolation Theory}}}},\
  \bibinfo {editor} {edited by\ \bibinfo {editor} {\bibfnamefont
  {M.}~\bibnamefont {Sahimi}}\ and\ \bibinfo {editor} {\bibfnamefont {A.~G.}\
  \bibnamefont {Hunt}}}\ (\bibinfo  {publisher} {Springer},\ \bibinfo {address}
  {New York},\ \bibinfo {year} {2021})\ p.~\bibinfo {pages} {61}\BibitemShut
  {NoStop}%
\bibitem [{\citenamefont {Swendsen}\ and\ \citenamefont
  {Wang}(1987)}]{SwendsenWang1987NonuniversalCriticalDynamicsInMonteCarloSimulations}%
  \BibitemOpen
  \bibfield  {author} {\bibinfo {author} {\bibfnamefont {R.~H.}\ \bibnamefont
  {Swendsen}}\ and\ \bibinfo {author} {\bibfnamefont {J.~S.}\ \bibnamefont
  {Wang}},\ }\bibfield  {title} {\bibinfo {title} {Nonuniversal critical
  dynamics in {M}onte {C}arlo simulations},\ }\href@noop {} {\bibfield
  {journal} {\bibinfo  {journal} {Phys. Rev. Lett.}\ }\textbf {\bibinfo
  {volume} {{58}}},\ \bibinfo {pages} {86} (\bibinfo {year}
  {1987})}\BibitemShut {NoStop}%
\bibitem [{\citenamefont {Wang}\ and\ \citenamefont
  {Swendsen}(1990)}]{SwendsenWang1990ClusterMonteCarloAlgorithms}%
  \BibitemOpen
  \bibfield  {author} {\bibinfo {author} {\bibfnamefont {J.~S.}\ \bibnamefont
  {Wang}}\ and\ \bibinfo {author} {\bibfnamefont {R.~H.}\ \bibnamefont
  {Swendsen}},\ }\bibfield  {title} {\bibinfo {title} {Cluster {M}onte {C}arlo
  algorithms},\ }\href@noop {} {\bibfield  {journal} {\bibinfo  {journal}
  {Physica A}\ }\textbf {\bibinfo {volume} {{167}}},\ \bibinfo {pages} {565}
  (\bibinfo {year} {1990})}\BibitemShut {NoStop}%
\bibitem [{\citenamefont {Binder}\ and\ \citenamefont
  {Young}(1986)}]{BinderYoung1986SpinGlasses}%
  \BibitemOpen
  \bibfield  {author} {\bibinfo {author} {\bibfnamefont {K.}~\bibnamefont
  {Binder}}\ and\ \bibinfo {author} {\bibfnamefont {A.~P.}\ \bibnamefont
  {Young}},\ }\bibfield  {title} {\bibinfo {title} {{Spin glasses: Experimental
  facts, theoretical concepts, and open questions}},\ }\href@noop {} {\bibfield
   {journal} {\bibinfo  {journal} {Rev. Mod. Phys.}\ }\textbf {\bibinfo
  {volume} {{58}}},\ \bibinfo {pages} {801} (\bibinfo {year}
  {1986})}\BibitemShut {NoStop}%
\bibitem [{\citenamefont {Stein}\ and\ \citenamefont
  {Newman}(2013)}]{NewmanStein2013SpinGlassesAndComplexity}%
  \BibitemOpen
  \bibfield  {author} {\bibinfo {author} {\bibfnamefont {D.~L.}\ \bibnamefont
  {Stein}}\ and\ \bibinfo {author} {\bibfnamefont {C.~M.}\ \bibnamefont
  {Newman}},\ }\href@noop {} {\emph {\bibinfo {title} {{Spin Glasses and
  Complexity}}}}\ (\bibinfo  {publisher} {Princeton University Press},\
  \bibinfo {address} {Princeton},\ \bibinfo {year} {2013})\BibitemShut
  {NoStop}%
\bibitem [{\citenamefont {M{\'e}zard}\ \emph {et~al.}(1987)\citenamefont
  {M{\'e}zard}, \citenamefont {Parisi},\ and\ \citenamefont
  {Virasoro}}]{MezardParisiVirasoro1987SpinGlassTheoryAndBeyond}%
  \BibitemOpen
  \bibfield  {author} {\bibinfo {author} {\bibfnamefont {M.}~\bibnamefont
  {M{\'e}zard}}, \bibinfo {author} {\bibfnamefont {G.}~\bibnamefont {Parisi}},\
  and\ \bibinfo {author} {\bibfnamefont {M.}~\bibnamefont {Virasoro}},\
  }\href@noop {} {\emph {\bibinfo {title} {{Spin Glass Theory and Beyond}}}}\
  (\bibinfo  {publisher} {World Scientific},\ \bibinfo {address} {Singapore},\
  \bibinfo {year} {1987})\BibitemShut {NoStop}%
\bibitem [{\citenamefont {Bolthausen}\ and\ \citenamefont
  {Bovier}(2007)}]{BolthausenBovier2007SpinGlasses}%
  \BibitemOpen
  \bibinfo {editor} {\bibfnamefont {E.}~\bibnamefont {Bolthausen}}\ and\
  \bibinfo {editor} {\bibfnamefont {A.}~\bibnamefont {Bovier}},\ eds.,\
  \href@noop {} {\emph {\bibinfo {title} {{Spin Glasses}}}}\ (\bibinfo
  {publisher} {Springer},\ \bibinfo {address} {Berlin, Heidelberg},\ \bibinfo
  {year} {2007})\BibitemShut {NoStop}%
\bibitem [{\citenamefont {De~Arcangelis}\ \emph {et~al.}(1991)\citenamefont
  {De~Arcangelis}, \citenamefont {Coniglio},\ and\ \citenamefont
  {Peruggi}}]{DeArcangelis1991PercolationTransitionInSpinGlasses}%
  \BibitemOpen
  \bibfield  {author} {\bibinfo {author} {\bibfnamefont {L.}~\bibnamefont
  {De~Arcangelis}}, \bibinfo {author} {\bibfnamefont {A.}~\bibnamefont
  {Coniglio}},\ and\ \bibinfo {author} {\bibfnamefont {F.}~\bibnamefont
  {Peruggi}},\ }\bibfield  {title} {\bibinfo {title} {Percolation transition in
  spin glasses},\ }\href@noop {} {\bibfield  {journal} {\bibinfo  {journal}
  {EPL}\ }\textbf {\bibinfo {volume} {{14}}},\ \bibinfo {pages} {515} (\bibinfo
  {year} {1991})}\BibitemShut {NoStop}%
\bibitem [{\citenamefont
  {Houdayer}(2001)}]{Houdayer2001ClusterMonteCarloAlgorithmFor2DimensionalSpinGlasses}%
  \BibitemOpen
  \bibfield  {author} {\bibinfo {author} {\bibfnamefont {J.}~\bibnamefont
  {Houdayer}},\ }\bibfield  {title} {\bibinfo {title} {A cluster monte carlo
  algorithm for 2-dimensional spin glasses},\ }\href@noop {} {\bibfield
  {journal} {\bibinfo  {journal} {Eur. Phys. J. B}\ }\textbf {\bibinfo {volume}
  {{22}}},\ \bibinfo {pages} {479} (\bibinfo {year} {2001})}\BibitemShut
  {NoStop}%
\bibitem [{\citenamefont {Chayes}\ \emph {et~al.}(1998)\citenamefont {Chayes},
  \citenamefont {Machta},\ and\ \citenamefont
  {Redner}}]{CMR1998GraphicalRepresentationsForIsingSystemsInExternalFields}%
  \BibitemOpen
  \bibfield  {author} {\bibinfo {author} {\bibfnamefont {L.}~\bibnamefont
  {Chayes}}, \bibinfo {author} {\bibfnamefont {J.}~\bibnamefont {Machta}},\
  and\ \bibinfo {author} {\bibfnamefont {O.}~\bibnamefont {Redner}},\
  }\bibfield  {title} {\bibinfo {title} {Graphical representations for {I}sing
  systems in external fields},\ }\href@noop {} {\bibfield  {journal} {\bibinfo
  {journal} {J. Stat. Phys.}\ }\textbf {\bibinfo {volume} {{93}}},\ \bibinfo
  {pages} {17} (\bibinfo {year} {1998})}\BibitemShut {NoStop}%
\bibitem [{\citenamefont {Redner}\ \emph {et~al.}(1998)\citenamefont {Redner},
  \citenamefont {Machta},\ and\ \citenamefont
  {Chayes}}]{CMR1998GraphicalRepresentationsAndClusterAlgorithmsForCriticalPointsWithFields}%
  \BibitemOpen
  \bibfield  {author} {\bibinfo {author} {\bibfnamefont {O.}~\bibnamefont
  {Redner}}, \bibinfo {author} {\bibfnamefont {J.}~\bibnamefont {Machta}},\
  and\ \bibinfo {author} {\bibfnamefont {L.~F.}\ \bibnamefont {Chayes}},\
  }\bibfield  {title} {\bibinfo {title} {{Graphical representations and cluster
  algorithms for critical points with fields}},\ }\href@noop {} {\bibfield
  {journal} {\bibinfo  {journal} {Phys. Rev. E}\ }\textbf {\bibinfo {volume}
  {{58}}},\ \bibinfo {pages} {2749} (\bibinfo {year} {1998})}\BibitemShut
  {NoStop}%
\bibitem [{\citenamefont {Machta}\ \emph {et~al.}(2008)\citenamefont {Machta},
  \citenamefont {Newman},\ and\ \citenamefont
  {Stein}}]{MachtaNewmanStein2008ThePercolationSignatureOfTheSpinGlassTransition}%
  \BibitemOpen
  \bibfield  {author} {\bibinfo {author} {\bibfnamefont {J.}~\bibnamefont
  {Machta}}, \bibinfo {author} {\bibfnamefont {C.~M.}\ \bibnamefont {Newman}},\
  and\ \bibinfo {author} {\bibfnamefont {D.~L.}\ \bibnamefont {Stein}},\
  }\bibfield  {title} {\bibinfo {title} {The percolation signature of the spin
  glass transition},\ }\href@noop {} {\bibfield  {journal} {\bibinfo  {journal}
  {J. Stat. Phys.}\ }\textbf {\bibinfo {volume} {{130}}},\ \bibinfo {pages}
  {113} (\bibinfo {year} {2008})}\BibitemShut {NoStop}%
\bibitem [{\citenamefont
  {J{\"o}rg}(2005)}]{Joerg2005ClusterMonteCarloAlgorithmsForDilutedSG}%
  \BibitemOpen
  \bibfield  {author} {\bibinfo {author} {\bibfnamefont {T.}~\bibnamefont
  {J{\"o}rg}},\ }\bibfield  {title} {\bibinfo {title} {Cluster {M}onte {C}arlo
  algorithms for diluted spin glasses},\ }\href@noop {} {\bibfield  {journal}
  {\bibinfo  {journal} {Prog. Theor. Phys. Suppl.}\ }\textbf {\bibinfo {volume}
  {{157}}},\ \bibinfo {pages} {349} (\bibinfo {year} {2005})}\BibitemShut
  {NoStop}%
\bibitem [{\citenamefont {Newman}\ and\ \citenamefont
  {Stein}(2007{\natexlab{a}})}]{NewmanStein2007ShortRangeSpinGlassesResultsAndSpeculations}%
  \BibitemOpen
  \bibfield  {author} {\bibinfo {author} {\bibfnamefont {C.~M.}\ \bibnamefont
  {Newman}}\ and\ \bibinfo {author} {\bibfnamefont {D.~L.}\ \bibnamefont
  {Stein}},\ }\bibfield  {title} {\bibinfo {title} {Short-range spin glasses:
  Results and speculations},\ }in\ \href@noop {} {\emph {\bibinfo {booktitle}
  {Spin Glasses}}},\ \bibinfo {editor} {edited by\ \bibinfo {editor}
  {\bibfnamefont {E.}~\bibnamefont {Bolthausen}}\ and\ \bibinfo {editor}
  {\bibfnamefont {A.}~\bibnamefont {Bovier}}}\ (\bibinfo  {publisher}
  {Springer},\ \bibinfo {address} {Berlin, Heidelberg},\ \bibinfo {year}
  {2007})\ p.\ \bibinfo {pages} {159}\BibitemShut {NoStop}%
\bibitem [{\citenamefont {Kumar}\ \emph {et~al.}(2020)\citenamefont {Kumar},
  \citenamefont {Gross}, \citenamefont {Janke},\ and\ \citenamefont
  {Weigel}}]{KumarEtALMassivelyParallelSimulationsForDisorderedSystems}%
  \BibitemOpen
  \bibfield  {author} {\bibinfo {author} {\bibfnamefont {R.}~\bibnamefont
  {Kumar}}, \bibinfo {author} {\bibfnamefont {J.}~\bibnamefont {Gross}},
  \bibinfo {author} {\bibfnamefont {W.}~\bibnamefont {Janke}},\ and\ \bibinfo
  {author} {\bibfnamefont {M.}~\bibnamefont {Weigel}},\ }\bibfield  {title}
  {\bibinfo {title} {{Massively parallel simulations for disordered systems}},\
  }\href@noop {} {\bibfield  {journal} {\bibinfo  {journal} {Eur. Phys. J. B}\
  }\textbf {\bibinfo {volume} {{93}}},\ \bibinfo {pages} {1} (\bibinfo {year}
  {2020})}\BibitemShut {NoStop}%
\bibitem [{\citenamefont {Hasenbusch}\ \emph {et~al.}(2008)\citenamefont
  {Hasenbusch}, \citenamefont {Pelissetto},\ and\ \citenamefont
  {Vicari}}]{HasenbuschEtAl2008TheCriticalBehaviorOf3DIsingGlassModels}%
  \BibitemOpen
  \bibfield  {author} {\bibinfo {author} {\bibfnamefont {M.}~\bibnamefont
  {Hasenbusch}}, \bibinfo {author} {\bibfnamefont {A.}~\bibnamefont
  {Pelissetto}},\ and\ \bibinfo {author} {\bibfnamefont {E.}~\bibnamefont
  {Vicari}},\ }\bibfield  {title} {\bibinfo {title} {Critical behavior of
  three-dimensional {I}sing spin glass models},\ }\href@noop {} {\bibfield
  {journal} {\bibinfo  {journal} {Phys. Rev. B}\ }\textbf {\bibinfo {volume}
  {{78}}},\ \bibinfo {pages} {214205} (\bibinfo {year} {2008})}\BibitemShut
  {NoStop}%
\bibitem [{\citenamefont
  {Yamaguchi}(2013)}]{Yamaguchi2013ConjecturedExactPercolationThresholdsOfTheFortuinKasteleynClusters}%
  \BibitemOpen
  \bibfield  {author} {\bibinfo {author} {\bibfnamefont {C.}~\bibnamefont
  {Yamaguchi}},\ }\bibfield  {title} {\bibinfo {title} {Conjectured exact
  percolation thresholds of the {F}ortuin-{K}asteleyn cluster for the {$\pm J$}
  {I}sing spin glass model},\ }\href@noop {} {\bibfield  {journal} {\bibinfo
  {journal} {Physica A}\ }\textbf {\bibinfo {volume} {{392}}},\ \bibinfo
  {pages} {1263} (\bibinfo {year} {2013})}\BibitemShut {NoStop}%
\bibitem [{\citenamefont {Young}\ and\ \citenamefont
  {Stinchcombe}(1976)}]{YoungStinchcombe1976RGRenormalizationGroupForSpinGlassesAndDilutedMagnets}%
  \BibitemOpen
  \bibfield  {author} {\bibinfo {author} {\bibfnamefont {A.~P.}\ \bibnamefont
  {Young}}\ and\ \bibinfo {author} {\bibfnamefont {R.~B.}\ \bibnamefont
  {Stinchcombe}},\ }\bibfield  {title} {\bibinfo {title} {{Real-space
  renormalization group calculations for spin glasses and dilute magnets}},\
  }\href@noop {} {\bibfield  {journal} {\bibinfo  {journal} {J. Phys. C}\
  }\textbf {\bibinfo {volume} {{9}}},\ \bibinfo {pages} {4419} (\bibinfo {year}
  {1976})}\BibitemShut {NoStop}%
\bibitem [{\citenamefont {Bray}\ and\ \citenamefont
  {Moore}(1987)}]{BrayMoore1987ScalingTheoryOfTheOrderedPhaseOfSpinGlasses}%
  \BibitemOpen
  \bibfield  {author} {\bibinfo {author} {\bibfnamefont {A.}~\bibnamefont
  {Bray}}\ and\ \bibinfo {author} {\bibfnamefont {M.}~\bibnamefont {Moore}},\
  }\bibfield  {title} {\bibinfo {title} {{Scaling theory of the ordered phase
  of spin glasses}},\ }in\ \href@noop {} {\emph {\bibinfo {booktitle}
  {Heidelberg Colloquium on Glassy Dynamics}}},\ \bibinfo {editor} {edited by\
  \bibinfo {editor} {\bibfnamefont {J.}~\bibnamefont {{van Hemmen}}}\ and\
  \bibinfo {editor} {\bibfnamefont {I.}~\bibnamefont {Morgenstern}}}\ (\bibinfo
   {publisher} {Springer},\ \bibinfo {address} {Berlin, Heidelberg},\ \bibinfo
  {year} {1987})\ p.\ \bibinfo {pages} {121}\BibitemShut {NoStop}%
\bibitem [{\citenamefont {Hartmann}\ and\ \citenamefont
  {Young}(2001)}]{HartmannYoung2001LowerCriticalDimensionOfIsingSpinGlasses}%
  \BibitemOpen
  \bibfield  {author} {\bibinfo {author} {\bibfnamefont {A.~K.}\ \bibnamefont
  {Hartmann}}\ and\ \bibinfo {author} {\bibfnamefont {A.~P.}\ \bibnamefont
  {Young}},\ }\bibfield  {title} {\bibinfo {title} {{Lower critical dimension
  of Ising spin glasses}},\ }\href@noop {} {\bibfield  {journal} {\bibinfo
  {journal} {Phys. Rev. B}\ }\textbf {\bibinfo {volume} {{64}}},\ \bibinfo
  {pages} {180404} (\bibinfo {year} {2001})}\BibitemShut {NoStop}%
\bibitem [{\citenamefont {Houdayer}\ and\ \citenamefont
  {Hartmann}(2004)}]{HoudayerHartmann2004LowTemperatureBehaviorOfTwoDimensionalGaussianIsingSpinGlasses}%
  \BibitemOpen
  \bibfield  {author} {\bibinfo {author} {\bibfnamefont {J.}~\bibnamefont
  {Houdayer}}\ and\ \bibinfo {author} {\bibfnamefont {A.~K.}\ \bibnamefont
  {Hartmann}},\ }\bibfield  {title} {\bibinfo {title} {{Low-temperature
  behavior of two-dimensional Gaussian Ising spin glasses}},\ }\href@noop {}
  {\bibfield  {journal} {\bibinfo  {journal} {Phys. Rev. B}\ }\textbf {\bibinfo
  {volume} {{70}}},\ \bibinfo {pages} {014418} (\bibinfo {year}
  {2004})}\BibitemShut {NoStop}%
\bibitem [{\citenamefont {Katzgraber}\ \emph {et~al.}(2004)\citenamefont
  {Katzgraber}, \citenamefont {Lee},\ and\ \citenamefont
  {Young}}]{KatzgraberLeeYoung2004CorrelationLengthOftheTwoDimensionalIsingSpinGlassWithGaussianInteractions}%
  \BibitemOpen
  \bibfield  {author} {\bibinfo {author} {\bibfnamefont {H.~G.}\ \bibnamefont
  {Katzgraber}}, \bibinfo {author} {\bibfnamefont {L.~W.}\ \bibnamefont
  {Lee}},\ and\ \bibinfo {author} {\bibfnamefont {A.~P.}\ \bibnamefont
  {Young}},\ }\bibfield  {title} {\bibinfo {title} {{Correlation length of the
  two-dimensional Ising spin glass with Gaussian interactions}},\ }\href@noop
  {} {\bibfield  {journal} {\bibinfo  {journal} {Phys. Rev. B}\ }\textbf
  {\bibinfo {volume} {{70}}},\ \bibinfo {pages} {014417} (\bibinfo {year}
  {2004})}\BibitemShut {NoStop}%
\bibitem [{\citenamefont {Fernandez}\ \emph {et~al.}(2016)\citenamefont
  {Fernandez}, \citenamefont {Marinari}, \citenamefont {Martin-Mayor},
  \citenamefont {Parisi},\ and\ \citenamefont
  {Ruiz-Lorenzo}}]{FernandezEtAl2016UniversalCriticalBehaviorOfTheTwoDimensionalIsingSpinGlass}%
  \BibitemOpen
  \bibfield  {author} {\bibinfo {author} {\bibfnamefont {L.}~\bibnamefont
  {Fernandez}}, \bibinfo {author} {\bibfnamefont {E.}~\bibnamefont {Marinari}},
  \bibinfo {author} {\bibfnamefont {V.}~\bibnamefont {Martin-Mayor}}, \bibinfo
  {author} {\bibfnamefont {G.}~\bibnamefont {Parisi}},\ and\ \bibinfo {author}
  {\bibfnamefont {J.}~\bibnamefont {Ruiz-Lorenzo}},\ }\bibfield  {title}
  {\bibinfo {title} {{Universal critical behavior of the two-dimensional Ising
  spin glass}},\ }\href@noop {} {\bibfield  {journal} {\bibinfo  {journal}
  {Phys. Rev. B}\ }\textbf {\bibinfo {volume} {{94}}},\ \bibinfo {pages}
  {024402} (\bibinfo {year} {2016})}\BibitemShut {NoStop}%
\bibitem [{\citenamefont {Khoshbakht}\ and\ \citenamefont
  {Weigel}(2018)}]{KhoshbakhtWeigel2018DomainWallExcitationsInTheTwoDimensionalIsingSpinGlass}%
  \BibitemOpen
  \bibfield  {author} {\bibinfo {author} {\bibfnamefont {H.}~\bibnamefont
  {Khoshbakht}}\ and\ \bibinfo {author} {\bibfnamefont {M.}~\bibnamefont
  {Weigel}},\ }\bibfield  {title} {\bibinfo {title} {{Domain-wall excitations
  in the two-dimensional Ising spin glass}},\ }\href@noop {} {\bibfield
  {journal} {\bibinfo  {journal} {Phys. Rev. B}\ }\textbf {\bibinfo {volume}
  {{97}}},\ \bibinfo {pages} {064410} (\bibinfo {year} {2018})}\BibitemShut
  {NoStop}%
\bibitem [{\citenamefont {Katzgraber}\ \emph {et~al.}(2001)\citenamefont
  {Katzgraber}, \citenamefont {Palassini},\ and\ \citenamefont
  {Young}}]{KatzgraberPalassiniYoung2001MonteCarloSimulationsOfSpinGlassesAtLowTemperatures}%
  \BibitemOpen
  \bibfield  {author} {\bibinfo {author} {\bibfnamefont {H.~G.}\ \bibnamefont
  {Katzgraber}}, \bibinfo {author} {\bibfnamefont {M.}~\bibnamefont
  {Palassini}},\ and\ \bibinfo {author} {\bibfnamefont {A.}~\bibnamefont
  {Young}},\ }\bibfield  {title} {\bibinfo {title} {{Monte Carlo simulations of
  spin glasses at low temperatures}},\ }\href@noop {} {\bibfield  {journal}
  {\bibinfo  {journal} {Phys. Rev. B}\ }\textbf {\bibinfo {volume} {{63}}},\
  \bibinfo {pages} {184422} (\bibinfo {year} {2001})}\BibitemShut {NoStop}%
\bibitem [{\citenamefont
  {Contucci}(2003)}]{Contucci2003ReplicaEquivalenceInTheEaModel}%
  \BibitemOpen
  \bibfield  {author} {\bibinfo {author} {\bibfnamefont {P.}~\bibnamefont
  {Contucci}},\ }\bibfield  {title} {\bibinfo {title} {{Replica equivalence in
  the Edwards-Anderson model}},\ }\href@noop {} {\bibfield  {journal} {\bibinfo
   {journal} {J. Phys. A}\ }\textbf {\bibinfo {volume} {{36}}},\ \bibinfo
  {pages} {10961} (\bibinfo {year} {2003})}\BibitemShut {NoStop}%
\bibitem [{Note1()}]{Note1}%
  \BibitemOpen
  \bibinfo {note} {Note that similar properties can be derived for the
  SK-model, but in case of the SK-model the covariance is given by the square
  power of the overlap \cite
  {BrayMoore1980SomeObservationsOnTheMeanFieldTheoryOfSG,ContucciEtAl2003ThermodynamicalLimitForCorrelatedGaussianRandomEnergyModels}.}\BibitemShut
  {Stop}%
\bibitem [{\citenamefont {Contucci}\ and\ \citenamefont
  {Giardina}(2005{\natexlab{a}})}]{ContucciGiardina2005SpinGlassStochasticStability}%
  \BibitemOpen
  \bibfield  {author} {\bibinfo {author} {\bibfnamefont {P.}~\bibnamefont
  {Contucci}}\ and\ \bibinfo {author} {\bibfnamefont {C.}~\bibnamefont
  {Giardina}},\ }\bibfield  {title} {\bibinfo {title} {Spin-glass stochastic
  stability: a rigorous proof},\ }\href@noop {} {\bibfield  {journal} {\bibinfo
   {journal} {Ann. Henri Poincare}\ }\textbf {\bibinfo {volume} {{6}}},\
  \bibinfo {pages} {915} (\bibinfo {year} {2005}{\natexlab{a}})}\BibitemShut
  {NoStop}%
\bibitem [{\citenamefont {Contucci}\ and\ \citenamefont
  {Giardina}(2005{\natexlab{b}})}]{Contucci2005FactorizationPropertiesInTheThreeDimensionalEAModel}%
  \BibitemOpen
  \bibfield  {author} {\bibinfo {author} {\bibfnamefont {P.}~\bibnamefont
  {Contucci}}\ and\ \bibinfo {author} {\bibfnamefont {C.}~\bibnamefont
  {Giardina}},\ }\bibfield  {title} {\bibinfo {title} {{Factorization
  properties in the three-dimensional Edwards-Anderson model}},\ }\href@noop {}
  {\bibfield  {journal} {\bibinfo  {journal} {Phys. Rev. B}\ }\textbf {\bibinfo
  {volume} {{72}}},\ \bibinfo {pages} {014456} (\bibinfo {year}
  {2005}{\natexlab{b}})}\BibitemShut {NoStop}%
\bibitem [{\citenamefont {Contucci}\ \emph {et~al.}(2006)\citenamefont
  {Contucci}, \citenamefont {Giardina}, \citenamefont {Giberti},\ and\
  \citenamefont {Vernia}}]{Contucci2006OverlapEquivalenceInTheEAModel}%
  \BibitemOpen
  \bibfield  {author} {\bibinfo {author} {\bibfnamefont {P.}~\bibnamefont
  {Contucci}}, \bibinfo {author} {\bibfnamefont {C.}~\bibnamefont {Giardina}},
  \bibinfo {author} {\bibfnamefont {C.}~\bibnamefont {Giberti}},\ and\ \bibinfo
  {author} {\bibfnamefont {C.}~\bibnamefont {Vernia}},\ }\bibfield  {title}
  {\bibinfo {title} {Overlap equivalence in the {E}dwards-{A}nderson model},\
  }\href@noop {} {\bibfield  {journal} {\bibinfo  {journal} {Phys. Rev. Lett.}\
  }\textbf {\bibinfo {volume} {{96}}},\ \bibinfo {pages} {217204} (\bibinfo
  {year} {2006})}\BibitemShut {NoStop}%
\bibitem [{\citenamefont {Newman}\ and\ \citenamefont
  {Stein}(2007{\natexlab{b}})}]{NewmanStein2007LocalVsGlobalVariablesForSpinGlasses}%
  \BibitemOpen
  \bibfield  {author} {\bibinfo {author} {\bibfnamefont {C.~M.}\ \bibnamefont
  {Newman}}\ and\ \bibinfo {author} {\bibfnamefont {D.~L.}\ \bibnamefont
  {Stein}},\ }\bibfield  {title} {\bibinfo {title} {Local vs. global variables
  for spin glasses},\ }in\ \href@noop {} {\emph {\bibinfo {booktitle} {Spin
  Glasses}}},\ \bibinfo {editor} {edited by\ \bibinfo {editor} {\bibfnamefont
  {E.}~\bibnamefont {Bolthausen}}\ and\ \bibinfo {editor} {\bibfnamefont
  {A.}~\bibnamefont {Bovier}}}\ (\bibinfo  {publisher} {Springer},\ \bibinfo
  {address} {Berlin, Heidelberg},\ \bibinfo {year} {2007})\ p.\ \bibinfo
  {pages} {145}\BibitemShut {NoStop}%
\bibitem [{\citenamefont
  {Wolff}(1989{\natexlab{a}})}]{Wolff1988CollectiveMCUpdatingForSpinSystems}%
  \BibitemOpen
  \bibfield  {author} {\bibinfo {author} {\bibfnamefont {U.}~\bibnamefont
  {Wolff}},\ }\bibfield  {title} {\bibinfo {title} {Collective {M}onte {C}arlo
  updating for spin systems},\ }\href@noop {} {\bibfield  {journal} {\bibinfo
  {journal} {Phys. Rev. Lett.}\ }\textbf {\bibinfo {volume} {{62}}},\ \bibinfo
  {pages} {361} (\bibinfo {year} {1989}{\natexlab{a}})}\BibitemShut {NoStop}%
\bibitem [{\citenamefont {Swendsen}\ and\ \citenamefont
  {Wang}(1986)}]{SwendsenWang1986replicaReplicaMonteCarloSimulationOfSpinGlasses}%
  \BibitemOpen
  \bibfield  {author} {\bibinfo {author} {\bibfnamefont {R.~H.}\ \bibnamefont
  {Swendsen}}\ and\ \bibinfo {author} {\bibfnamefont {J.-S.}\ \bibnamefont
  {Wang}},\ }\bibfield  {title} {\bibinfo {title} {Replica {M}onte {C}arlo
  simulation of spin-glasses},\ }\href@noop {} {\bibfield  {journal} {\bibinfo
  {journal} {Phys. Rev. Lett.}\ }\textbf {\bibinfo {volume} {{57}}},\ \bibinfo
  {pages} {2607} (\bibinfo {year} {1986})}\BibitemShut {NoStop}%
\bibitem [{\citenamefont
  {Yamaguchi}(2010)}]{Yamaguchi2010PercolationThresholdsOfTheFortuinKasteleynClusterForTheEAIsingModelOnComplexNetworks}%
  \BibitemOpen
  \bibfield  {author} {\bibinfo {author} {\bibfnamefont {C.}~\bibnamefont
  {Yamaguchi}},\ }\bibfield  {title} {\bibinfo {title} {Percolation thresholds
  of the {F}ortuin-{K}asteleyn cluster for the {E}dwards-{A}nderson {I}sing
  model on complex networks: --{A}nalytical results on the {N}ishimori
  line--},\ }\href@noop {} {\bibfield  {journal} {\bibinfo  {journal} {Prog.
  Theor. Phys.}\ }\textbf {\bibinfo {volume} {{124}}},\ \bibinfo {pages} {399}
  (\bibinfo {year} {2010})}\BibitemShut {NoStop}%
\bibitem [{\citenamefont
  {Wolff}(1989{\natexlab{b}})}]{Wolff1989ComparisonBetweenClusterMCAlgorithmsInTheIsingModel}%
  \BibitemOpen
  \bibfield  {author} {\bibinfo {author} {\bibfnamefont {U.}~\bibnamefont
  {Wolff}},\ }\bibfield  {title} {\bibinfo {title} {Comparison between cluster
  {M}onte {C}arlo algorithms in the ising model},\ }\href@noop {} {\bibfield
  {journal} {\bibinfo  {journal} {Phys. Lett. B}\ }\textbf {\bibinfo {volume}
  {{228}}},\ \bibinfo {pages} {379} (\bibinfo {year}
  {1989}{\natexlab{b}})}\BibitemShut {NoStop}%
\bibitem [{\citenamefont {Newman}\ and\ \citenamefont
  {Barkema}(1999)}]{NewmanBarkema1999MonteCarloMethodsInStatisticalPhysics}%
  \BibitemOpen
  \bibfield  {author} {\bibinfo {author} {\bibfnamefont {M.~E.}\ \bibnamefont
  {Newman}}\ and\ \bibinfo {author} {\bibfnamefont {G.~T.}\ \bibnamefont
  {Barkema}},\ }\href@noop {} {\emph {\bibinfo {title} {{Monte Carlo Methods in
  Statistical Physics}}}}\ (\bibinfo  {publisher} {Oxford University Press},\
  \bibinfo {address} {Oxford},\ \bibinfo {year} {1999})\BibitemShut {NoStop}%
\bibitem [{\citenamefont {Coniglio}\ \emph {et~al.}(1991)\citenamefont
  {Coniglio}, \citenamefont {di~Liberto}, \citenamefont {Monroy},\ and\
  \citenamefont
  {Peruggi}}]{ConiglioEtAl1991ClusterApproachToSpinGlassesAndTheFrustratedPercolationProblem}%
  \BibitemOpen
  \bibfield  {author} {\bibinfo {author} {\bibfnamefont {A.}~\bibnamefont
  {Coniglio}}, \bibinfo {author} {\bibfnamefont {F.}~\bibnamefont
  {di~Liberto}}, \bibinfo {author} {\bibfnamefont {G.}~\bibnamefont {Monroy}},\
  and\ \bibinfo {author} {\bibfnamefont {F.}~\bibnamefont {Peruggi}},\
  }\bibfield  {title} {\bibinfo {title} {{Cluster approach to spin glasses and
  the frustrated-percolation problem}},\ }\href@noop {} {\bibfield  {journal}
  {\bibinfo  {journal} {Phys. Rev. B}\ }\textbf {\bibinfo {volume} {{44}}},\
  \bibinfo {pages} {12605} (\bibinfo {year} {1991})}\BibitemShut {NoStop}%
\bibitem [{\citenamefont {Cataudella}\ \emph {et~al.}(1994)\citenamefont
  {Cataudella}, \citenamefont {Franzese}, \citenamefont {Nicodemi},
  \citenamefont {Scala},\ and\ \citenamefont
  {Coniglio}}]{CataudellaEtAL1994CriticalClustersAndEfficientDynamicsForFrustratedSpinModels}%
  \BibitemOpen
  \bibfield  {author} {\bibinfo {author} {\bibfnamefont {V.}~\bibnamefont
  {Cataudella}}, \bibinfo {author} {\bibfnamefont {G.}~\bibnamefont
  {Franzese}}, \bibinfo {author} {\bibfnamefont {M.}~\bibnamefont {Nicodemi}},
  \bibinfo {author} {\bibfnamefont {A.}~\bibnamefont {Scala}},\ and\ \bibinfo
  {author} {\bibfnamefont {A.}~\bibnamefont {Coniglio}},\ }\bibfield  {title}
  {\bibinfo {title} {Critical clusters and efficient dynamics for frustrated
  spin models},\ }\href@noop {} {\bibfield  {journal} {\bibinfo  {journal}
  {Phys. Rev. Lett.}\ }\textbf {\bibinfo {volume} {{72}}},\ \bibinfo {pages}
  {1541} (\bibinfo {year} {1994})}\BibitemShut {NoStop}%
\bibitem [{\citenamefont {De~Santis}\ and\ \citenamefont
  {Gandolfi}(1999)}]{DeSantis1999BondPercolationInFrustratedSystems}%
  \BibitemOpen
  \bibfield  {author} {\bibinfo {author} {\bibfnamefont {E.}~\bibnamefont
  {De~Santis}}\ and\ \bibinfo {author} {\bibfnamefont {A.}~\bibnamefont
  {Gandolfi}},\ }\bibfield  {title} {\bibinfo {title} {Bond percolation in
  frustrated systems},\ }\href@noop {} {\bibfield  {journal} {\bibinfo
  {journal} {The Annals of Probability}\ }\textbf {\bibinfo {volume} {27}},\
  \bibinfo {pages} {1781} (\bibinfo {year} {1999})}\BibitemShut {NoStop}%
\bibitem [{\citenamefont {Lundow}\ and\ \citenamefont
  {Campbell}(2012)}]{LundowCampbell2012FortuinKasteleynAndDamageSpreadingTransitionsInRandomBondIsingLattices}%
  \BibitemOpen
  \bibfield  {author} {\bibinfo {author} {\bibfnamefont {P.~H.}\ \bibnamefont
  {Lundow}}\ and\ \bibinfo {author} {\bibfnamefont {I.~A.}\ \bibnamefont
  {Campbell}},\ }\bibfield  {title} {\bibinfo {title} {{F}ortuin-{K}asteleyn
  and damage-spreading transitions in random-bond {I}sing lattices},\
  }\href@noop {} {\bibfield  {journal} {\bibinfo  {journal} {Phys. Rev. E}\
  }\textbf {\bibinfo {volume} {{86}}},\ \bibinfo {pages} {041121} (\bibinfo
  {year} {2012})}\BibitemShut {NoStop}%
\bibitem [{\citenamefont {Binder}\ and\ \citenamefont
  {Heermann}(2010)}]{Binder2010FSSOfPercolation}%
  \BibitemOpen
  \bibfield  {author} {\bibinfo {author} {\bibfnamefont {K.}~\bibnamefont
  {Binder}}\ and\ \bibinfo {author} {\bibfnamefont {D.~W.}\ \bibnamefont
  {Heermann}},\ }\bibinfo {title} {Theoretical foundations of the {M}onte
  {C}arlo method and its applications in statistical physics},\ in\ \href@noop
  {} {\emph {\bibinfo {booktitle} {{Monte Carlo Simulation in Statistical
  Physics: An Introduction}}}}\ (\bibinfo  {publisher} {Springer},\ \bibinfo
  {address} {Berlin, Heidelberg},\ \bibinfo {year} {2010})\ p.~\bibinfo {pages}
  {5},\ \bibinfo {edition} {5th}\ ed.\BibitemShut {Stop}%
\bibitem [{\citenamefont {Melchert}()}]{Melchert2009Autoscale}%
  \BibitemOpen
  \bibfield  {author} {\bibinfo {author} {\bibfnamefont {O.}~\bibnamefont
  {Melchert}},\ }\bibfield  {title} {\bibinfo {title} {{autoScale.py -- A
  program for automatic finite-size scaling analyses: A user's guide}},\ }\href
  {https://arxiv.org/abs/0910.5403} {\bibinfo  {journal} {arXiv:0910.5403}\
  }\BibitemShut {NoStop}%
\bibitem [{Note2()}]{Note2}%
  \BibitemOpen
\bibfield  {journal} {  }\bibinfo {note} {Note that we could have used such
  estimates from different observables for improving the accuracy of the
  estimates for $\beta _\protect \mathrm {FK}$ and the critical exponents by
  taking the cross correlations into account~\cite {weigel:09}, but we found
  that even without such extra effort the agreement with the percolation
  universality class is rather clear-cut.}\BibitemShut {Stop}%
\bibitem [{\citenamefont {Janke}\ and\ \citenamefont
  {Schakel}(2005)}]{JankeSchakel2005FractalStructureOfSpinClustersAndDomainWallsInThe2DIsingModel}%
  \BibitemOpen
  \bibfield  {author} {\bibinfo {author} {\bibfnamefont {W.}~\bibnamefont
  {Janke}}\ and\ \bibinfo {author} {\bibfnamefont {A.~M.}\ \bibnamefont
  {Schakel}},\ }\bibfield  {title} {\bibinfo {title} {{Fractal structure of
  spin clusters and domain walls in the two-dimensional Ising model}},\
  }\href@noop {} {\bibfield  {journal} {\bibinfo  {journal} {Phys. Rev. E}\
  }\textbf {\bibinfo {volume} {{71}}},\ \bibinfo {pages} {036703} (\bibinfo
  {year} {2005})}\BibitemShut {NoStop}%
\bibitem [{\citenamefont {Akritidis}\ \emph {et~al.}(2022)\citenamefont
  {Akritidis}, \citenamefont {Fytas},\ and\ \citenamefont
  {Weigel}}]{AkritidisFytasWeigel2022CorrectionsToScalingInGeometricalClustersOfThe2DIsingModel}%
  \BibitemOpen
  \bibfield  {author} {\bibinfo {author} {\bibfnamefont {M.}~\bibnamefont
  {Akritidis}}, \bibinfo {author} {\bibfnamefont {N.~G.}\ \bibnamefont
  {Fytas}},\ and\ \bibinfo {author} {\bibfnamefont {M.}~\bibnamefont
  {Weigel}},\ }\bibfield  {title} {\bibinfo {title} {{Corrections to scaling in
  geometrical clusters of the 2D Ising model}},\ }\href@noop {} {\bibfield
  {journal} {\bibinfo  {journal} {J. Phys. Conf. Ser.}\ }\textbf {\bibinfo
  {volume} {{2207}}},\ \bibinfo {pages} {012004} (\bibinfo {year}
  {2022})}\BibitemShut {NoStop}%
\bibitem [{\citenamefont {Imaoka}\ \emph {et~al.}(1997)\citenamefont {Imaoka},
  \citenamefont {Ikeda},\ and\ \citenamefont
  {Kasai}}]{ImaokaEtAl1997PercolationTransitionIn2DJIsingSG}%
  \BibitemOpen
  \bibfield  {author} {\bibinfo {author} {\bibfnamefont {H.}~\bibnamefont
  {Imaoka}}, \bibinfo {author} {\bibfnamefont {H.}~\bibnamefont {Ikeda}},\ and\
  \bibinfo {author} {\bibfnamefont {Y.}~\bibnamefont {Kasai}},\ }\bibfield
  {title} {\bibinfo {title} {Percolation transition in two-dimensional {$\pm
  J$} {I}sing spin glasses},\ }\href@noop {} {\bibfield  {journal} {\bibinfo
  {journal} {Physica A}\ }\textbf {\bibinfo {volume} {{246}}},\ \bibinfo
  {pages} {18} (\bibinfo {year} {1997})}\BibitemShut {NoStop}%
\bibitem [{\citenamefont {Fajen}\ \emph {et~al.}(2020)\citenamefont {Fajen},
  \citenamefont {Hartmann},\ and\ \citenamefont
  {Young}}]{FajenHartmannYoung2020percolationPercolationOfFortuinKasteleynClustersForTheRandomBondIsingModel}%
  \BibitemOpen
  \bibfield  {author} {\bibinfo {author} {\bibfnamefont {H.}~\bibnamefont
  {Fajen}}, \bibinfo {author} {\bibfnamefont {A.~K.}\ \bibnamefont
  {Hartmann}},\ and\ \bibinfo {author} {\bibfnamefont {A.~P.}\ \bibnamefont
  {Young}},\ }\bibfield  {title} {\bibinfo {title} {Percolation of
  {F}ortuin-{K}asteleyn clusters for the random-bond ising model},\ }\href@noop
  {} {\bibfield  {journal} {\bibinfo  {journal} {Phys. Rev. E}\ }\textbf
  {\bibinfo {volume} {{102}}},\ \bibinfo {pages} {012131} (\bibinfo {year}
  {2020})}\BibitemShut {NoStop}%
\bibitem [{\citenamefont
  {Nishimori}(1981)}]{Nishimori1981IInternalEnergySpecificHeatAndCorrelationFunctionOftheBondRandomIsingModel}%
  \BibitemOpen
  \bibfield  {author} {\bibinfo {author} {\bibfnamefont {H.}~\bibnamefont
  {Nishimori}},\ }\bibfield  {title} {\bibinfo {title} {Internal energy,
  specific heat and correlation function of the bond-random {I}sing model},\
  }\href@noop {} {\bibfield  {journal} {\bibinfo  {journal} {Prog. Theor.
  Phys.}\ }\textbf {\bibinfo {volume} {{66}}},\ \bibinfo {pages} {1169}
  (\bibinfo {year} {1981})}\BibitemShut {NoStop}%
\bibitem [{\citenamefont
  {Mazza}(1999)}]{Mazza1999GaugeSymmetriesAndPercolationInJIsingSpinGlasses}%
  \BibitemOpen
  \bibfield  {author} {\bibinfo {author} {\bibfnamefont {C.}~\bibnamefont
  {Mazza}},\ }\bibfield  {title} {\bibinfo {title} {Gauge symmetries and
  percolation in {$\pm J$} {I}sing spin glasses},\ }\href@noop {} {\bibfield
  {journal} {\bibinfo  {journal} {Probab. Theory Relat. Fields}\ }\textbf
  {\bibinfo {volume} {{113}}},\ \bibinfo {pages} {171} (\bibinfo {year}
  {1999})}\BibitemShut {NoStop}%
\bibitem [{\citenamefont
  {Gandolfi}(1999)}]{Gandolfi1999ARemarkOnGaugeSymmetriesInIsingSGs}%
  \BibitemOpen
  \bibfield  {author} {\bibinfo {author} {\bibfnamefont {A.}~\bibnamefont
  {Gandolfi}},\ }\bibfield  {title} {\bibinfo {title} {A remark on gauge
  symmetries in {I}sing spin glasses},\ }\href@noop {} {\bibfield  {journal}
  {\bibinfo  {journal} {Probab. Theory Relat. Fields}\ }\textbf {\bibinfo
  {volume} {114}},\ \bibinfo {pages} {419} (\bibinfo {year}
  {1999})}\BibitemShut {NoStop}%
\bibitem [{\citenamefont
  {Nishimori}(1980)}]{Nishimori1980ExactResultsAndCriticalPropertiesOfTheIsingModelWithCompetingInteractions}%
  \BibitemOpen
  \bibfield  {author} {\bibinfo {author} {\bibfnamefont {H.}~\bibnamefont
  {Nishimori}},\ }\bibfield  {title} {\bibinfo {title} {{Exact results and
  critical properties of the Ising model with competing interactions}},\
  }\href@noop {} {\bibfield  {journal} {\bibinfo  {journal} {J. Phys. C}\
  }\textbf {\bibinfo {volume} {{13}}},\ \bibinfo {pages} {4071} (\bibinfo
  {year} {1980})}\BibitemShut {NoStop}%
\bibitem [{Note3()}]{Note3}%
  \BibitemOpen
  \bibinfo {note} {For the $\pm J$ Ising spin glass it was proven by Gandolfi
  that FKCK percolation on the Nishimori line reduces to a random-bond
  percolation problem~\cite {Gandolfi1999ARemarkOnGaugeSymmetriesInIsingSGs},
  such that the percolation temperatures predicted by Yamaguchi are exact in
  this case.}\BibitemShut {Stop}%
\bibitem [{\citenamefont {Bai}\ and\ \citenamefont
  {Breen}(2008)}]{BaiBreen2008CalculatingCenterOfMassInAnUnbounded2DEnvironment}%
  \BibitemOpen
  \bibfield  {author} {\bibinfo {author} {\bibfnamefont {L.}~\bibnamefont
  {Bai}}\ and\ \bibinfo {author} {\bibfnamefont {D.}~\bibnamefont {Breen}},\
  }\bibfield  {title} {\bibinfo {title} {Calculating center of mass in an
  unbounded {2D} environment},\ }\href@noop {} {\bibfield  {journal} {\bibinfo
  {journal} {J. Graphics Tools}\ }\textbf {\bibinfo {volume} {{13}}},\ \bibinfo
  {pages} {53} (\bibinfo {year} {2008})}\BibitemShut {NoStop}%
\bibitem [{Note4()}]{Note4}%
  \BibitemOpen
  \bibinfo {note} {Note that this implies that they are satisfied in {\protect
  \em all\/} (i.e., both) replicas for $I=2$.}\BibitemShut {Stop}%
\bibitem [{\citenamefont
  {J\"org}(2006)}]{Joerg2006CriticalBehaviorOfThe3DBondDilutedIsingSG}%
  \BibitemOpen
  \bibfield  {author} {\bibinfo {author} {\bibfnamefont {T.}~\bibnamefont
  {J\"org}},\ }\bibfield  {title} {\bibinfo {title} {{Critical behavior of the
  three-dimensional bond-diluted Ising spin glass: Finite-size scaling
  functions and universality}},\ }\href@noop {} {\bibfield  {journal} {\bibinfo
   {journal} {Phys. Rev. B}\ }\textbf {\bibinfo {volume} {{73}}},\ \bibinfo
  {pages} {224431} (\bibinfo {year} {2006})}\BibitemShut {NoStop}%
\bibitem [{\citenamefont {Vaezi}\ \emph {et~al.}(2018)\citenamefont {Vaezi},
  \citenamefont {Ortiz}, \citenamefont {Weigel},\ and\ \citenamefont
  {Nussinov}}]{vaezi:17}%
  \BibitemOpen
  \bibfield  {author} {\bibinfo {author} {\bibfnamefont {M.-S.}\ \bibnamefont
  {Vaezi}}, \bibinfo {author} {\bibfnamefont {G.}~\bibnamefont {Ortiz}},
  \bibinfo {author} {\bibfnamefont {M.}~\bibnamefont {Weigel}},\ and\ \bibinfo
  {author} {\bibfnamefont {Z.}~\bibnamefont {Nussinov}},\ }\bibfield  {title}
  {\bibinfo {title} {Binomial spin glass},\ }\href
  {https://doi.org/10.1103/PhysRevLett.121.080601} {\bibfield  {journal}
  {\bibinfo  {journal} {Phys. Rev. Lett.}\ }\textbf {\bibinfo {volume} {121}},\
  \bibinfo {pages} {080601} (\bibinfo {year} {2018})}\BibitemShut {NoStop}%
\bibitem [{\citenamefont {Arguin}\ and\ \citenamefont
  {Damron}(2014)}]{ArguinDamron2014OnTheNumberOfGroundStatesOfTheEAModel}%
  \BibitemOpen
  \bibfield  {author} {\bibinfo {author} {\bibfnamefont {L.-P.}\ \bibnamefont
  {Arguin}}\ and\ \bibinfo {author} {\bibfnamefont {M.}~\bibnamefont
  {Damron}},\ }\bibfield  {title} {\bibinfo {title} {On the number of ground
  states of the {E}dwards-{A}nderson spin glass model},\ }\href@noop {}
  {\bibfield  {journal} {\bibinfo  {journal} {Ann. I. H. Poincar\'e-PR}\
  }\textbf {\bibinfo {volume} {{50}}},\ \bibinfo {pages} {28} (\bibinfo {year}
  {2014})}\BibitemShut {NoStop}%
\bibitem [{\citenamefont {Thomas}\ \emph {et~al.}(2011)\citenamefont {Thomas},
  \citenamefont {Huse},\ and\ \citenamefont
  {Middleton}}]{CreightonHuseMiddleton2011ZeroAndLowTemperatureBehaviorOfThe2DPmJIsingSG}%
  \BibitemOpen
  \bibfield  {author} {\bibinfo {author} {\bibfnamefont {C.~K.}\ \bibnamefont
  {Thomas}}, \bibinfo {author} {\bibfnamefont {D.~A.}\ \bibnamefont {Huse}},\
  and\ \bibinfo {author} {\bibfnamefont {A.~A.}\ \bibnamefont {Middleton}},\
  }\bibfield  {title} {\bibinfo {title} {Zero- and low-temperature behavior of
  the two-dimensional {$\pm J$} {I}sing spin glass},\ }\href@noop {} {\bibfield
   {journal} {\bibinfo  {journal} {Phys. Rev. Lett.}\ }\textbf {\bibinfo
  {volume} {{107}}},\ \bibinfo {pages} {047203} (\bibinfo {year}
  {2011})}\BibitemShut {NoStop}%
\bibitem [{\citenamefont {Parisen~Toldin}\ \emph {et~al.}(2011)\citenamefont
  {Parisen~Toldin}, \citenamefont {Pelissetto},\ and\ \citenamefont
  {Vicari}}]{ParisenEtAl2011FSSIn2DIsingSG}%
  \BibitemOpen
  \bibfield  {author} {\bibinfo {author} {\bibfnamefont {F.}~\bibnamefont
  {Parisen~Toldin}}, \bibinfo {author} {\bibfnamefont {A.}~\bibnamefont
  {Pelissetto}},\ and\ \bibinfo {author} {\bibfnamefont {E.}~\bibnamefont
  {Vicari}},\ }\bibfield  {title} {\bibinfo {title} {Finite-size scaling in
  two-dimensional {I}sing spin-glass models},\ }\href@noop {} {\bibfield
  {journal} {\bibinfo  {journal} {Phys. Rev. E}\ }\textbf {\bibinfo {volume}
  {{84}}},\ \bibinfo {pages} {051116} (\bibinfo {year} {2011})}\BibitemShut
  {NoStop}%
\bibitem [{\citenamefont {J\"org}\ \emph {et~al.}(2006)\citenamefont {J\"org},
  \citenamefont {Lukic}, \citenamefont {Marinari},\ and\ \citenamefont
  {Martin}}]{JoergEtAl2006StrongUniversalityAndAlgebraicScalingIn2DIsingSG}%
  \BibitemOpen
  \bibfield  {author} {\bibinfo {author} {\bibfnamefont {T.}~\bibnamefont
  {J\"org}}, \bibinfo {author} {\bibfnamefont {J.}~\bibnamefont {Lukic}},
  \bibinfo {author} {\bibfnamefont {E.}~\bibnamefont {Marinari}},\ and\
  \bibinfo {author} {\bibfnamefont {O.~C.}\ \bibnamefont {Martin}},\ }\bibfield
   {title} {\bibinfo {title} {Strong universality and algebraic scaling in
  two-dimensional {I}sing spin glasses},\ }\href@noop {} {\bibfield  {journal}
  {\bibinfo  {journal} {Phys. Rev. Lett.}\ }\textbf {\bibinfo {volume}
  {{96}}},\ \bibinfo {pages} {237205} (\bibinfo {year} {2006})}\BibitemShut
  {NoStop}%
\bibitem [{\citenamefont
  {Young}(2015)}]{Young2015EverythingYouWantedToKnowAboutDataAnalysis}%
  \BibitemOpen
  \bibfield  {author} {\bibinfo {author} {\bibfnamefont {A.~P.}\ \bibnamefont
  {Young}},\ }\href@noop {} {\emph {\bibinfo {title} {{Everything You Wanted to
  Know About Data Analysis and Fitting but Were Afraid to Ask}}}}\ (\bibinfo
  {publisher} {Springer},\ \bibinfo {address} {Cham},\ \bibinfo {year}
  {2015})\BibitemShut {NoStop}%
\bibitem [{\citenamefont {Pei}\ and\ \citenamefont
  {Di~Ventra}()}]{PeiDiVentra2021AFiniteTemperaturePhaseTransitionForTheSpinGlassIn2D}%
  \BibitemOpen
  \bibfield  {author} {\bibinfo {author} {\bibfnamefont {Y.~R.}\ \bibnamefont
  {Pei}}\ and\ \bibinfo {author} {\bibfnamefont {M.}~\bibnamefont
  {Di~Ventra}},\ }\bibfield  {title} {\bibinfo {title} {A finite-temperature
  phase transition for the {I}sing spin-glass in $d\geq 2$},\ }\href
  {https://arxiv.org/abs/2105.01188} {\bibinfo  {journal} {arXiv:2105.01188}\
  }\BibitemShut {NoStop}%
\bibitem [{Note5()}]{Note5}%
  \BibitemOpen
\bibfield  {journal} {  }\bibinfo {note} {Note that using more replicas in
  Eq.~\protect \textup {\hbox {\mathsurround \z@ \protect \normalfont
  (\ignorespaces \ref {eq:i_replica_fkck}\unskip \@@italiccorr )}} notably
  reduces the occupation probability and thus lowers the onset temperature for
  percolation. As a consequence, it might be interesting to see if for a
  certain $I$ the percolation transition is directly linked to the finite
  temperature spin-glass transition for $d>2$.}\BibitemShut {Stop}%
\bibitem [{\citenamefont {Katzgraber}\ \emph {et~al.}(2006)\citenamefont
  {Katzgraber}, \citenamefont {K{\"o}rner},\ and\ \citenamefont
  {Young}}]{KatzgraberEtAl2006Universality3DIsingSpinGlasses}%
  \BibitemOpen
  \bibfield  {author} {\bibinfo {author} {\bibfnamefont {H.~G.}\ \bibnamefont
  {Katzgraber}}, \bibinfo {author} {\bibfnamefont {M.}~\bibnamefont
  {K{\"o}rner}},\ and\ \bibinfo {author} {\bibfnamefont {A.}~\bibnamefont
  {Young}},\ }\bibfield  {title} {\bibinfo {title} {{Universality in
  three-dimensional Ising spin glasses: A Monte Carlo study}},\ }\href@noop {}
  {\bibfield  {journal} {\bibinfo  {journal} {Phys. Rev. B}\ }\textbf {\bibinfo
  {volume} {{73}}},\ \bibinfo {pages} {224432} (\bibinfo {year}
  {2006})}\BibitemShut {NoStop}%
\bibitem [{\citenamefont {Baity-Jesi}\ \emph {et~al.}(2013)\citenamefont
  {Baity-Jesi}, \citenamefont {Ba{\~n}os}, \citenamefont {Cruz}, \citenamefont
  {Fernandez}, \citenamefont {Gil-Narvion}, \citenamefont {Gordillo-Guerrero},
  \citenamefont {Iniguez}, \citenamefont {Maiorano}, \citenamefont {Mantovani},
  \citenamefont {Marinari} \emph
  {et~al.}}]{BaityJesiEtAl2013CriticalParametersOfThe3DIsingSG}%
  \BibitemOpen
  \bibfield  {author} {\bibinfo {author} {\bibfnamefont {M.}~\bibnamefont
  {Baity-Jesi}}, \bibinfo {author} {\bibfnamefont {R.}~\bibnamefont
  {Ba{\~n}os}}, \bibinfo {author} {\bibfnamefont {A.}~\bibnamefont {Cruz}},
  \bibinfo {author} {\bibfnamefont {L.~A.}\ \bibnamefont {Fernandez}}, \bibinfo
  {author} {\bibfnamefont {J.~M.}\ \bibnamefont {Gil-Narvion}}, \bibinfo
  {author} {\bibfnamefont {A.}~\bibnamefont {Gordillo-Guerrero}}, \bibinfo
  {author} {\bibfnamefont {D.}~\bibnamefont {Iniguez}}, \bibinfo {author}
  {\bibfnamefont {A.}~\bibnamefont {Maiorano}}, \bibinfo {author}
  {\bibfnamefont {F.}~\bibnamefont {Mantovani}}, \bibinfo {author}
  {\bibfnamefont {E.}~\bibnamefont {Marinari}}, \emph {et~al.},\ }\bibfield
  {title} {\bibinfo {title} {{Critical parameters of the three-dimensional
  Ising spin glass}},\ }\href@noop {} {\bibfield  {journal} {\bibinfo
  {journal} {Phys. Rev. B}\ }\textbf {\bibinfo {volume} {{88}}},\ \bibinfo
  {pages} {224416} (\bibinfo {year} {2013})}\BibitemShut {NoStop}%
\bibitem [{\citenamefont {Machta}\ \emph {et~al.}(2009)\citenamefont {Machta},
  \citenamefont {Newman},\ and\ \citenamefont
  {Stein}}]{MachtaNewmanStein2009APercolationTheoreticApproachToSGPhaseTransitions}%
  \BibitemOpen
  \bibfield  {author} {\bibinfo {author} {\bibfnamefont {J.}~\bibnamefont
  {Machta}}, \bibinfo {author} {\bibfnamefont {C.~M.}\ \bibnamefont {Newman}},\
  and\ \bibinfo {author} {\bibfnamefont {D.~L.}\ \bibnamefont {Stein}},\
  }\bibfield  {title} {\bibinfo {title} {A percolation-theoretic approach to
  spin glass phase transitions},\ }in\ \href@noop {} {\emph {\bibinfo
  {booktitle} {Spin Glasses: Statics and Dynamics}}},\ \bibinfo {editor}
  {edited by\ \bibinfo {editor} {\bibfnamefont {A.~B.}\ \bibnamefont
  {de~Monvel}}\ and\ \bibinfo {editor} {\bibfnamefont {A.}~\bibnamefont
  {Bovier}}}\ (\bibinfo  {publisher} {Birkh{\"a}user},\ \bibinfo {address}
  {Basel},\ \bibinfo {year} {2009})\ p.\ \bibinfo {pages} {205}\BibitemShut
  {NoStop}%
\bibitem [{\citenamefont {Coniglio}\ \emph {et~al.}(1976)\citenamefont
  {Coniglio}, \citenamefont {Nappi}, \citenamefont {Peruggi},\ and\
  \citenamefont
  {Russo}}]{ConiglioEtAl1976PercolationAndPhaseTransitionsInTheIsingModel}%
  \BibitemOpen
  \bibfield  {author} {\bibinfo {author} {\bibfnamefont {A.}~\bibnamefont
  {Coniglio}}, \bibinfo {author} {\bibfnamefont {C.~R.}\ \bibnamefont {Nappi}},
  \bibinfo {author} {\bibfnamefont {F.}~\bibnamefont {Peruggi}},\ and\ \bibinfo
  {author} {\bibfnamefont {L.}~\bibnamefont {Russo}},\ }\bibfield  {title}
  {\bibinfo {title} {{Percolation and phase transitions in the Ising model}},\
  }\href@noop {} {\bibfield  {journal} {\bibinfo  {journal} {Comm. Math.
  Phys.}\ }\textbf {\bibinfo {volume} {{51}}},\ \bibinfo {pages} {315}
  (\bibinfo {year} {1976})}\BibitemShut {NoStop}%
\bibitem [{\citenamefont {Zhu}\ \emph {et~al.}(2015)\citenamefont {Zhu},
  \citenamefont {Ochoa},\ and\ \citenamefont
  {Katzgraber}}]{ZhuOchoaKatzgraber2015EfficientClusterAlgorithmForSpinGlassesInAnySpaceDimension}%
  \BibitemOpen
  \bibfield  {author} {\bibinfo {author} {\bibfnamefont {Z.}~\bibnamefont
  {Zhu}}, \bibinfo {author} {\bibfnamefont {A.~J.}\ \bibnamefont {Ochoa}},\
  and\ \bibinfo {author} {\bibfnamefont {H.~G.}\ \bibnamefont {Katzgraber}},\
  }\bibfield  {title} {\bibinfo {title} {Efficient cluster algorithm for spin
  glasses in any space dimension},\ }\href@noop {} {\bibfield  {journal}
  {\bibinfo  {journal} {Phys. Rev. Lett.}\ }\textbf {\bibinfo {volume}
  {{115}}},\ \bibinfo {pages} {077201} (\bibinfo {year} {2015})}\BibitemShut
  {NoStop}%
\bibitem [{\citenamefont {Vandenbroucque}\ \emph {et~al.}()\citenamefont
  {Vandenbroucque}, \citenamefont {Chiacchio},\ and\ \citenamefont
  {Munro}}]{VandenbroucqueChiacchioMunro2022TheHoudayerAlgorithm}%
  \BibitemOpen
  \bibfield  {author} {\bibinfo {author} {\bibfnamefont {A.}~\bibnamefont
  {Vandenbroucque}}, \bibinfo {author} {\bibfnamefont {E.~I.~R.}\ \bibnamefont
  {Chiacchio}},\ and\ \bibinfo {author} {\bibfnamefont {E.}~\bibnamefont
  {Munro}},\ }\bibfield  {title} {\bibinfo {title} {{The Houdayer Algorithm:
  Overview, Extensions, and Applications}},\ }\href
  {https://arxiv.org/abs/2211.11556} {\bibinfo  {journal} {arXiv:2211.11556}\
  }\BibitemShut {NoStop}%
\bibitem [{\citenamefont {Akritidis}\ \emph {et~al.}()\citenamefont
  {Akritidis}, \citenamefont {Fytas},\ and\ \citenamefont
  {Weigel}}]{akritidis:prep}%
  \BibitemOpen
\bibfield  {journal} {  }\bibfield  {author} {\bibinfo {author} {\bibfnamefont
  {M.}~\bibnamefont {Akritidis}}, \bibinfo {author} {\bibfnamefont {N.~G.}\
  \bibnamefont {Fytas}},\ and\ \bibinfo {author} {\bibfnamefont
  {M.}~\bibnamefont {Weigel}},\ }\href@noop {} {\ }\bibinfo {note} {{i}n
  preparation.}\BibitemShut {Stop}%
\bibitem [{\citenamefont {Kandel}\ and\ \citenamefont
  {Domany}(1991)}]{KandelDomany1991GeneralClusterMonteCarloDynamics}%
  \BibitemOpen
  \bibfield  {author} {\bibinfo {author} {\bibfnamefont {D.}~\bibnamefont
  {Kandel}}\ and\ \bibinfo {author} {\bibfnamefont {E.}~\bibnamefont
  {Domany}},\ }\bibfield  {title} {\bibinfo {title} {{General cluster Monte
  Carlo dynamics}},\ }\href@noop {} {\bibfield  {journal} {\bibinfo  {journal}
  {Phys. Rev. B}\ }\textbf {\bibinfo {volume} {{43}}},\ \bibinfo {pages} {8539}
  (\bibinfo {year} {1991})}\BibitemShut {NoStop}%
\bibitem [{\citenamefont {Cataudella}\ \emph {et~al.}(1996)\citenamefont
  {Cataudella}, \citenamefont {Franzese}, \citenamefont {Nicodemi},
  \citenamefont {Scala},\ and\ \citenamefont
  {Coniglio}}]{Cataudella1996PercolationAndClusterMonteCarloDynamicsForSpinModels}%
  \BibitemOpen
  \bibfield  {author} {\bibinfo {author} {\bibfnamefont {V.}~\bibnamefont
  {Cataudella}}, \bibinfo {author} {\bibfnamefont {G.}~\bibnamefont
  {Franzese}}, \bibinfo {author} {\bibfnamefont {M.}~\bibnamefont {Nicodemi}},
  \bibinfo {author} {\bibfnamefont {A.}~\bibnamefont {Scala}},\ and\ \bibinfo
  {author} {\bibfnamefont {A.}~\bibnamefont {Coniglio}},\ }\bibfield  {title}
  {\bibinfo {title} {{Percolation and cluster Monte Carlo dynamics for spin
  models}},\ }\href@noop {} {\bibfield  {journal} {\bibinfo  {journal} {Phys.
  Rev. E}\ }\textbf {\bibinfo {volume} {{54}}},\ \bibinfo {pages} {175}
  (\bibinfo {year} {1996})}\BibitemShut {NoStop}%
\bibitem [{\citenamefont {Liu}\ \emph {et~al.}(2017)\citenamefont {Liu},
  \citenamefont {Qi}, \citenamefont {Meng},\ and\ \citenamefont
  {Fu}}]{LiuEtAl2017SelfLearningMonteCarloMethod}%
  \BibitemOpen
  \bibfield  {author} {\bibinfo {author} {\bibfnamefont {J.}~\bibnamefont
  {Liu}}, \bibinfo {author} {\bibfnamefont {Y.}~\bibnamefont {Qi}}, \bibinfo
  {author} {\bibfnamefont {Z.~Y.}\ \bibnamefont {Meng}},\ and\ \bibinfo
  {author} {\bibfnamefont {L.}~\bibnamefont {Fu}},\ }\bibfield  {title}
  {\bibinfo {title} {{Self-learning Monte Carlo method}},\ }\href@noop {}
  {\bibfield  {journal} {\bibinfo  {journal} {Phys. Rev. B}\ }\textbf {\bibinfo
  {volume} {{95}}},\ \bibinfo {pages} {041101} (\bibinfo {year}
  {2017})}\BibitemShut {NoStop}%
\bibitem [{\citenamefont {McNaughton}\ \emph {et~al.}(2020)\citenamefont
  {McNaughton}, \citenamefont {Milo\ifmmode \check{s}\else
  \v{s}\fi{}evi\ifmmode~\acute{c}\else \'{c}\fi{}}, \citenamefont {Perali},\
  and\ \citenamefont
  {Pilati}}]{McNaughtonMilosevicPeraliPilati2020BoostingMCSimulationsOfSGUsingAutoregressiveNeuralNetworks}%
  \BibitemOpen
  \bibfield  {author} {\bibinfo {author} {\bibfnamefont {B.}~\bibnamefont
  {McNaughton}}, \bibinfo {author} {\bibfnamefont {M.~V.}\ \bibnamefont
  {Milo\ifmmode \check{s}\else \v{s}\fi{}evi\ifmmode~\acute{c}\else
  \'{c}\fi{}}}, \bibinfo {author} {\bibfnamefont {A.}~\bibnamefont {Perali}},\
  and\ \bibinfo {author} {\bibfnamefont {S.}~\bibnamefont {Pilati}},\
  }\bibfield  {title} {\bibinfo {title} {{Boosting Monte Carlo simulations of
  spin glasses using autoregressive neural networks}},\ }\href@noop {}
  {\bibfield  {journal} {\bibinfo  {journal} {Phys. Rev. E}\ }\textbf {\bibinfo
  {volume} {{101}}},\ \bibinfo {pages} {053312} (\bibinfo {year}
  {2020})}\BibitemShut {NoStop}%
\bibitem [{\citenamefont
  {Wang}(2017)}]{Wang2017ExploringClusterMCUpdatesWithBoltzmannMachines}%
  \BibitemOpen
  \bibfield  {author} {\bibinfo {author} {\bibfnamefont {L.}~\bibnamefont
  {Wang}},\ }\bibfield  {title} {\bibinfo {title} {{Exploring cluster Monte
  Carlo updates with Boltzmann machines}},\ }\href@noop {} {\bibfield
  {journal} {\bibinfo  {journal} {Phys. Rev. E}\ }\textbf {\bibinfo {volume}
  {{96}}},\ \bibinfo {pages} {051301} (\bibinfo {year} {2017})}\BibitemShut
  {NoStop}%
\bibitem [{\citenamefont {Pei}\ and\ \citenamefont {{Di
  Ventra}}(2022)}]{PeiDiVentra2022NonEquilibriumCriticalityAndEfficientExplorationOfGlassyLandscapesWithMemoryDynamics}%
  \BibitemOpen
  \bibfield  {author} {\bibinfo {author} {\bibfnamefont {Y.~R.}\ \bibnamefont
  {Pei}}\ and\ \bibinfo {author} {\bibfnamefont {M.}~\bibnamefont {{Di
  Ventra}}},\ }\bibfield  {title} {\bibinfo {title} {{Non-equilibrium
  criticality and efficient exploration of glassy landscapes with memory
  dynamics}},\ }\href@noop {} {\bibfield  {journal} {\bibinfo  {journal}
  {Physica A}\ }\textbf {\bibinfo {volume} {{591}}},\ \bibinfo {pages} {126727}
  (\bibinfo {year} {2022})}\BibitemShut {NoStop}%
\bibitem [{\citenamefont
  {Krauth}(2004)}]{Krauth2004ClusterMonteCarloAlgorithms}%
  \BibitemOpen
  \bibfield  {author} {\bibinfo {author} {\bibfnamefont {W.}~\bibnamefont
  {Krauth}},\ }\bibfield  {title} {\bibinfo {title} {{Cluster Monte Carlo
  Algorithms}},\ }in\ \href@noop {} {\emph {\bibinfo {booktitle} {New
  Optimization Algorithms in Physics}}},\ \bibinfo {editor} {edited by\
  \bibinfo {editor} {\bibfnamefont {A.~K.}\ \bibnamefont {Hartmann}}\ and\
  \bibinfo {editor} {\bibfnamefont {H.}~\bibnamefont {Rieger}}}\ (\bibinfo
  {publisher} {Wiley-VCH},\ \bibinfo {address} {Weinheim},\ \bibinfo {year}
  {2004})\ p.~\bibinfo {pages} {7}\BibitemShut {NoStop}%
\bibitem [{\citenamefont {Bray}\ and\ \citenamefont
  {Moore}(1980)}]{BrayMoore1980SomeObservationsOnTheMeanFieldTheoryOfSG}%
  \BibitemOpen
  \bibfield  {author} {\bibinfo {author} {\bibfnamefont {A.~J.}\ \bibnamefont
  {Bray}}\ and\ \bibinfo {author} {\bibfnamefont {M.~A.}\ \bibnamefont
  {Moore}},\ }\bibfield  {title} {\bibinfo {title} {Some observations on the
  mean-field theory of spin glasses},\ }\href@noop {} {\bibfield  {journal}
  {\bibinfo  {journal} {J. Phys. C}\ }\textbf {\bibinfo {volume} {{13}}},\
  \bibinfo {pages} {419} (\bibinfo {year} {1980})}\BibitemShut {NoStop}%
\bibitem [{\citenamefont {Contucci}\ \emph {et~al.}(2003)\citenamefont
  {Contucci}, \citenamefont {Esposti}, \citenamefont {Giardina},\ and\
  \citenamefont
  {Graffi}}]{ContucciEtAl2003ThermodynamicalLimitForCorrelatedGaussianRandomEnergyModels}%
  \BibitemOpen
  \bibfield  {author} {\bibinfo {author} {\bibfnamefont {P.}~\bibnamefont
  {Contucci}}, \bibinfo {author} {\bibfnamefont {M.~D.}\ \bibnamefont
  {Esposti}}, \bibinfo {author} {\bibfnamefont {C.}~\bibnamefont {Giardina}},\
  and\ \bibinfo {author} {\bibfnamefont {S.}~\bibnamefont {Graffi}},\
  }\bibfield  {title} {\bibinfo {title} {Thermodynamical limit for correlated
  {G}aussian random energy models},\ }\href@noop {} {\bibfield  {journal}
  {\bibinfo  {journal} {Comm. Math. Phys.}\ }\textbf {\bibinfo {volume}
  {{236}}},\ \bibinfo {pages} {55} (\bibinfo {year} {2003})}\BibitemShut
  {NoStop}%
\bibitem [{\citenamefont {Weigel}\ and\ \citenamefont
  {Janke}(2009)}]{weigel:09}%
  \BibitemOpen
  \bibfield  {author} {\bibinfo {author} {\bibfnamefont {M.}~\bibnamefont
  {Weigel}}\ and\ \bibinfo {author} {\bibfnamefont {W.}~\bibnamefont {Janke}},\
  }\bibfield  {title} {\bibinfo {title} {Cross correlations in scaling analyses
  of phase transitions},\ }\href
  {https://doi.org/10.1103/PhysRevLett.102.100601} {\bibfield  {journal}
  {\bibinfo  {journal} {Phys. Rev. Lett.}\ }\textbf {\bibinfo {volume} {102}},\
  \bibinfo {pages} {100601} (\bibinfo {year} {2009})}\BibitemShut {NoStop}%
\end{thebibliography}%
\end{document}